\documentclass{aa}
\usepackage[varg]{txfonts}
\bibpunct{(}{)}{;}{a}{}{,}
\usepackage{graphicx}

\begin{document}

\title{Near-infrared transmission spectrum of TRAPPIST-1 h using Hubble WFC3 G141 observations}

\author{A.~Gressier\inst{\ref{inst1},\ref{inst2},\ref{inst3}}\and M.~Mori\inst{\ref{inst4}}\and Q.~Changeat\inst{\ref{inst5}}\and B.~Edwards\inst{\ref{inst5},\ref{inst6}}\and J.P.~Beaulieu\inst{\ref{inst2},\ref{inst7}}\and E.~Marcq\inst{\ref{inst1}}\and B.~Charnay\inst{\ref{inst3}}}

\institute{LATMOS, CNRS, Sorbonne Universit\'e UVSQ, 11 boulevard d’Alembert, F-78280 Guyancourt, France \email{amelie.gressier@latmos.ipsl.fr}\label{inst1}
\and
Sorbonne Universit\'es, UPMC Universit\'e Paris 6 et CNRS, UMR 7095, Institut d'Astrophysique de Paris, 98 bis bd Arago, 75014 Paris, France\label{inst2}
\and
LESIA, Observatoire de Paris, Universit\'e PSL, CNRS, Sorbonne Universit\'e, Universit\'e de Paris, 5 place Jules Janssen, 92195 Meudon, France\label{inst3}
\and
Department of Astronomy, University of Tokyo, Tokyo, Japan\label{inst4}
\and
Department of Physics and Astronomy, University College London, London, United Kingdom\label{inst5}
\and
AIM, CEA, CNRS, Universit\'e Paris-Saclay, Universit\'e de Paris, F-91191 Gif-sur-Yvette, France\label{inst6}
\and
School of Physical Sciences, University of Tasmania, Private Bag 37 Hobart, Tasmania 7001 Australia\label{inst7} }

\date{Received 3 September 2021/ Accepted 27 November 2021}

\abstract {The TRAPPIST-1 planetary system is favourable for transmission spectroscopy and offers the unique opportunity to study rocky planets with possibly non-primary envelopes. We present here the transmission spectrum of the seventh planet of the TRAPPIST-1 system, TRAPPIST-1 h (R$_{\rm P}$=0.752 R$_{\oplus}$, T$_{\rm eq}$=173K) using Hubble Space Telescope (HST), Wide Field Camera 3 Grism 141 (WFC3/G141) data.} 
{Our purpose is to reduce the HST observations of the seventh planet of the TRAPPIST-1 system and, by testing a simple atmospheric hypothesis, to put a new constraint on the composition and the nature of the planet.} 
{First we extracted and corrected the raw data to obtain a transmission spectrum in the near-infrared (NIR) band (1.1-1.7$\mu$m). TRAPPIST-1 is a cold M-dwarf and its activity could affect the transmission spectrum. We corrected for stellar modulations using three different stellar contamination models; while some fit the data better, they are statistically not significant and the conclusion remains unchanged concerning the presence or lack thereof of an atmosphere. Finally, using a Bayesian atmospheric retrieval code, we put new constraints on the atmosphere composition of TRAPPIST-1h.}{According to the retrieval analysis, there is no evidence of molecular absorption in the NIR spectrum. This suggests the presence of a high cloud deck or a layer of photochemical hazes in either a primary atmosphere or a secondary atmosphere dominated by heavy species such as nitrogen. This result could even be the consequence of the lack of an atmosphere as the spectrum is better fitted using a flat line. Variations in the transit depth around 1.3$\mu$m are likely due to remaining scattering noise and the results do not improve while changing the spectral resolution. TRAPPIST-1 h has probably lost its atmosphere or possesses a layer of clouds and hazes blocking the NIR signal. We cannot yet distinguish between a primary cloudy or a secondary clear envelope using HST/WFC3 data; however, in most cases with more than 3$\sigma$ confidence, we can reject the hypothesis of a clear atmosphere dominated by hydrogen and helium. By testing the forced secondary atmospheric scenario, we find that a CO-rich atmosphere (i.e. with a volume mixing ratio of 0.2) is one of the best fits to the spectrum with a Bayes factor of 1.01, corresponding to a 2.1$\sigma$ detection.} {}


\maketitle
\section{Introduction}
The TRAPPIST-1 planetary system was discovered by \citet{Gillon_2016} and \citet{Gillon_2017}, using the Transiting Planets and PlanetIsimals Small Telescope \citep{Gillon_2011, Gillon_2013}. TRAPPIST-1 h is the most outer planets detected in this system, its detection was first suggested in \citet{Gillon_2017}, but later confirmed in \citet{Luger_2017b}. Further observations using Spitzer and K2 photometry followed the discovery to better constrain planetary parameters \citep{Delrez_2018, Ducrot_2018, Burdanov_2019, Ducrot_2020}. Since then, important scientific efforts have been carried out to observe, characterise, and model the seven planets orbiting this M8-type star. This is motivated by the fact that the TRAPPIST-1 system offers the most favourable conditions to study rocky planets in the habitable zone, that is to say planets that could harbour liquid water on their surface as defined in \citet{Kasting_1993}.

TRAPPIST-1 is close (39.14 light years), cool (2559 K), and small (0.117 R$_\odot$), making it favourable for observations \citep{Gillon_2017}. On the other hand, the star is also the limiting factor in studying the atmosphere of TRAPPIST-1 planets. M-type stars stay for millions of years in the pre-main sequence (PMS) phase, during which planets are exposed to strong non-thermal extreme UV (EUV) and far-UV irradiation, which is expected to lead to atmospheric hydrodynamical escape \citep{Vidal-Madjar_2003, Bourrier_2017a} and a runaway greenhouse effect \citep{Ramirez_2014}. TRAPPIST-1 is a very cold M-dwarf, but it is supposedly very active with strong flaring events \citep{Vida_2017} and EUV flux \citep{Wheatley_2017}. Atmospheric erosion might have stripped all planets in the TRAPPIST-1 system of their atmospheres \citep{lammer_2003, Bolmont_2017a}. Whether or not an atmosphere was sustained depends on the initial amount of accreted volatiles during the planetary formation phase, and the intensity of the atmospheric escape due to the star activity.

\begin{table}
    \caption{Stellar and planetary parameters used in this work}
    \centering
    \begin{tabular}{l | c} \hline \hline
    Parameter & Value \\ \hline 
    Spectral type & M8-V \\
    R$_{\rm s}$ (R$_\odot$) & 0.1170 $\pm$ 0.0036 \\
    M$_{\rm s}$ (M$_\odot$) & 0.0802 $\pm$ 0.0073 \\
    T$_{\rm s}$ (K) & 2559 $\pm$ 50 \\
    $\log$(g) & 5.21 \\
    Fe/H & 0.040 $\pm$ 0.080 \\ \hline
    R$_{\rm p}$ (R$_{\oplus}$) & 0.752 $\pm$ 0.032\\
    M$_{\rm p}$ (M$_{\oplus}$) & 0.331 $^{+0.056}_{-0.049}$ \\
    a (AU) & 0.059 $\pm$ 0.004\\ 
    T$_{\rm eff}$ (K) & 173 $\pm$ 4\\
    S (S$_\oplus$) & 0.165 $\pm$ 0.025 \\ 
    a/R$_{\rm s}$ & 109 $\pm$ 4 \\
    i (deg) & 89.76$^{+0.05}_{-0.03}$\\
    e\tablefootmark{a} & 0  \\
    b & 0.45 \\
    P$_{\rm orb}$ (days) & 18.767 $^{+0.004}_{-0.003}$\\
    T$_{\rm mid}$ (BJD$_{\rm TDB}$) & 2\,458\,751.06983 $\pm$ 0.00021\tablefootmark{b} \\ \hline
    \end{tabular}
    \tablefoot{Values are from \citet{Gillon_2017} and \citet{Luger_2017a}.
    \tablefoottext{a}{Fixed to zero}
    \tablefoottext{b}{Obtained in this work}}
    \label{table1:parameter}
\end{table}

The TRAPPIST-1 planetary system is very compact, all the planets are within 0.06 AU and they are co-planar \citep{Luger_2017a, Luger_2017b, Delrez_2018}. In addition to this, they all have a circularised orbit with eccentricities below 0.01 \citep{Gillon_2017, Luger_2017b} and present gravitational interactions forming a resonant chain, thus suggesting that the system had a relatively peaceful history. TRAPPIST-1 h is the furthest and the smallest known planet of this planetary system. It has a radius of 0.752$\pm$0.032 R$\oplus$ and a mass of 0.331$^{+0.056}_{-0049}$ M$\oplus$ \citep{Luger_2017b, Gillon_2017}, which suggests a density similar to that of Mars ($\sim$ 4000 kg/m$^3$). The planetary parameters are detailed in Table \ref{table1:parameter} along with stellar and orbital parameters of the system. 

Two possible formation scenarios have been proposed for the TRAPPIST-1 system and in particular for TRAPPIST-1 h. The first one suggests that all the planets that formed beyond the water frost line migrated inwards, causing the resonance, and they are now located between planets g and h. This possibility was proposed in the discovery papers \citet{Gillon_2017} and \citet{Luger_2017b}, but also detailed in \citet{Ormel_2017}, \citet{Tamayo_2017}, and \citet{Coleman_2019}. If TRAPPIST-1 h formed far from the host star, it could be volatiles-rich because the atmospheric escape would only remove between 1 and 10\% of the total planet mass \citep{Tian_Ida_2015, Bolmont_2017a, Bourrier_2017a, Turbet_2020b}. TRAPPIST-1 h could also have formed in situ, and a short migration or an eccentricity damping could have caused the resonant chain \citep{MacDonald_2018}. In this case, the planet is probably dry \citep{Turbet_2020b} because of the strong atmospheric erosion. 

On the other hand, TRAPPIST-1 h, being the furthest planet of the system, might have had a more important quantity of initial gas than inner planets. It could have formed with TRAPPIST-1 f and g in a different part of the proto-planetary disk leading to a different bulk composition \citep{Papaloizou_2018, Turbet_2020b}. Volatiles could also have been brought after by cometary impacts or degassing \citep{Kral_2018, Dencs_2019, Turbet_2020b, Kimura_2020}, and this is favoured for outer planets because volatiles' impacts dominate over the impact erosion mechanism \citep{Kral_2018}. 

For close-in planetary systems, the effects of gravitational tides by the star on the planets are important and shape the orbital dynamics, that is to say they slow down the rotation rate, reduce the obliquity, and circularise the orbit. As shown in \citet{Turbet_2018}, the evolution timescales for TRAPPIST-1 h is 7 million years for the rotation and 80 million years for the obliquity. Knowing the age of the TRAPPIST-1 system, which is 8 billion years \citep{Burgasser_2017}, it is likely that TRAPPIST-1 h is in a synchronous rotation state. However, tidal heating is unlikely to be the dominant interior heating process for outer planets \citep{Turbet_2018, Makarov_2018, Dobos_2019} as compared with direct atmospheric warming. The received stellar flux is indeed two orders of magnitude higher than the tidal heating for TRAPPIST-1 h \citep{Turbet_2020b}. It is then unlikely that TRAPPIST-1 h tidal heating caused the melting of the mantle leading to the out-gassing of volcanic gases \citep{Turbet_2020b}.

As of today, TRAPPIST-1 h is the only planet of the system for which the near-infrared (NIR) spectrum (1.1-1.7$\mu$m) from the Hubble Space Telescope (HST) Wide Field Camera 3 Grism 141 (WFC3/G141) has not been published. The other planets' spectra have already been studied with different pipelines and stellar contamination models in \citet{de_Wit_2016} and \citet{de_Wit_2018}, \citet{Zhang_2018}, and \citet{Wakeford_2019}. From these analyses, we learned that the TRAPPIST-1 planets probably do not have an H$_2$, He extended atmosphere. However, it was impossible to rule out this hypothesis using only HST/WFC3 \citep{de_Wit_2018,Moran_2018}. All spectra are consistent with flat spectra and could be fitted with different models including a high altitude cloud cover and/or a high metallicity hydrogen-rich atmosphere. A featureless spectrum could also be the result of the absence of an atmosphere around these planets. However, \citet{Bourrier_2017a} and \citet{Bourrier_2017b} analysed Lyman-$\alpha$ HST/STIS transits of TRAPPIST-1 b and c and detected a decrease in the flux, which might hint at the presence of an extended hydrogen exosphere. 

We present the first attempt to characterise the atmosphere of the seventh planet of the system, TRAPPIST-1 h. In Sec. \ref{sec:1.1}, we analyse the HST/WFC3 G141 raw data using the python package Iraclis \citep{Tsiaras_2016b,Tsiaras_2016a, Tsiaras_2018} and detail the stellar contamination models used to correct our spectrum in Sec. \ref{sec:1.2}. Section \ref{sec:1.3} presents the atmospheric characterisation of TRAPPIST-1 h. First, different atmospheric scenarios are discussed based on the recent review by \citet{Turbet_2020b}. Then, we detail the atmospheric retrieval set-ups we performed using the Bayesian radiative transfer code TauREx3 \citep{Alrefaie2019}\footnote{\url{https://github.com/ucl-exoplanets/TauREx3_public}}. Finally, we discuss our findings in Sec. \ref{sec:3}. 

\section {Data analysis}\label{sec:1}
\subsection{Hubble WFC3 data reduction and extraction}\label{sec:1.1}
We used the raw spatially scanned spectroscopic images obtained from Proposal 15\,304 (PI: Julien de Wit) in the Mikulski Archive for Space Telescope\footnote{\url{https://archive.stsc"i.e"du/hst/}}. Three transit observations of TRAPPIST-1 h were acquired using the Grism 256 aperture and 256 x 256 sub-array with an exposure time of 112.08 s. We refer to the data taken in July 2017, September 2019, and July 2020 as Observations 1, 2, and 3, respectively. Each visit is made up of four HST orbits, with 60 exposures in Observation 2 and 50 exposures for Observations 1 and 3, each being made in the forward spatial scan mode.

\begin{figure}
    \centering
    \resizebox{\hsize}{!}{\includegraphics{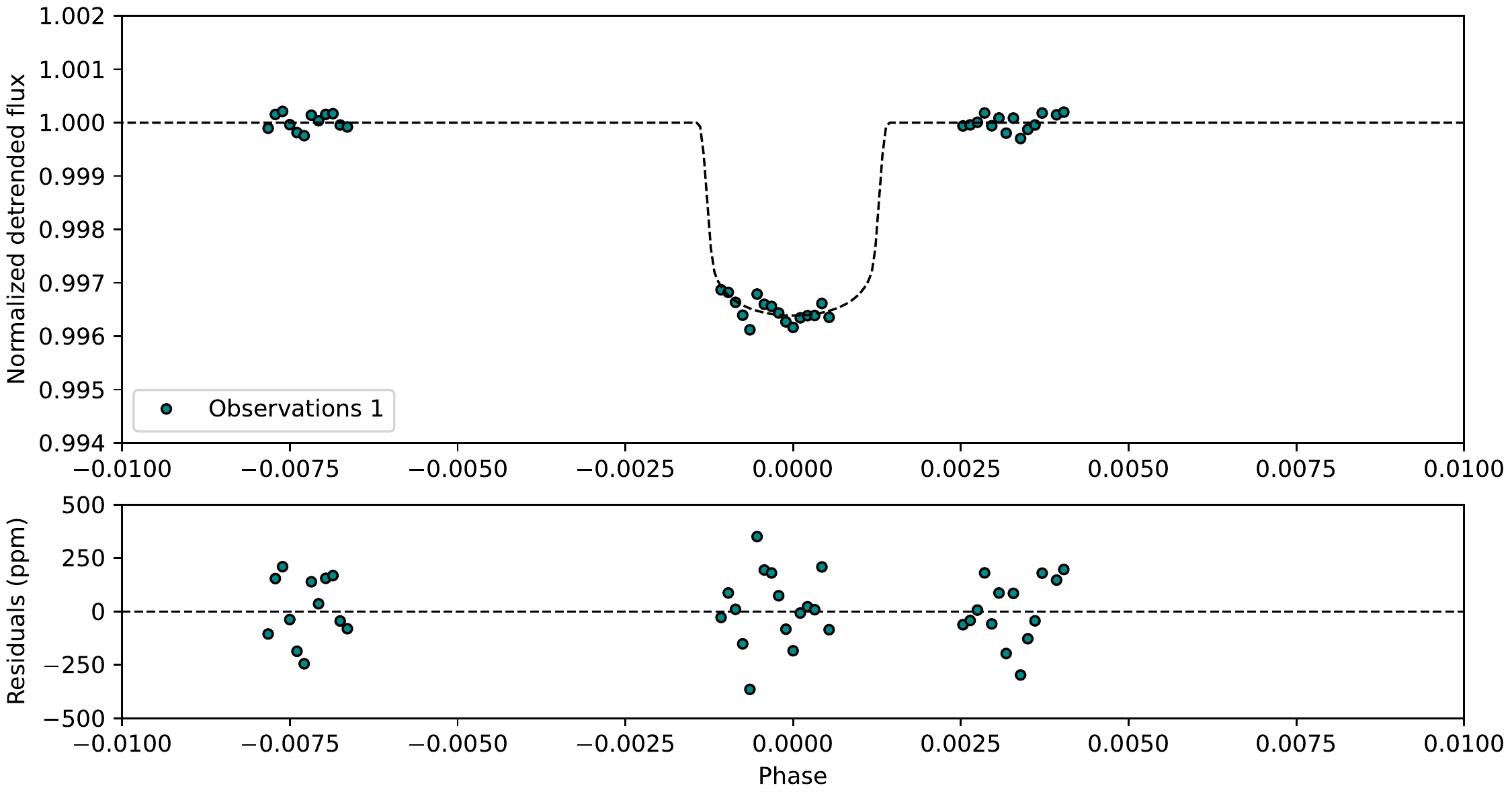}}
    \resizebox{\hsize}{!}{\includegraphics{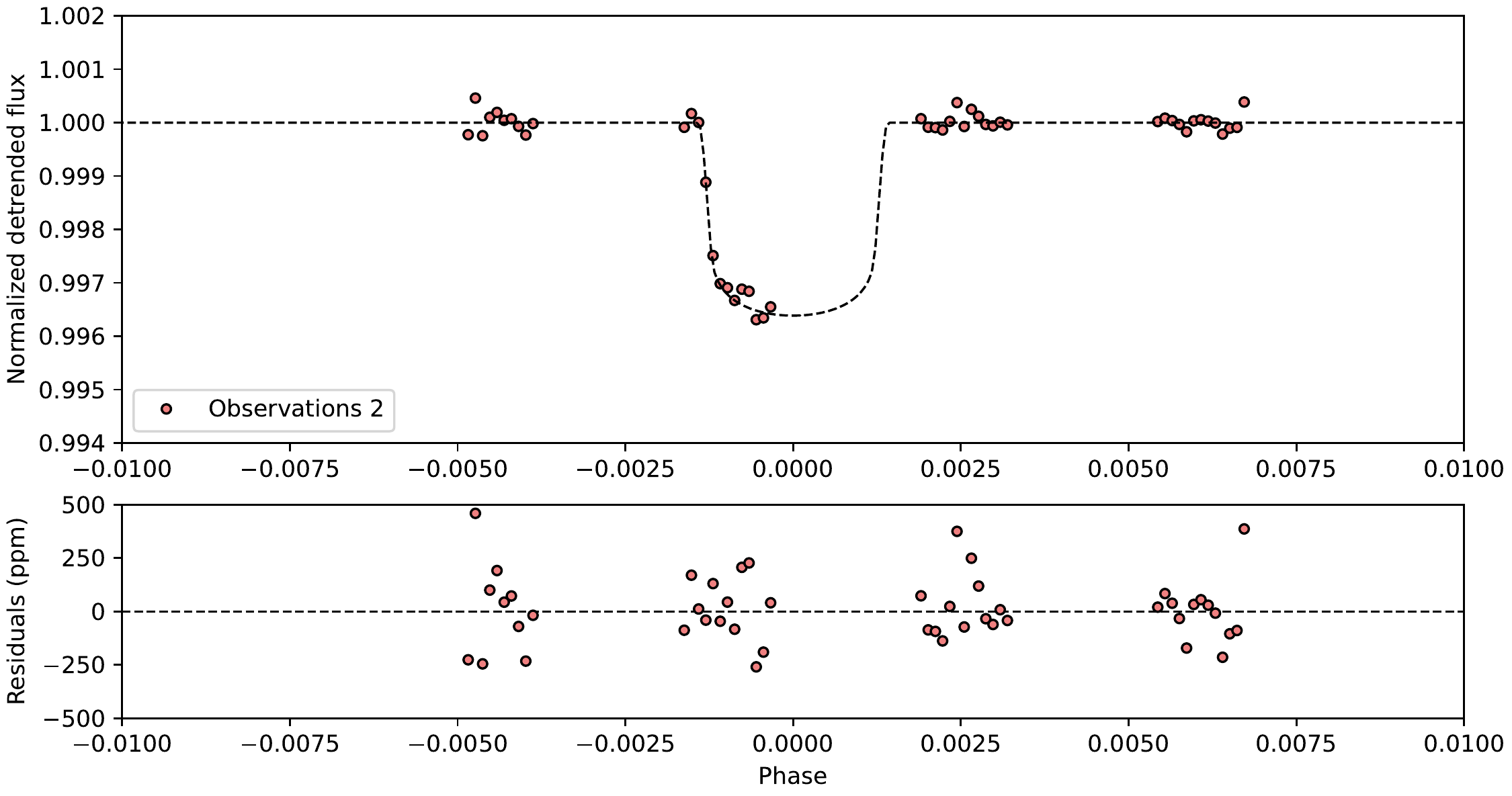}}
    \resizebox{\hsize}{!}{\includegraphics{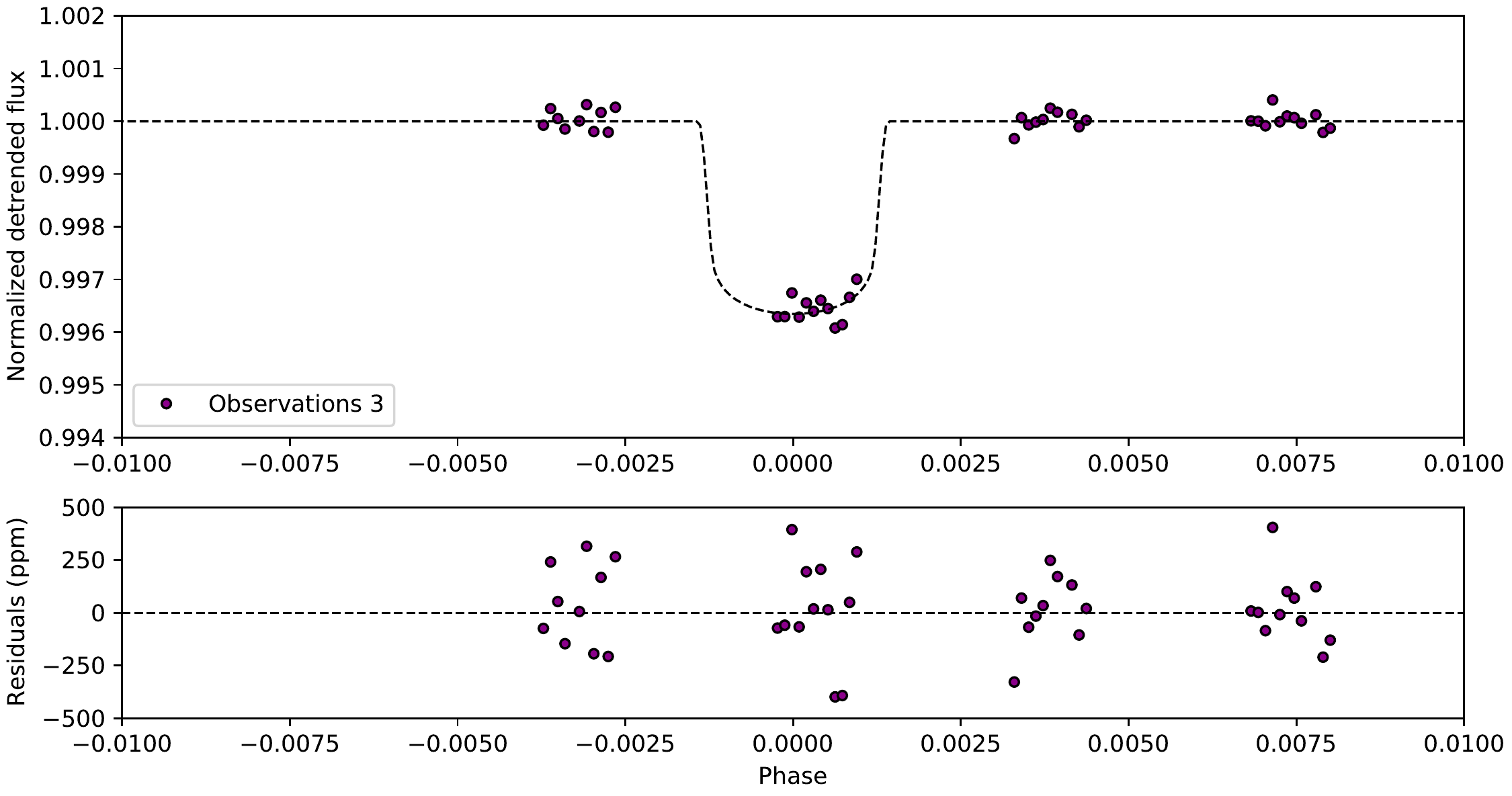}}
    \caption{White light curve fits for the three visits on TRAPPIST-1h (top: July 2017, middle: September 2019, and bottom: July 2020). For every observation, we show the de-trended flux (colour points) and the best fit model (dotted lines) along with the residuals from the best fit model.}
    \label{fig:wlc_fitting}
\end{figure}

\begin{figure*}[htpb]
    \centering
    \includegraphics[width=17cm]{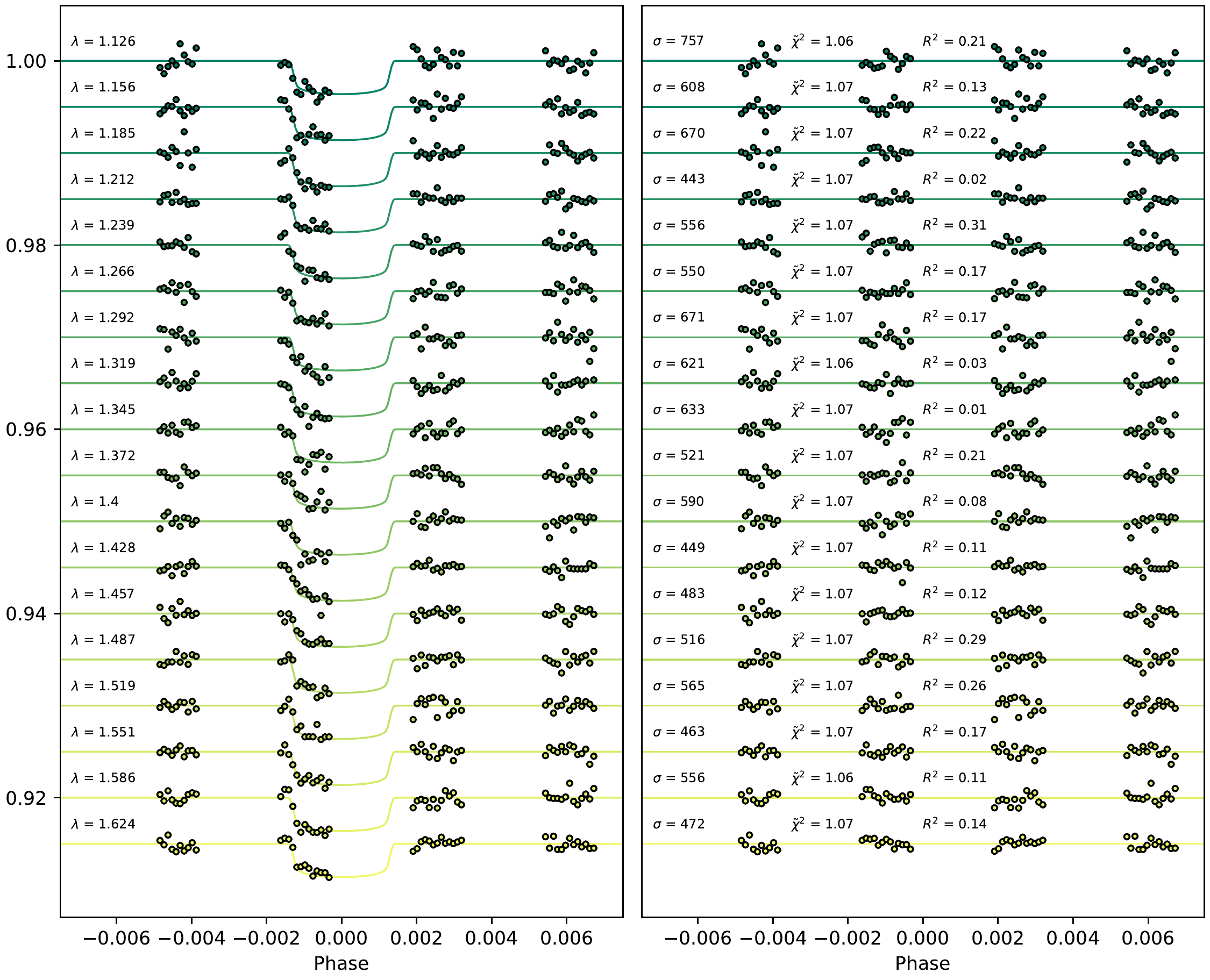}
    \caption{Spectral light curve fits of the September 2019 visit (Observation 2) for the transmission spectra of TRAPPIST-1 h. An artificial offset in the y-axis was applied for clarity. For each light curve, the left panel shows the de-trended spectral light curves with the best fit model in dotted lines with the centred wavelength and the right panel shows the residuals and values for the standard deviation ($\sigma$) in ppm, the reduced Chi-squared ($\tilde{\chi}^2$), and the auto-correlation (R$^2$).}
    \label{fig:spclc_obs2}
\end{figure*}

\begin{table*}[htpb!]
    \caption{Combined transit depth, associated uncertainties and limb-darkening coefficients.}
    \centering
    \begin{tabular}{cccccccc} \hline \hline
    Wavelength ($\mu$m)   & Bandwidth ($\mu$m)  & Transit depth (ppm) & Error (ppm) &  \multicolumn{4}{c}{Limb-darkening coefficients}\\ 
       &   &  &  & a1 & a2 & a3 & a4 \\
    \hline 
    1.1262 & 0.0308 & 3128.22 & 129.30 & 2.0139 &-1.6261 & 0.8709 &-2.0398\\
    1.1563 & 0.0293 & 2981.61 & 132.61 & 2.1956 & -2.0725 & 1.2403 & -0.3147 \\
    1.1849 & 0.0279 & 3224.87 & 121.09 & 2.1292 &       -1.9036 & 1.0964 &-0.2708\\
    1.2123 & 0.0269 & 3275.06 & 112.69 & 1.9514 & -1.5303 & 0.8020 & -0.1849\\
    1.2390 & 0.0265 & 3476.20 & 95.97 & 1.9236 &        -1.5957 & 0.8838 & -0.2137\\
    1.2657 & 0.0269 & 3264.82 & 95.65 & 2.0255 & -1.8405 & 1.0765 & -0.2698\\
    1.2925 & 0.0267 & 3589.24 & 115.10 & 2.1105 &       -2.1495 & 1.3561 & -0.3578 \\
    1.3190 & 0.0263 & 3686.39 & 110.98 & 2.1650 & -2.2486 & 1.4262 &-0.3772\\
    1.3454 & 0.0265 & 3368.20 & 126.87 & 1.2204 &       -0.1088 & -0.1857 &       0.0789\\
    1.3723 & 0.0274 & 2900.07 & 108.34 & 1.0023 &       0.4493 & -0.6644 & 0.2195\\
    1.4000 & 0.0280 & 3271.69 & 109.38 & 0.9553 &       0.4582 &-0.6187 & 0.1988\\
    1.4283 & 0.0285 & 3321.59 & 103.01 & 0.7774 &       0.7086 &-0.7252 & 0.2128\\
    1.4572 & 0.0294 & 3111.09 & 113.41 & 0.9247 & 0.4694 &-0.6071 & 0.1921\\
    1.4873 & 0.0308 & 3070.67 & 113.98 & 1.0279 &       0.2998 &-0.530181 &       0.18063\\
    1.5186 & 0.0318 & 3037.95 & 112.45 & 1.2541 &       -0.1103 & -0.2727 & 0.1188\\
    1.5514 & 0.0337 & 3125.30 & 102.78 & 1.5025 & -0.6408 & 0.1247 &    0.0082\\
    1.5862 & 0.0360 & 3472.00  & 114.20 & 1.7942 & -1.3368 &    0.6809 &-0.1553\\
    1.6237 & 0.0390 & 3045.52 & 95.82 & 1.9296 & -1.7566 & 1.0358 & -0.2629 \\ \hline
    1.3750 & 0.5500 & 3268.70 & 51.38 & 2.009 & -1.7704 & 1.0225 &-0.2546 
 \\\hline 
    \end{tabular}
    \tablefoot{ The final transmission spectrum was computed in ppm using the three HST/WFC3 G141 transit observations from July 2017, September 2019, and July 2020 on TRAPPIST-1 h.}
    \label{table2:spectrum}
\end{table*}
To reduce and analyse the data, we used Iraclis \footnote{\url{https://github.com/ucl-exoplanets/Iraclis}} \citep{Tsiaras_2016b,Tsiaras_2016a, Tsiaras_2018}, a publicly available pipeline, dedicated to the analysis of the scanned spectroscopic observations obtained with the near-infrared grisms (G102, G141) of Hubble's Wide Field Camera 3. The reduction of the raw observations follows these steps: zero-read subtraction, reference pixels correction, non-linearity correction, dark current subtraction, gain conversion, sky background subtraction, flat-field correction, and corrections for bad pixels and cosmic rays. For all three observations, we used the reduced spatially scanned spectroscopic images to extract the white and spectral light curves. We used the default 'low' resolution from Iraclis for the spectral light curves bins, which correspond to a resolving power of around 50 at 1.4${\rm \mu}$m. 

Using the extracted light curves and the time of the observations, we first looked for contamination from others TRAPPIST-1 planets transits using the python package PyLightcurve \citep{Tsiaras_2016a} \footnote{\url{https://github.com/ucl-exoplanets/pylightcurve}}. The planets and transit parameters were set to those of  \citet{Gillon_2017}. TRAPPIST-1 c was also transiting during the second orbit of the first observation (July 2017) and we then suppressed this orbit from the rest of the analysis. We plot in the appendix \ref{appendix:predicted_transits} the extracted raw flux and the corresponding predicted transits of TRAPPIST-1 planets for the three visits. 

The first orbit always presents a stronger wavelength-dependent ramp than the other orbits and is usually suppressed from the analysis. However, we decided to keep the first HST orbit in every transit observation in order to conserve an out-of-transit baseline and correctly fit the transit parameters. Indeed, every attempt was made to keep as many exposures as possible. For Observations 1 and 2, we removed the first two exposures of these first orbits, but kept all exposures of every subsequent orbit. However, for Observation 3, an adequate fit could only be obtained by removing the first exposure of every orbit, a practice which is normal as these exposures present significantly lower counts than the following exposures \citep[e.g.][]{Deming_2013, Tsiaras_2016b, Edwards_2021}.

\begin{figure}
    \centering
    \resizebox{\hsize}{!}{\includegraphics{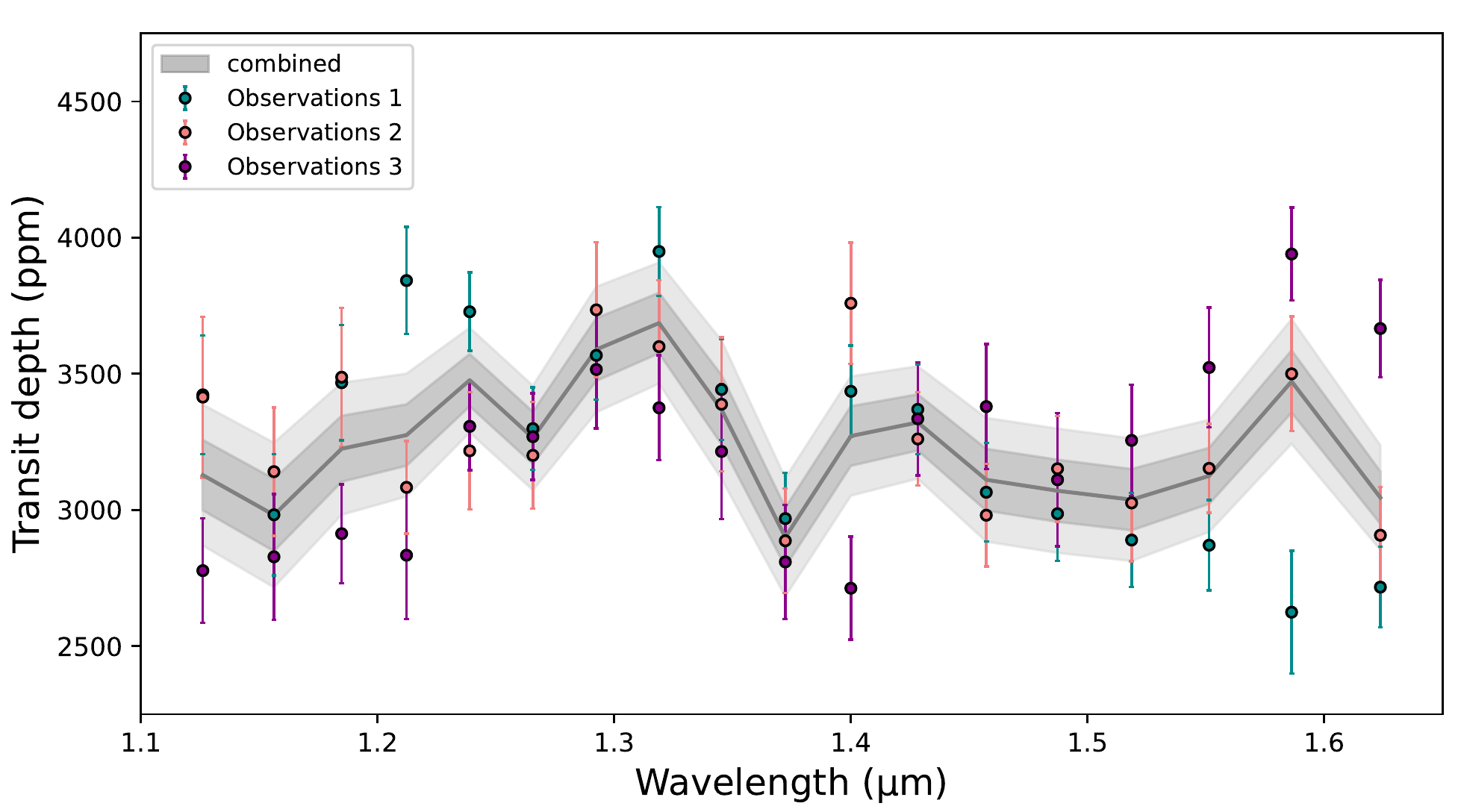}}
    \resizebox{\hsize}{!}{\includegraphics{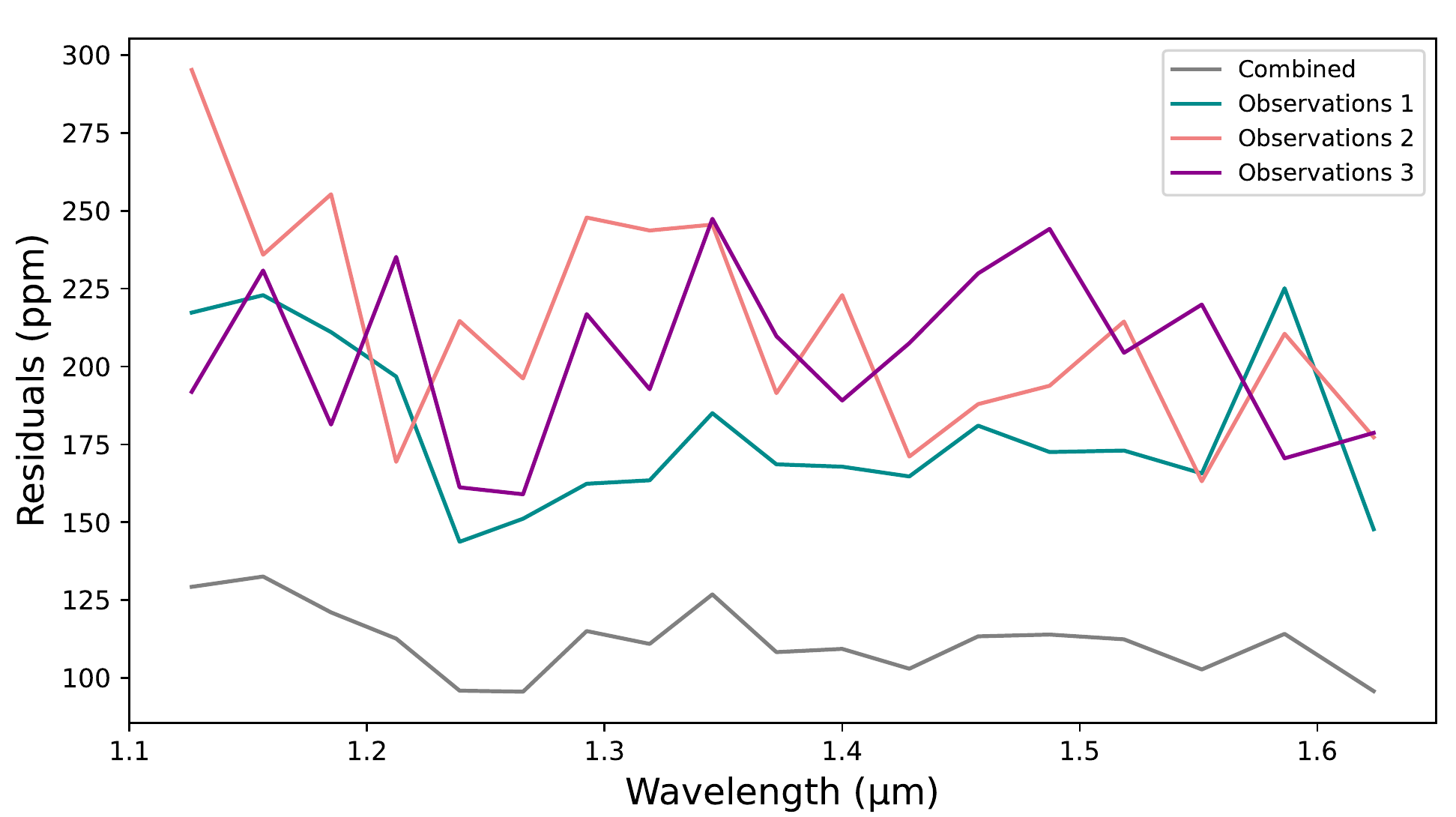}}
    \caption{Recovered transit depths for the three observations and combined transmission spectrum with 1 and 2$\sigma$ uncertainty ranges (top). First, we suppressed the white light curve values from each visit raw flux, then, we computed the weighted mean, and finally we added the mean white light curve value to obtain the transit depth. Residuals are from the spectral light curves analysis and the combined spectrum (bottom).}
    \label{fig:all_transits}
\end{figure}

We fitted the white light curves and the spectral light curves using the transit model from PyLightcurve \citep{Tsiaras_2016a} and the Markov chain Monte Carlo (MCMC) method implemented in emcee \citep{Foreman_Mackey_2013}. For the white light curve fitting of all the observations, the only free planetary parameters are the mid-transit time and the planet-to-star radius ratio. The other planetary parameters were fixed to the values from \citet{Luger_2017a} (a/R$_{\rm s}$=109$\pm$4 and i=89.76$^{\circ}$) and stellar parameters are from \citet{Gillon_2017} (T$_{\rm s}$=2559$\pm$50 K, log(g)=5.21, Fe/H=0.04). We also fitted for the coefficients ra, r$b_1$, and r$b_2$. We adopted the parameterisation of \citet{Claret_2012} and \citet{Claret_2013} with four parameters to describe the limb-darkening coefficient. We used the PHOENIX database \citep{Claret_2018} and  ExoTETHyS package \citep{Morello_2020} to obtain the limb-darkening coefficients for the white light curve analysis but also in every wavelength  bin for the spectral curves fitting (see Table \ref{table2:spectrum}).
We accounted for the ramp time-dependent systematic effect in the white light curve fitting using the following formula with t being the time, t$_0$ the beginning time of each HST orbit, T$_0$ the mid-transit time, n$_{\rm scan}$ a normalisation factor, ra the slope of a linear systematic trend, and (rb$_1$ , rb$_2$ ) the coefficients of the exponential systematic trend along each HST orbit:
\begin{equation}
R_w(t)=n_w^{scan}(t)(1-r_a(t-T_0))(1-r_{b_1}e^{r_{b_2}(t-t_0))}
        \label{eq1}
.\end{equation}
We then fitted for the planet-to-star radius ratio in every wavelength band. We used the white light curve divide method \citep{Kreidberg_2014a} along with a spectral-dependent visit-long slope \citep{Tsiaras_2018} model to account for the systematic effects as follows, with $\chi_\lambda$ being the slope for the wavelength-dependent systematic effects along each orbit, LC$_{\rm w}$ the white light curve signal, and M$_{\rm w}$ the white light curve best fit model:
\begin{equation}
R_\lambda(t)=n_\lambda^{scan}(t)(1-\chi_\lambda(t-T_0))\frac{LC_w}{M_w}
        \label{eq2}
.\end{equation}
The white light curve fits for the three different observations are shown in Fig. \ref{fig:wlc_fitting}. The planet-to-star radius ratio are found to be compatible with 0.0575$\pm$0.0006 for Observation 1, 0.0565$\pm$0.0009 for Observation 2, and 0.0575$\pm$0.0012 for Observation 3. We found the following mid-transit times in BJD$_{\rm TDB}$: 2\,458\,319.4282 $\pm$ 0.00020 for Observation 1, 2\,458\,751.06983 $\pm$ 0.00021 for Observation 2, and 2\,459\,051.3428 $\pm$ 0.00021 for Observation 3. The spectral light curve fit for the second observation is presented in Fig. \ref{fig:spclc_obs2}, while the two other spectral light curve fits are in appendix \ref{appendix:spcl_obs2_obs3}. We computed the final transmission spectrum by combining the three spectral fits using a weighted mean of the transmission spectra. After the initial white light curve fit, the errors on each exposure were scaled to match the root mean square of the residuals. The white fitting was then performed a second time with these scaled errors. A similar scaling was also applied to the spectral light curves. This method ensures that the recovered uncertainties on the transit depth are not underestimated \citep{Tsiaras_2016b}. The transmission spectra and the recovered final transit depth are overplotted in Fig. \ref{fig:all_transits}, along with the corresponding residuals. We note a rise in the transit depth around 1.3$\mu$m. All three observations exhibit similar features over these regions, suggesting this is of astrophysical origin and part of the transit spectrum and not a contamination, or poor fitting, of a single visit. We also present in appendix \ref{appendix:wlc_spcl_all} the three white light curve fits in the same plot using a planet-to-star radius ratio weighted by the mean of the three white light curve best fits for the transit model. The combined extracted spectrum and the uncertainties are presented in Table \ref{table2:spectrum}.

\subsection{Modelling the stellar contamination}\label{sec:1.2}
TRAPPIST-1 is known for presenting a heterogeneous photosphere that can lead to a misinterpretation of the transmission spectra. The goal of this section is to use different existing models to correct our spectrum for a stellar contribution. The star presents a $\sim$ 1$\%$ photometric variability in the I+z bandpass interpreted as active regions rotating in and out of view \citep{Gillon_2016}. 
\citet{Rackham_2018} show that it would cause a non-negligible effect (transit light source effect, TLSE) on the transmission spectrum if the variability is consistent with rotational modulations. Several previous studies have examined the stellar surface models using a variety of methods, but their results are not consistent. In the present study, three stellar models from three studies, \citet{Zhang_2018}, \citet{Morris_2018a}, and \citet{Wakeford_2019}, were introduced and examined. 

Table \ref{tab:stellar_model} shows the temperature and the covering fraction of each component for each model. We note that $T_i$ is the temperature, $f_i$ is the covering fraction at the photosphere, and $f'_i$ is the covering fraction at the transit chord. The M18 model is the best fit model from \citet{Morris_2018a}. Z18 is the best fit contamination model taken from Table 16 in \citet{Zhang_2018}. W19 is the $3T_{c+m}$ model from \citet{Wakeford_2019}. We note that what we call the W19 model here is not the best fit model in their analysis, as they conclude that TLSE is not significant in their data, but they did not exclude $3T_{c+m}$. 
\begin{table}[htpb!]
    \caption{Summary of the adopted TRAPPIST-1 stellar models.}
    \centering
    \begin{tabular}{c|ccc}\hline \hline
        Model & Z18 & M18 & W19 \\ \hline
        $T_1$(K) & $2000$ & $2500$ & $2400$ \\      
        $T_2$(K) & $2400$ & $5300$ & $3000$ \\
        $T_3$(K) & $3000$ & $-$ & $5800$ \\ \hline
        $f_1$& $0.38$ & $0.999952$ & $0.64$ \\
        $f_2$ &$0.14$ & $4.8\times10^{-5}$ & $0.35$\\\hline
        $f'_1$ & $0.10$ & $1.0$ & $0.646$\\
        $f'_2$ &  $0.45$ & $0.0$ & $0.354$\\\hline
    \end{tabular}
    \label{tab:stellar_model}
\end{table}

We define the wavelength-dependent contamination factor $\varepsilon_\lambda$ as
\begin{equation}
    \delta_\lambda = \varepsilon_\lambda \times \delta_{real,\lambda}
,\end{equation}
where $\delta_\lambda$ is the measured transit depth and $\delta_{real,\lambda}$ is the actual transit depth. For each stellar surface model, $\varepsilon_\lambda$ was calculated as  
\begin{eqnarray}
    \varepsilon_\lambda &= \frac{f'_1 S_{1,\lambda} + f'_2 S_{2,\lambda} + f'_3 S_{3,\lambda}}{f_1 S_{1,\lambda} + f_2 S_{2,\lambda} + f_3 S_{3,\lambda}}\\
    f_3 &= 1-f_1-f_2 \\
    f'_3 &= 1-f'_1-f'_2
,\end{eqnarray}
where $S_{i,\lambda}$ is the stellar flux of each temperature component. We used the BT-Settl model for each temperature, with the metallicity
[Fe/H]= 0 dex and the stellar surface gravity log g at 5.2, from the SVO theoretical spectra web server (\footnote{\url{http://svo2.cab.inta-csic.es/theory/newov2/}}). 

\subsection{Atmospheric modelling}\label{sec:1.3}
\subsubsection{Possible atmospheric scenarios}\label{sec:1.3.1}
\citet{Turbet_2020b} reviewed the different atmospheric scenarios for TRAPPIST-1 planets. We discuss the different possibilities mentioned for TRAPPIST-1 h, such as a H$_2$/He rich atmosphere, a H$_2$O envelope, as well as a O$_2$, a CO$_2$, a CH$_4$/NH$_3$, or a N$_2$ dominated atmosphere. First, numerical modelling using mass and radius measurements have shown that a H$_2$/He envelope is unlikely for all TRAPPIST-1 planets. \citet{Turbet_2020b} constructed a mass-radius relation using the \citet{Grimm_2018} atmospheric climate calculation and estimated that for a 'cold' scenario assuming 100 x solar metallicity and based on TRAPPIST-1 h irradiation, the maximum hydrogen to core mass fraction is $4\times10^{-4}$ for a clear atmosphere. Using the estimation of \citet{Wheatley_2017} for the EUV flux received by the planet ($10^{2}$ erg.s$^{-1}$.cm$^{-2}$) and the results from \citet{Bolmont_2017a}, \citet{Bourrier_2017a} and \citet{Bourrier_2017b}, they computed the equivalent mass loss over the age of the system (8 billion years) and found $10^{22}$ kg (i.e. $5\times10^{-3}$ mass fraction). A hydrogen-rich envelope could be ripped out in $\sim$100 million years for TRAPPIST-1 h \citep{Turbet_2020b}, meaning that this type of atmosphere is not completely impossible but unstable and unlikely to be sustained around this low mass planet. The recent publication by \citet{Hori_2020} has also shown that the total mass loss over the planet lifetime is supposedly higher than the initial amount of accreted gas. 

Regarding a water-rich atmosphere scenario, \citet{Turbet_2019a}, \citet{Turbet_2020a} and \citet{Turbet_2020b} estimated the water content in TRAPPIST-1 planets by taking the runaway greenhouse limit into account, while \citet{Bourrier_2017a} investigated the hydrodynamic water loss. Combining those two pieces of information leads to the conclusion that TRAPPIST-1 h could have lost less than three Earth oceans and could have retained water in its atmosphere or surface. \citet{Lincowski_2019} show that O$_2$ atmospheres would be the best candidate for TRAPPIST-1 planets as a remnant of H$_2$O erosion and \citet{Wordsworth_2018} determine that O$_2$ build-up is limited to one bar for TRAPPIST-1h. 

We note that NH$_3$ and CH$_4$ are highly sensitive to photo-dissociation \citep{Turbet_2018} and for TRAPPIST-1h to sustain a CH$_4$ or a NH$_3$ rich atmosphere would require an important source of those species. Assuming an Earth-like methane production rate, the planet could have a concentration up to 0.3\% \citep{Rugheimer_2015}. However, methane or ammonia photolysis rates could decrease via the formation of high altitude clouds or hazes \citep{Sagan_1997, Wolf_2010, Arney_2016}.

An Earth-like atmosphere, that is one bar and a N$_2$ rich atmosphere, might be stable against stellar wind for TRAPPIST-1 h if CO$_2$ is abundant \citep{Dong_2018, Dong_2019}; CO$_2$ could accumulate in TRAPPIST-1 planets \citep{Lincowski_2019} because it is less sensitive to atmospheric escape \citep{Dong_2017,Dong_2018, Dong_2019}. However, \citet{Turbet_2018} and \citet{Turbet_2020b} show that TRAPPIST-1 h would probably experience a CO$_2$ collapse. The planet is far from the star and probably tidally locked, favouring CO$_2$ surface condensation. Furthermore, CO and O$_2$ could also be found in the case of a CO$_2$ rich atmosphere due to the photo-dissociation of CO$_2$ and the low recombination of CO and O$_2$ \citep{Gao_2015, Hu_2020}. 

Finally, a water ocean at the surface of TRAPPIST-1 h, implying a potential habitability, is unlikely. As the planet is beyond the CO$_2$ collapse region, the atmosphere does not warm the surface \citep{Turbet_2020b}. To counterbalance the CO$_2$ condensation, the planet would require a very thick CO$_2$ atmosphere with volcanic gases such as H$_2$ and CH$_4$, but, as explained above, neither H$_2$ nor CH$_4$ are expected to be stable in the TRAPPIST-1 h atmosphere \citep{Pierrehumbert_2011, Wordsworth_2017, Ramirez_2017, Lincowski_2018, Turbet_2018,Turbet_2019b, Turbet_2020a}. Very few observational constraints have been brought on TRAPPIST-1 planets, leaving a wide range of atmospheric possibilities. The goal of the following section is to analyse the TRAPPIST-1 h IR spectrum with regards to the predictions mentioned above in order to bring new constraints and prepare further observations.

\subsubsection{Retrieval analysis set-up}\label{sec:1.3.2}
We used TauREx3 \citep{Alrefaie2019,Alrefaie2021} and the nested sampling algorithm Multinest \citep{Feroz_2009} with an evidence tolerance of 0.5 and 1500 live points to perform the atmospheric retrieval analysis. TauREx3 is a fully Bayesian code that maps the atmospheric parameters space to find the best fit model for the transmission spectrum. It includes the molecular line lists from the ExoMol project \citep{Tennyson_2016,Chubb_2021}, HITEMP \citep{Tennyson_2018}, and HITRAN \citep{Rothman_1987, Rothman_2013}. We simulated the atmosphere assuming a constant temperature-pressure profile and every layer of the simulated atmosphere is uniformly distributed in log spaced, with a total of 100 ranging from $10^{-2}$ to $10^5$ Pa. We included the collision-induced absorption (CIA) of H2-H2 \citep{abel_2011, fletcher_2018}, H$_2$-He \citep{abel_2012}, and Rayleigh scattering. We used a wide range of temperatures (50-1000K) to adjust the temperature of the planet, using the effective temperature ($\sim$173 K) as the initial value. The planetary radius was also fitted as a free parameter in the model and its value ranges from $\pm 50\%$ of the published value reported in Table \ref{table1:parameter}. The planetary radius fitted corresponds to the bottom of the atmosphere, that is the radius of the planet assumed to be at one bar here. Clouds were included using a simple grey opacity model and the top clouds pressure varies from $10^{-2}$ to $10^5$ Pa. We considered the following opacity sources: H$_2$O \citep{Polyansky_2018}, CO$_2$ \citep{Rothman_2010}, NH$_3$ \citep{Yurchenko_2011}, and CO \citep{Yurchenko_2014}. 

We performed two different atmospheric retrievals by forcing a primary and then a secondary atmosphere. We modelled the TRAPPIST-1 h atmosphere using H$_2$, He, and N$_2$ as fill gas and H$_2$O, CO, CO$_2$, NH$_3$, and CH$_4$ as trace gases. We note that H$_2$, He, and N$_2$ do not display features in the spectrum; they contribute to the continuum and shape the mean molecular weight. The ratio between H$_2$ and He abundances was fixed to the solar value of 0.17, while the ratio between the abundance of N$_2$ over the abundance of H$_2$ varied between $10^{-12}$ and $10^{-2}$ for the primary model and between $10^{-12}$ and $10^{4}$ for the secondary scenario. The mean molecular weight can then evolve towards higher values and we were able to test a Hydrogen rich and then a Nitrogen rich atmosphere. The abundance of the other molecular absorption sources were included in the fit as a volume mixing ratio, allowing us to vary between $10^{-12}$ and $10^{-2}$. 

A flat-line model, only including a cloud deck, was performed to assess the significance of the different scenarios compared to a baseline. A baseline is representative of the lack of an atmosphere (e.g. an atmosphere with no spectral features) or a flat spectrum that can only be fitted by a high altitude cloud deck. The significance was computed using a Bayes factor, that is the difference of logarithm evidence between the best fit model and the baseline model. The Bayesian evidence was computed using Bayes' theorem for a set of $\theta$ parameters in a model H for the data D \citep{Feroz_2009}
\begin{equation}
    P(\theta|D, H)=\frac{P(D|\theta, H)P(\theta|H)}{P(D|H)}
,\end{equation}
where P($\theta$|D, H) $\equiv$P($\theta$) is the posterior probability distribution, P(D|$\theta$, H) $\equiv$L($\theta$) is the likelihood, P($\theta$|H)$\equiv$ $\pi$($\theta$) is the prior, and P(D|H)$\equiv$ E is the Bayesian evidence. The nested sampling method estimates the Bayesian evidence of a given likelihood volume and the evidence can be expressed as follows:
\begin{equation}
 E=\int L(\theta) \pi(\theta) \, \mathrm{d}\theta
.\end{equation}
To compare the two H$_0$ and H$_1$ models, in our case the flat-line model and the primary or secondary scenario, we can compute the respective posterior probabilities, given the observed data set D,
\begin{equation}
 \frac{P(H_1|D)}{P(H_0|D)}= \frac{P(D|H_1)P(H_1)}{P(D|H_0)P(H_0)}=\frac{E_1 P(H_1)}{E_0 P(H_0)}
,\end{equation}
where P(H$_1$)/P(H$_0$) is the a priori probability ratio for the two models, which can often be set to unity \citep{Feroz_2009}. We used the logarithm version of the model selection to compute the Bayes factor, $\Delta$log(E) between the flat-line and the tested model. This factor is also called the atmospheric detectability index (ADI) in \citet{Tsiaras_2018} and defined as a positive value. The significance ($\sigma$) represents the strength of a detection and it was estimated using a \citet{Kass_1995}, \citet{Trotta_2008}, and \citet{Benneke_2013}  formalism. We used Table 2 in \citet{Trotta_2008} and Table 2 in \citet{Benneke_2013} to find the equivalence between the Bayes factor and the significance $\sigma$ and evaluate the strength of a detection. A Bayes factor of 1 corresponds to a 2.1$\sigma$ detection and is considered weak, a Bayes factor greater than 3 (3$\sigma$) is considered significant, and one superior to 11 (5$\sigma$) is considered as a strong detection. 
For the rest of the paper, we define $\Delta$log(E)=log($\textrm{E}_\textrm{Atmospheric Model}$)-log($\textrm{E}_\textrm {Flat line}$). The atmospheric model can be considered a better fit compared to the flat line if the $\Delta$log(E) is superior to 3.

\section{Results}\label{sec:2}
\subsection{Atmospheric retrieval results}\label{sec:2.1}
There is no evidence of molecular absorption in the recovered spectrum of TRAPPIST-1h from the two retrieval results. Both primary and secondary retrieval analyses have logarithm evidence (109.92 and 110.18, respectively) comparable to the one of the flat-line model, that is 110.55.
This result favours the scenario of a planet with no atmosphere, that is the presence of a high cloud layer in a primary atmosphere or a secondary envelope. It is consistent with previous work on other TRAPPIST-1 planets \citep{de_Wit_2018,Wakeford_2019,Zhang_2018}. 
\begin{figure}[htpb]
    \centering
    \resizebox{\hsize}{!}{\includegraphics{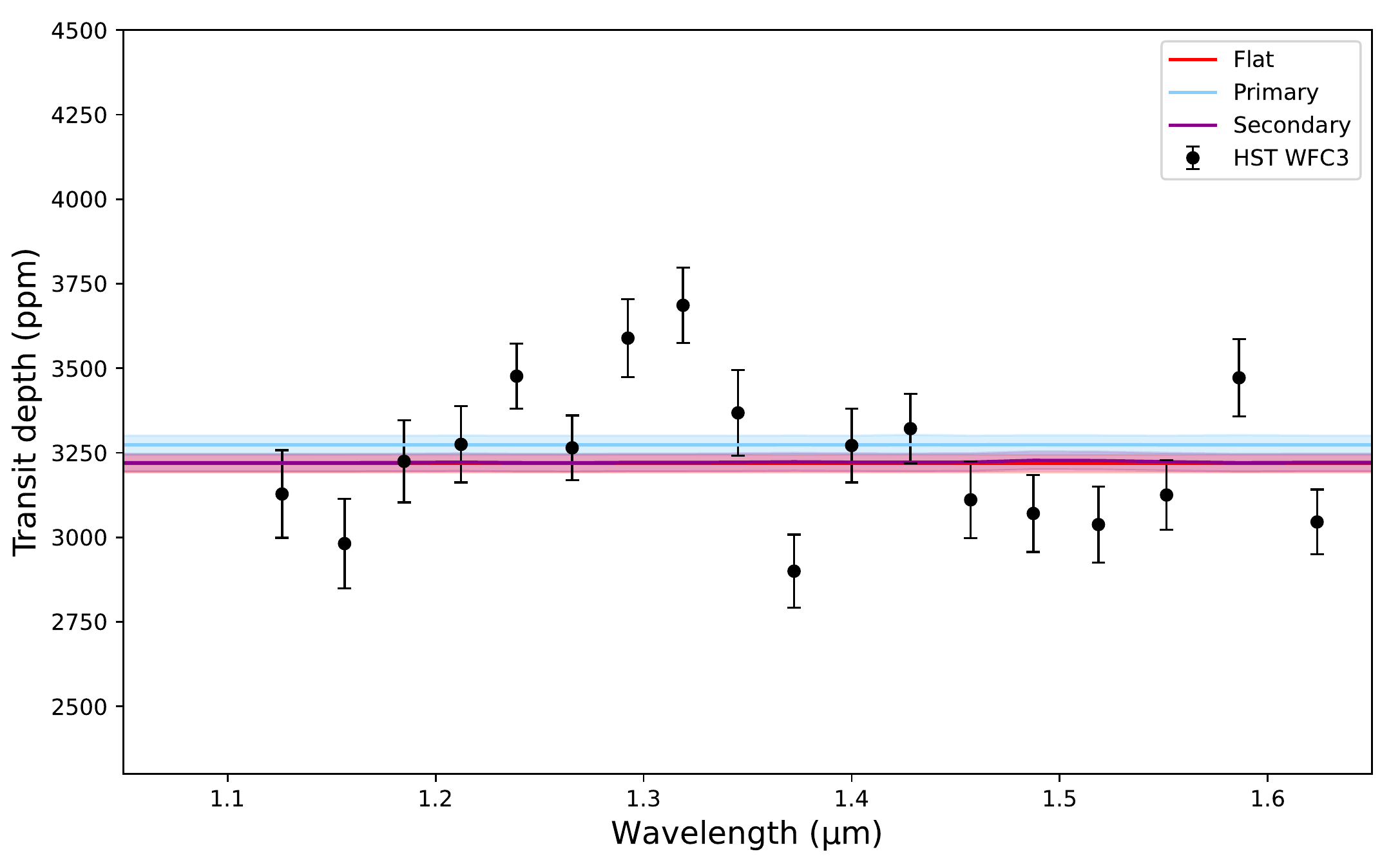}}
    \resizebox{\hsize}{!}{\includegraphics{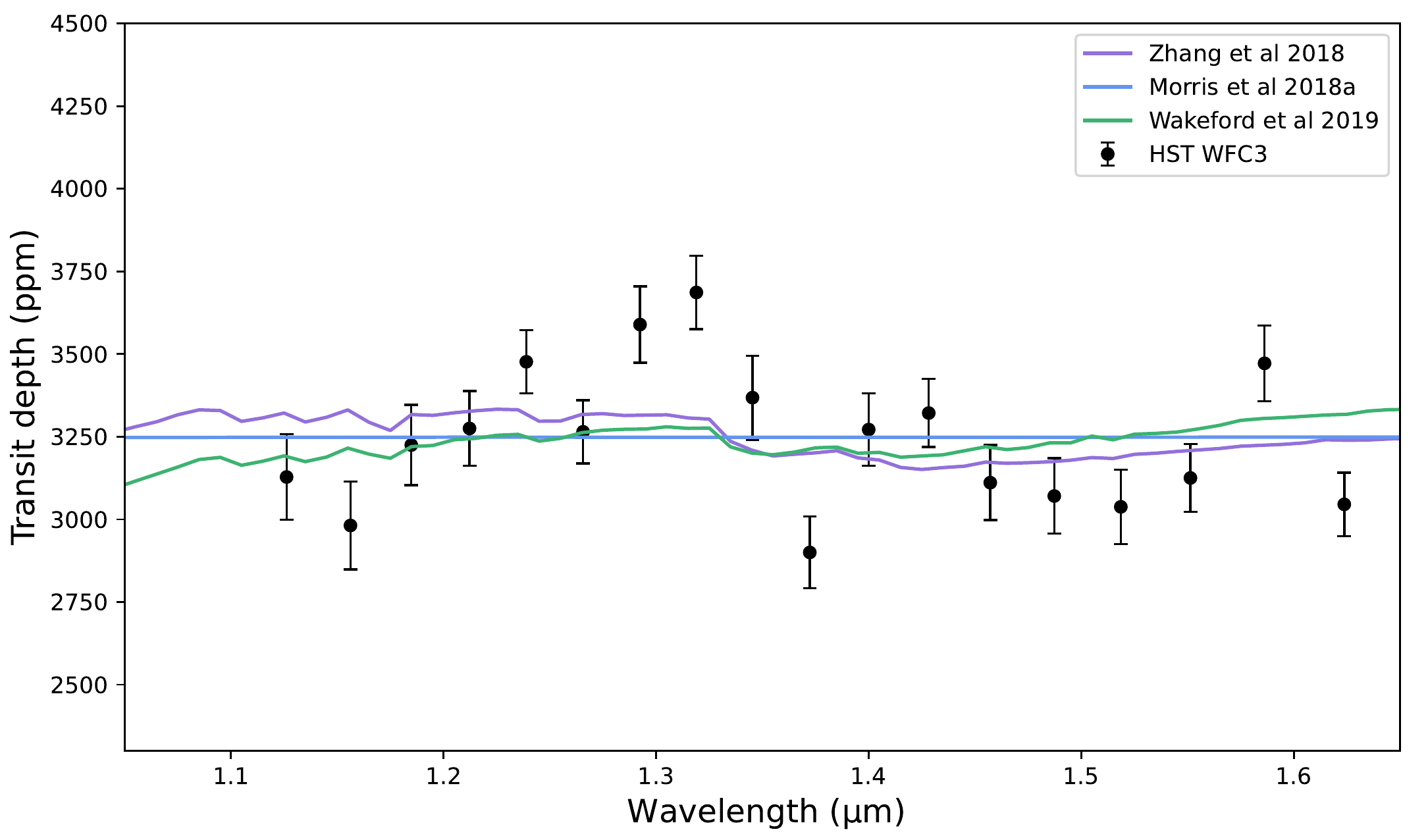}}
    \caption{Best-fit models to TRAPPIST-1 h HST WFC3 G141 data from atmosphere retrievals (top) and stellar contamination models based on \citet{Zhang_2018}, \citet{Wakeford_2019}, and \citet{Morris_2018a} (bottom).}
    \label{fig:best_fit}
\end{figure}
Figure \ref{fig:best_fit} (top) shows the extracted spectrum with the best-fit atmospheric results: flat-line (red), primary (blue), and secondary model (purple). The flat-line and the secondary best fit models are similarly flat with a transit depth around 3220 ppm, while the primary models are found around 3274 ppm. This difference is due to the different radius and temperature estimations depending on the scale height and the weight of the atmosphere. We present the correlations among parameters for the primary and the secondary model in Fig. \ref{fig:posteriors_taurex}. We overplotted the two posterior distributions for a direct comparison, but the values are displayed for the secondary best-fit model. The primary model posterior distribution alone is presented in Appendix \ref{appendix:posteriors_primary}.

The secondary atmospheric retrieval analysis estimates the radius to be 0.69$^{+0.03}_{-0.07}$R$_{\oplus}$ and the temperature reaches 345$^{+326}_{-196}$K. The mean molecular weight distribution is bi-modal, and the code is able to retrieve two solutions: a light atmosphere with a 2.3 g/mol mean molecular weight and a heavier solution with a mean molecular weight reaching 25.35$^{+2.46}_{-23.02}$ g/mol corresponding to a 16 km scale height. This is correlated to the abundance of N$_2$ retrieved as the ratio of inactive gases, that is log(N$_2$/H$_2$). When we allowed this ratio to increase beyond one, the best fit value was constrained to 1.01$^{+1.18}_{-6.13}$. Yet, we note the presence of a second solution, around seven, which corresponds to the primary analysis retrieved value and creates the bi-modal distribution in the mean molecular weight. Nitrogen is the only parameter that impacts the value of the mean molecular weight as no constraints can be put on H$_2$O, CH$_4$, CO, CO$_2$, and NH$_3$. 
\begin{table*}[htpb!]
\centering
\caption{Statistical results of the atmospheric retrieval analysis and the stellar contamination modelling on TRAPPIST-1h HST WFC3 G141 data.}
\label{table:stats}
\begin{tabular}{ l|c c c c } \hline\hline
Model & $\chi^2$ &$\tilde{\chi}^2$ & log(E) & $\Delta$log(E)\\ \hline
Flat-line & 64.95 & 3.61 & 110.55 & N/A\\
Atmosphere primary &65.20 & 3.62 & 109.92 & -0.63 \\
Atmosphere secondary &65.25 & 3.62 & 110.18 & -0.37\\
Stellar \citet{Zhang_2018} & 54.02 & 3.00 &N/A & N/A \\
Stellar \citet{Wakeford_2019} & 60.70 & 3.37 &N/A & N/A\\
Stellar \citet{Morris_2018a} &63.91 & 3.55 & N/A & N/A \\
\hline
Corrected by \citet{Zhang_2018} \\
Flat-line &54.94 & 3.05 & 115.48 & N/A\\
Atmosphere primary &50.54  & 2.81 & 115.02 & -0.46\\
Atmosphere secondary & 54.77 & 3.04 & 115.20 & -0.28\\
\hline
Corrected by \citet{Wakeford_2019} \\
Flat-line & 61.41 & 3.41 & 112.13 & N/A\\
Atmosphere primary & 61.91 & 3.44 &111.51 &-0.62\\
Atmosphere secondary & 61.80 &3.43 & 112.01 &-0.12\\
\hline
Corrected by \citet{Morris_2018a} \\
Flat-line & 65.11 & 3.62 &110.39& N/A \\
Atmosphere primary & 65.28 &3.63 &109.79 &-0.60\\
Atmosphere secondary & 65.23 &3.62 & 110.12 &-0.27\\
 \hline
\end{tabular}
\tablefoot{Chi-squared ($\chi^2$) and reduced chi-squared ($\tilde{\chi}^2$2) were computed using the result of the retrieval best-fit model and the stellar contamination models. Bayesian logarithm evidence (log(E)) and the Bayes factor ($\Delta$log(E)) were computed when applicable, i.e. only for the atmospheric retrieval analysis.}
\end{table*}
From both posteriors distributions, we found the anti-correlation between the radius, temperature, and layer for top clouds, that is the radius decreases with increasing temperature and decreasing layer for top clouds. The latter is found really high in the atmosphere, log(P$_{\rm clouds}$)=1.02$^{+1.90}_{-1.72}$, which corresponds to a cloud layer at approximately 10$^{-4}$ bar. Considering the pressure of this layer, it is likely that these clouds may not be condensation clouds, but rather photochemical mists or hazes with particles big enough not to have a spectral slope.
From those two retrievals analyses, we show that the atmosphere must be either secondary (probably dominated by nitrogen) or primary with a very high photochemical haze layer. The two retrieval analyses have similar statistical results so we cannot favour one solution. We cannot rule out the hypothesis of a lack of an atmosphere either as the log(E) of the flat-line model remains the highest. We can reject a primary clear atmosphere as expected for this planet as the primary model does indeed require a layer of clouds to correctly fit the spectrum (see Sec. \ref{sec:3.1}).
\begin{figure*}[htpb!]
    \centering
    \includegraphics[width=17cm]{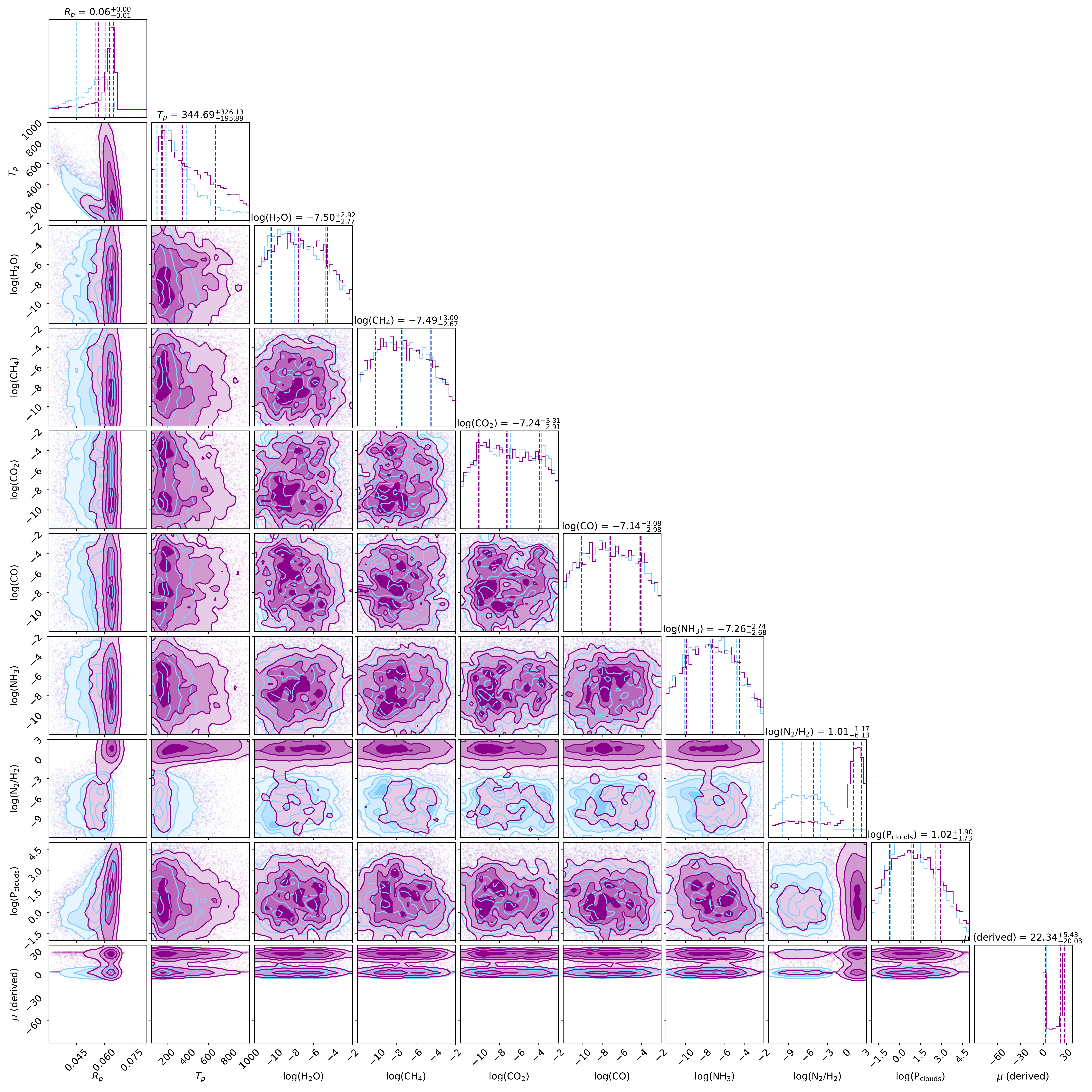}
    \caption{Posterior distributions for the primary (blue) and the secondary retrieval (purple) on the extracted TRAPPIST-1h spectrum. Only the values from the secondary best-fit analysis are displayed. }
    \label{fig:posteriors_taurex}
\end{figure*}
\subsection{Including the stellar contamination}\label{sec:2.2}
We present in Fig. \ref{fig:best_fit} (bottom) the stellar contamination models and in Table \ref{table:stats} the statistical results on both the atmosphere and stellar models. We computed the chi-squared ($\chi2$) and the reduced chi-squared ($\tilde{\chi}^2$2) for all models and indicate the logarithm evidence (log(E)) from the retrieval analysis.
The stellar contamination model of \citet{Zhang_2018} is favoured, according to the chi-squared computation but none of the models we tested here  are significant and can explain variations in the TRAPPIST-1h spectrum. In particular, the rise in the transit depth around 1.3$\mu$m is not reproduced. To account for stellar contamination, we corrected our HST/WFC3 extracted spectrum by subtracting the stellar contributions using the \citet{Zhang_2018}, \citet{Wakeford_2019}, and \citet{Morris_2018a} formalism. 
\begin{figure}[htpb!]
    \centering
    \resizebox{\hsize}{!}{\includegraphics{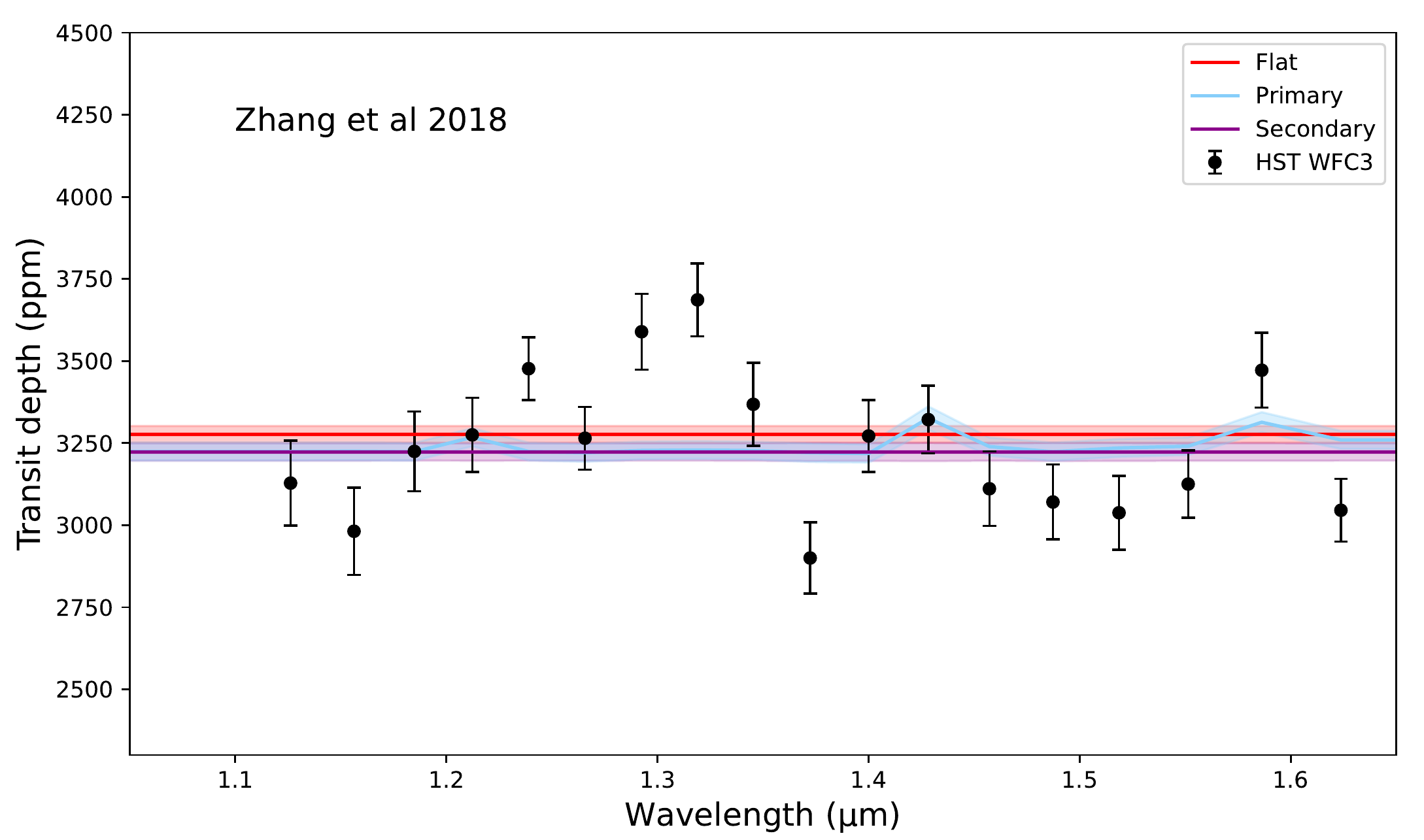}}
    \resizebox{\hsize}{!}{\includegraphics{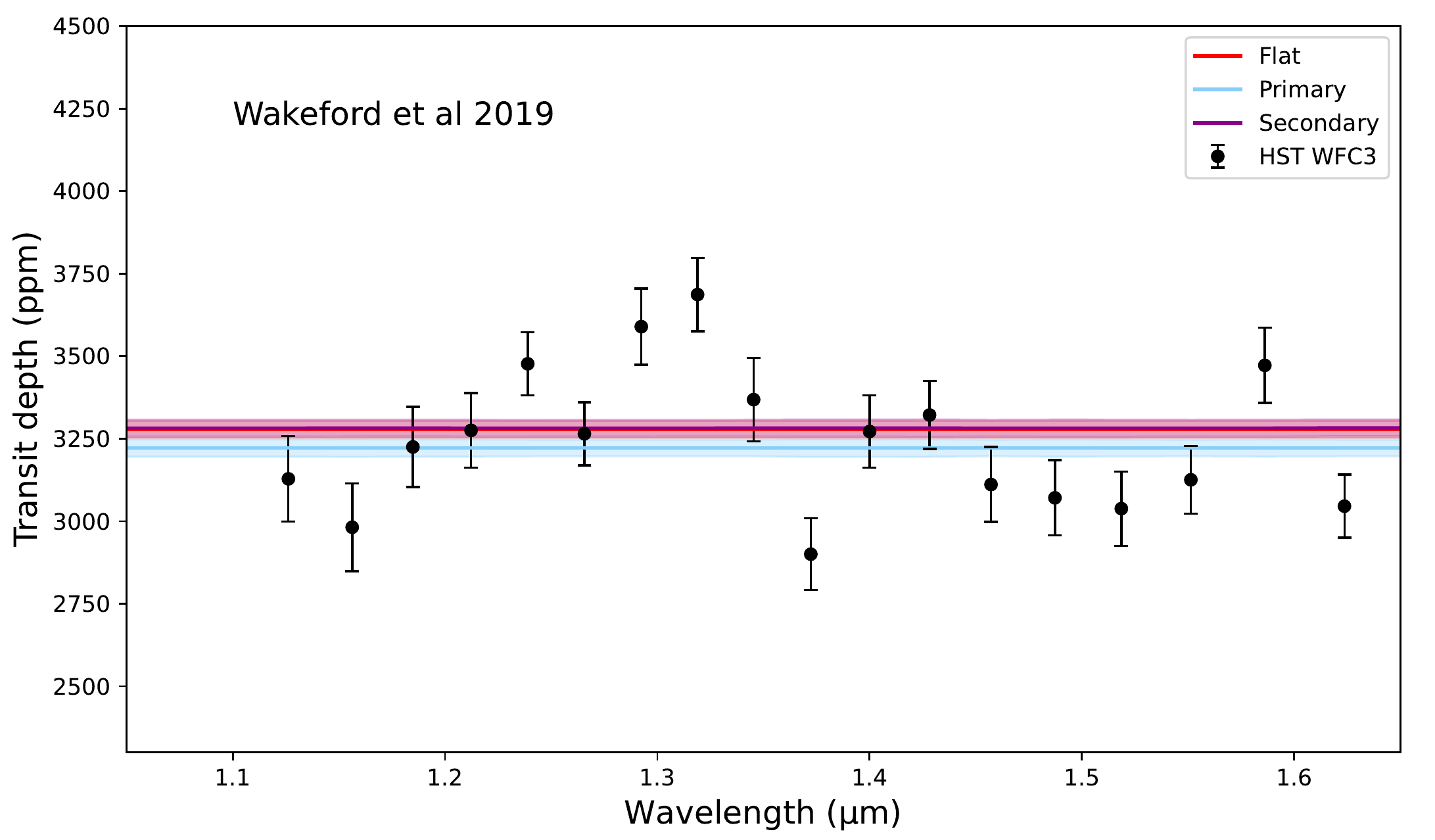}}
    \resizebox{\hsize}{!}{\includegraphics{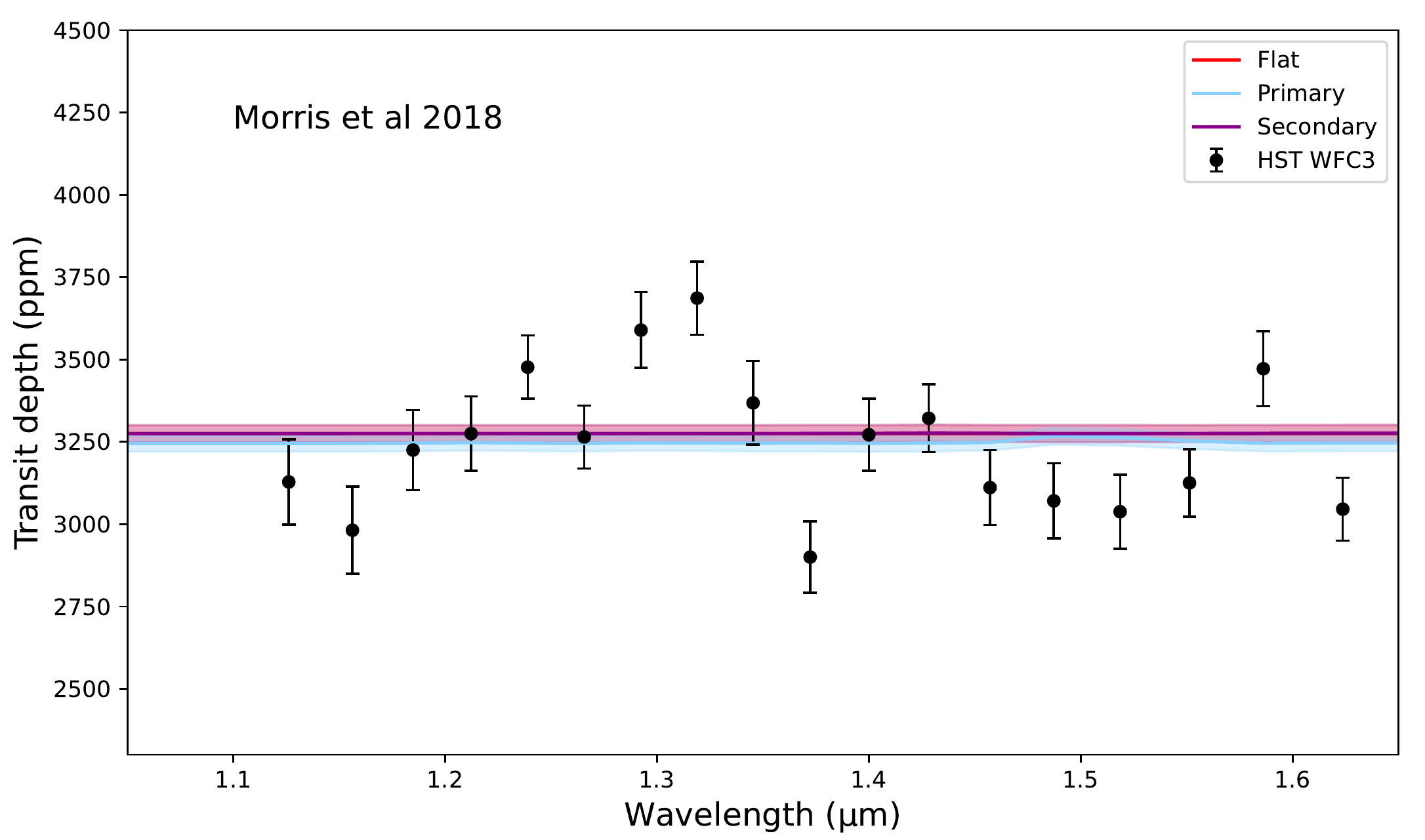}}
    \caption{Best fit models to TRAPPIST-1 h HST WFC3 G141 data after subtraction of stellar contamination contributions according to the \citet{Zhang_2018} (top), \citet{Wakeford_2019} (middle), and \citet{Morris_2018a} (bottom) formalism. }
    \label{fig:best_fit_corrected}
\end{figure}
We present in Table \ref{table4bis:spectra_corrected} the transit depth after subtraction of the stellar contamination for the three models and conduct the same retrieval analysis as in Sec. \ref{sec:2.1} on those corrected spectra. We overplotted all the different spectra as a comparison in appendix \ref{appendix:corrected_spectra}. Statistical results on the retrieved corrected spectra are detailed in Table \ref{table:stats}. For all the corrected spectra, the flat line is the favoured model, but the correction by \citet{Zhang_2018} leads to the highest log(E). We present in Fig. \ref{fig:best_fit_corrected} best-fit atmospheric retrieval results on the three spectra, while posterior distributions are in Appendix \ref{appendix:posteriors_taurex_Zhang}, \ref{appendix:posteriors_taurex_Wakeford}, and \ref{appendix:posteriors_taurex_Morris}. We note transit depth variations at 1.2 $\mu$m, 1.45$\mu$m, and 1.6$\mu$m on the primary best fit model on the spectrum corrected by \citet{Zhang_2018}. This is due to the contribution of CO$_2$ to the best-fit solution,  but the amount of CO$_2$ is not constrained (see posterior distributions in Appendix \ref{appendix:posteriors_taurex_Zhang}) and the model is not statistically significant. We also observe variations in the transit depth around 1.5$\mu$m on the primary best fit model on the spectrum corrected by \citet{Morris_2018a}. This is due to the absorption of ammonia. Once again, this absorption is not constrained in terms of abundance and the log(E) remains below the one of the flat line. As an indication, we put the best-fit opacity contributions from those two models in appendix \ref{appendix:contributions}. The correction made here to the spectra does improve the retrieval statistical results in the case of the \citet{Zhang_2018} correction, but it does not lead to molecular detection and does not allow us to provide further constraints on the atmosphere of TRAPPIST-1 h.

To better constrain the stellar contamination, we also tried to add the optical value found in \citet{Luger_2017a} using K2 photometry. As seen in the plot of Appendix \ref{appendix:stellar_contamination_K2} and in the $\chi^2$ computation results in Appendix \ref{appendix:stats_K2}, the existing stellar models discussed here fit the spectrum poorly. First, we cannot ensure inter-instrument calibration at this accuracy and combining a different transit depth could lead to misinterpretations of the spectrum \citep{Yip_2020}. In addition, it is possible that the stellar spot distribution has changed in the intervening time between the observations, but this is unlikely as they are not that far apart. K2 light curves were taken between 15 December 2016 and 4 March 2017, while the HST data were taken in July 2017, September 2019, and July 2020. A more likely explanation is that for both K2 and HST data, multiple transits were stacked, regardless of the phase of the star's rotation. If the stellar rotational phase and activity were different from time to time, the effect in the transmission spectrum would be suppressed when they are stacked. Adding this point does not further constrain the stellar contamination models in the case of TRAPPIST-1h.
\begin{table}[htpb!]
    \caption{Corrected transit depth in ppm using stellar contamination models.}
    \centering
    \begin{tabular}{p{2.5cm} p{1.5cm} p{1.5cm} p{1.5cm}} \hline \hline
    Wavelength ($\mu$m)  & \multicolumn{3}{c}{Transit depth (ppm)} \\
    \hline 
    1.1262 & 3072.99 & 3193.93 & 3129.83\\
    1.1563 & 2934.85 & 3032.81 & 2983.10 \\
    1.1849 & 3171.13 & 3259.82 & 3226.35\\
    1.2123 & 3202.19 & 3283.50 & 3276.38\\
    1.2390 & 3406.64 & 3477.46 & 3478.10\\
    1.2657 & 3207.19 & 3257.41 & 3266.07\\
    1.2925 & 3520.20 & 3564.16 & 3590.50 \\
    1.3190 & 3643.19 & 3673.80 & 3687.80\\
    1.3454 & 3388.59 & 3396.28 & 3369.86\\
    1.3723 & 2948.59 & 2939.83 & 2901.69\\
    1.4000 & 3341.90 & 3318.92 & 3273.55\\
    1.4283 & 3414.44 & 3373.99 & 3323.56\\
    1.4572 & 3193.82 & 3149.00 & 3112.83\\
    1.4873 & 3141.10 & 3090.15 & 3072.24\\
    1.5186 & 3093.99 & 3037.20 & 3039.34\\
    1.5514 & 3165.61 & 3101.84 & 3126.55\\
    1.5862 & 3499.04 & 3418.90 & 3473.19\\
    1.6237 & 3057.41 & 2980.84 & 3046.45
 \\\hline 
 References &  1 & 2 & 3\\ \hline
    \end{tabular}
    \label{table4bis:spectra_corrected}
    \tablebib{ (1) \citet{Zhang_2018}; (2)\citet{Wakeford_2019}; (3) \citet{Morris_2018a}}
\end{table}

\section{Discussion} \label{sec:3}
\begin{figure*}[htpb!]
    \centering
    \includegraphics[width=\columnwidth]{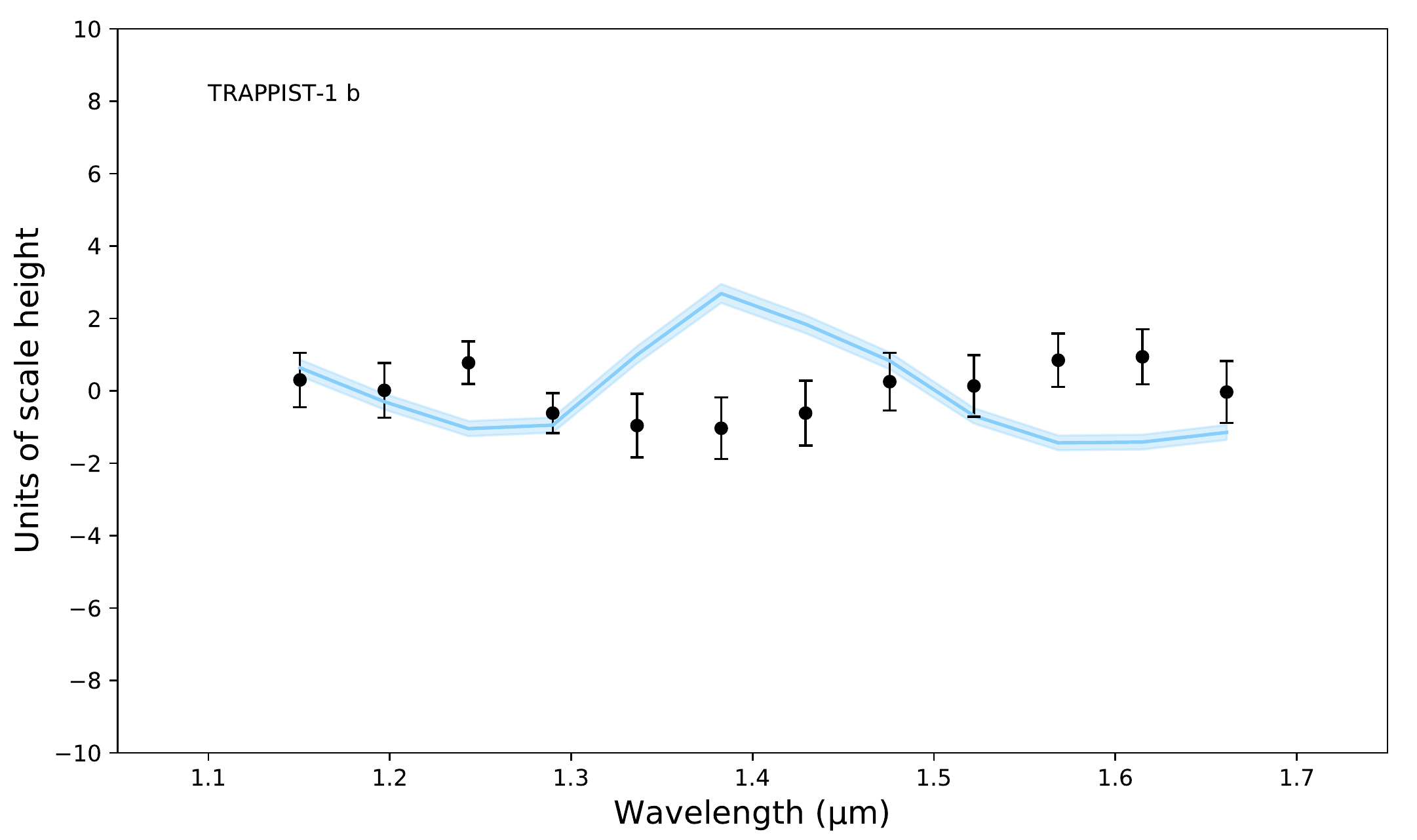}
    \includegraphics[width=\columnwidth]{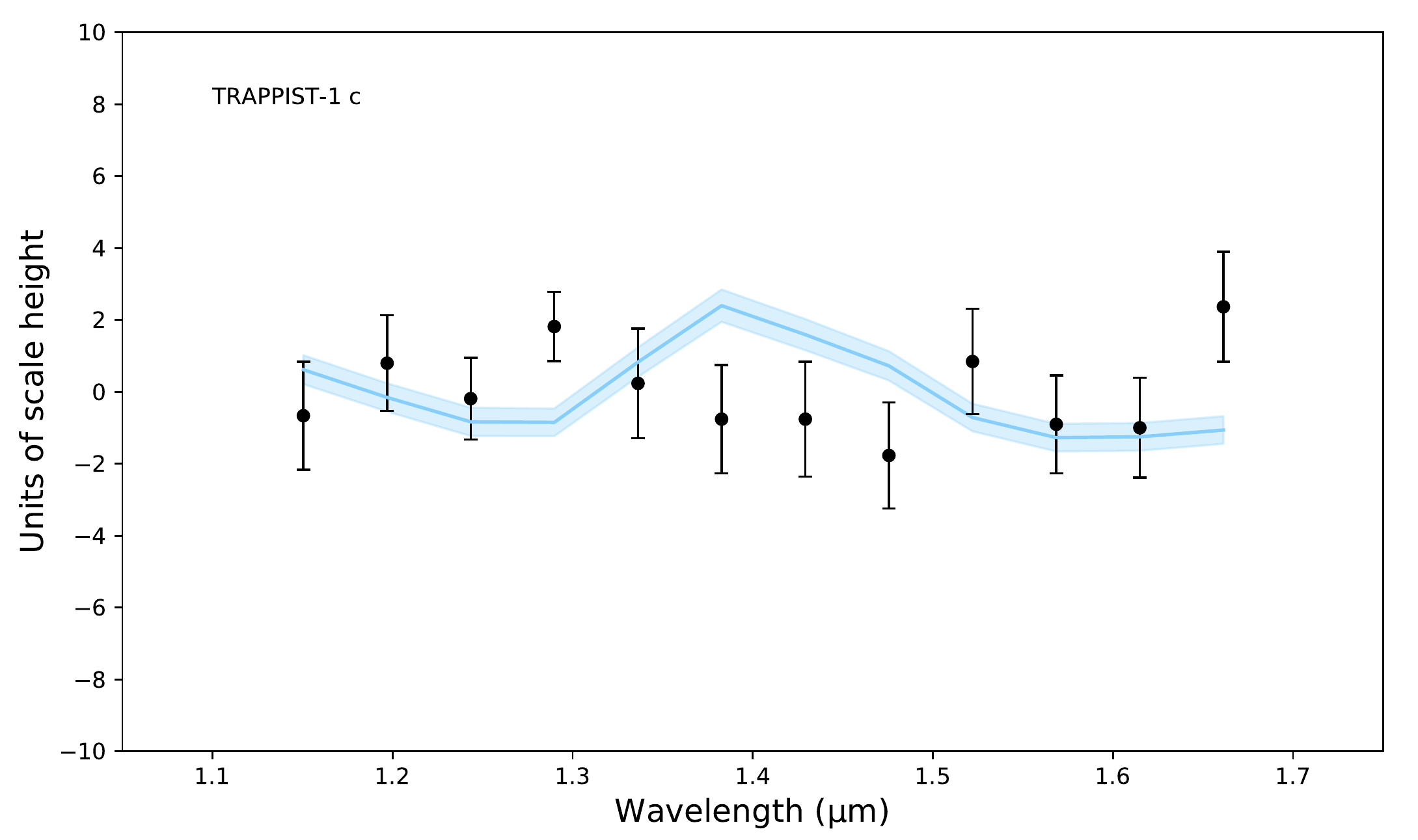}
    
    \includegraphics[width=\columnwidth]{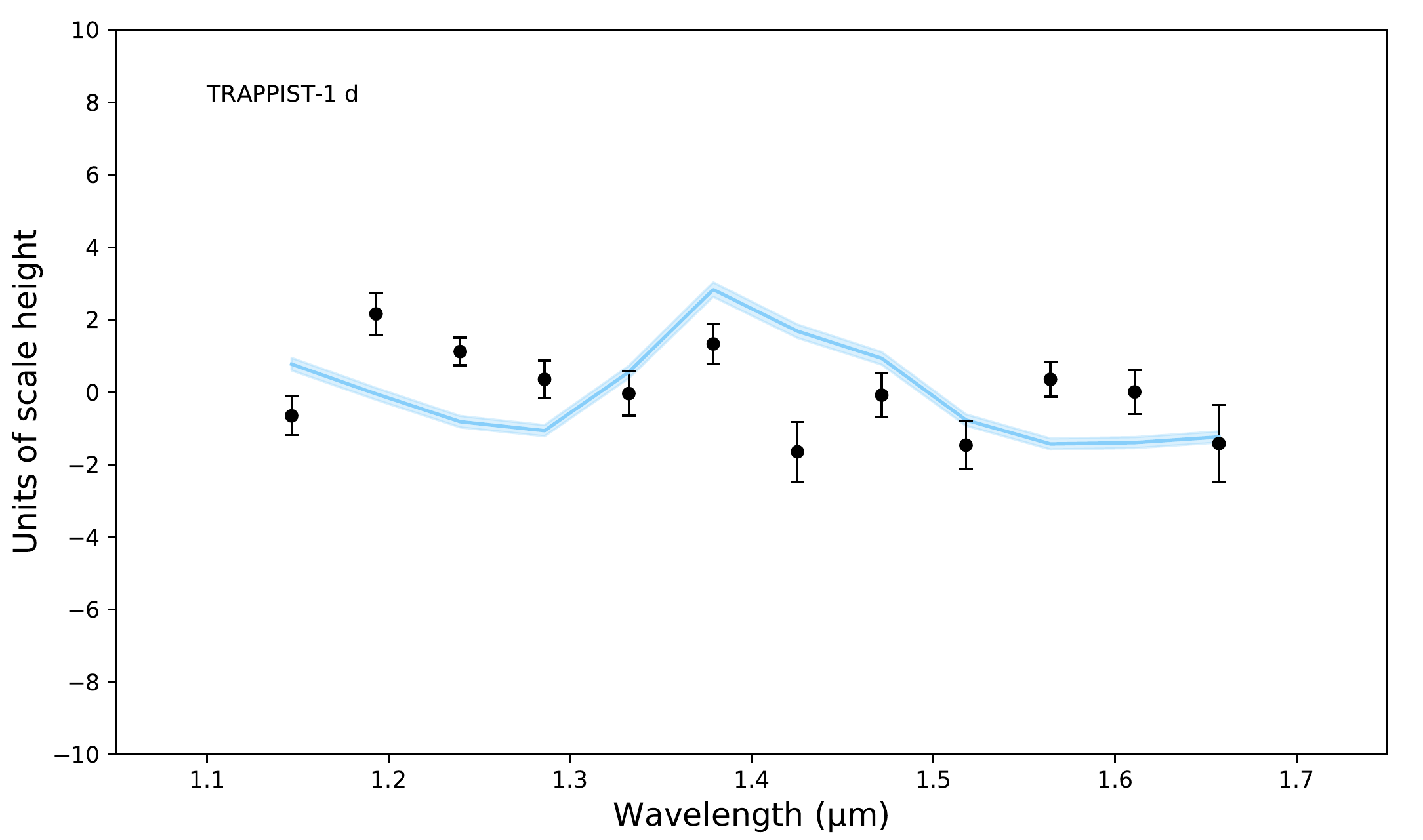}
    \includegraphics[width=\columnwidth]{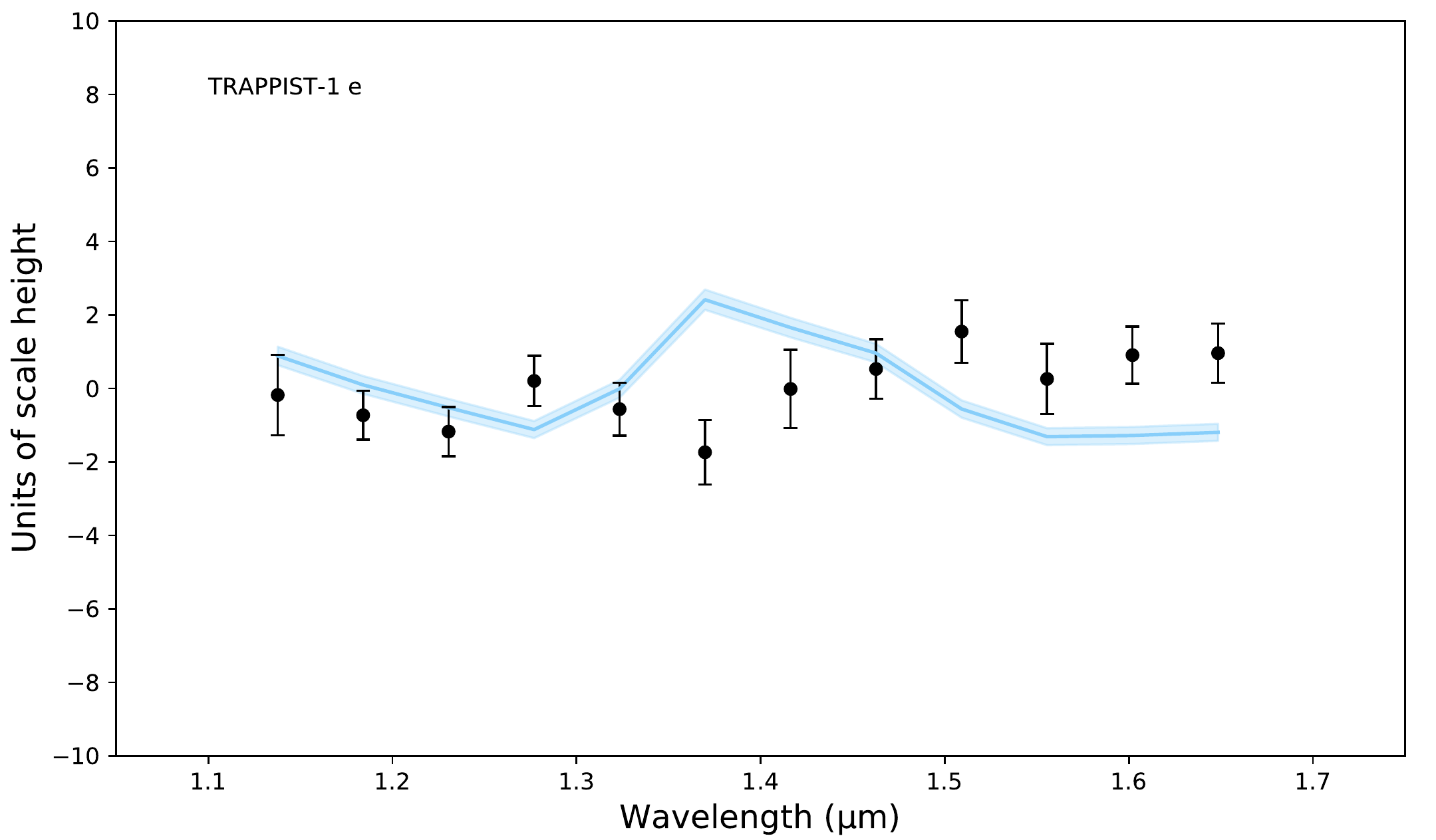}
    
    \includegraphics[width=\columnwidth]{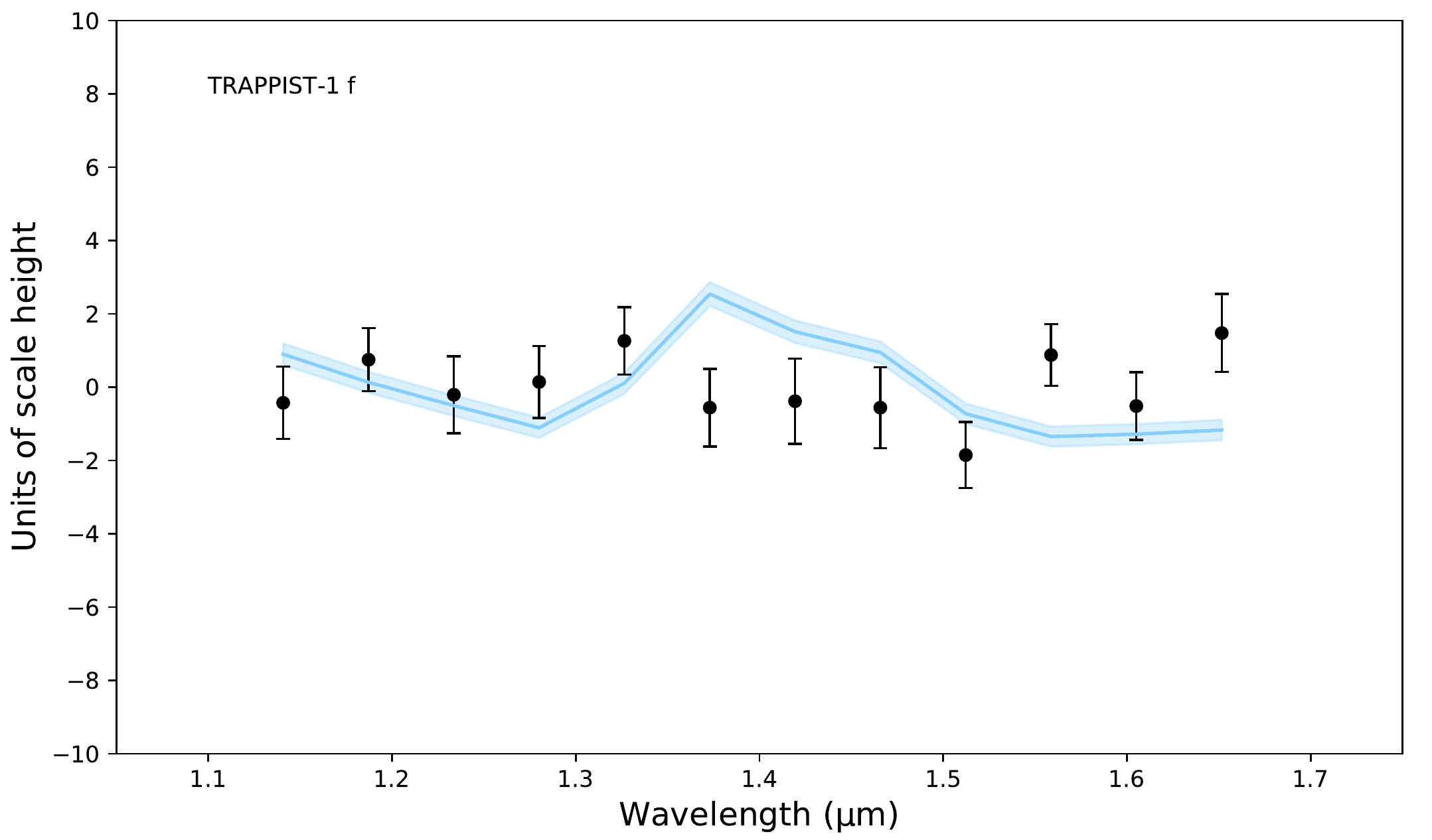}
    \includegraphics[width=\columnwidth]{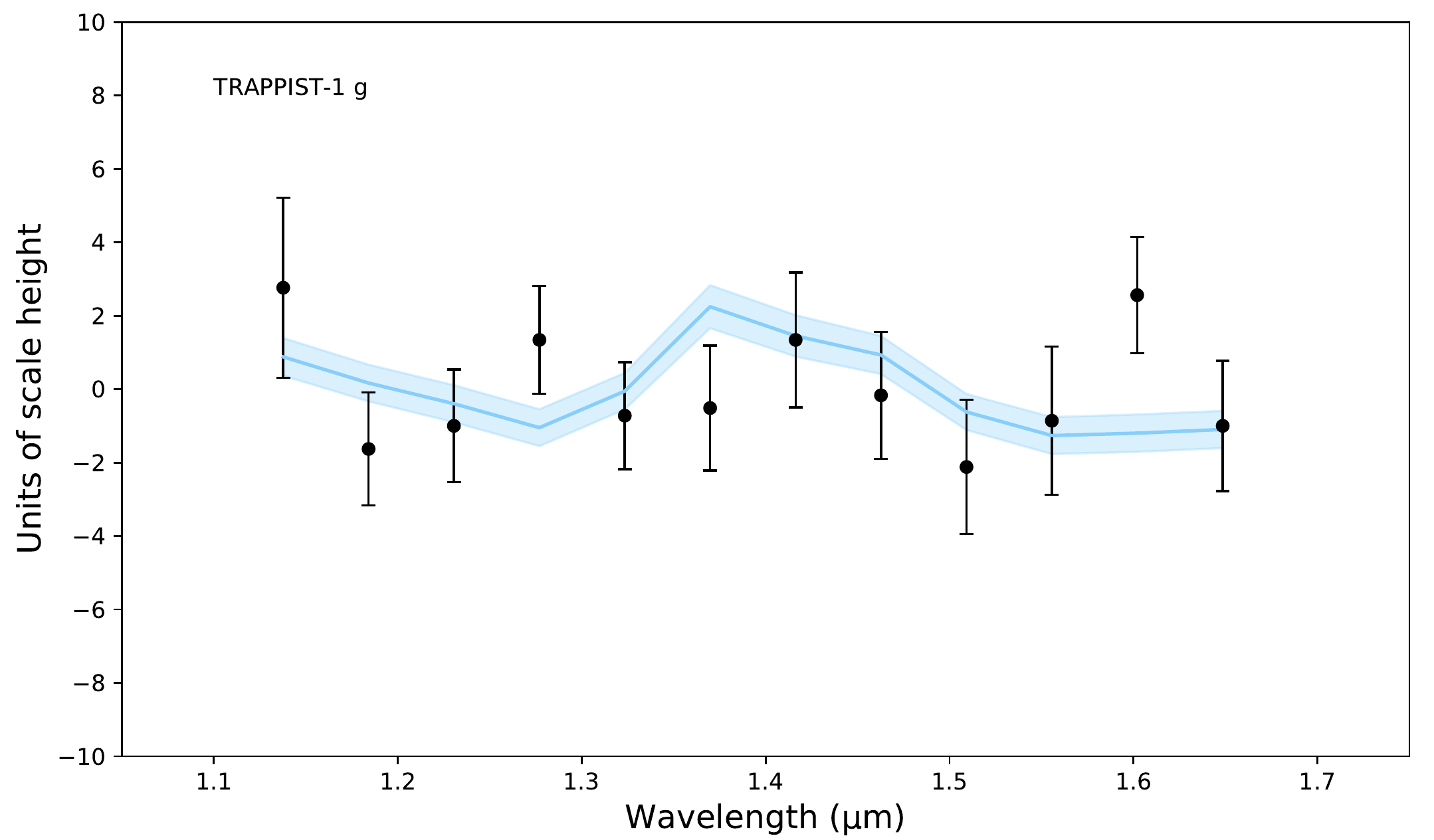}
    
    \includegraphics[width=\columnwidth]{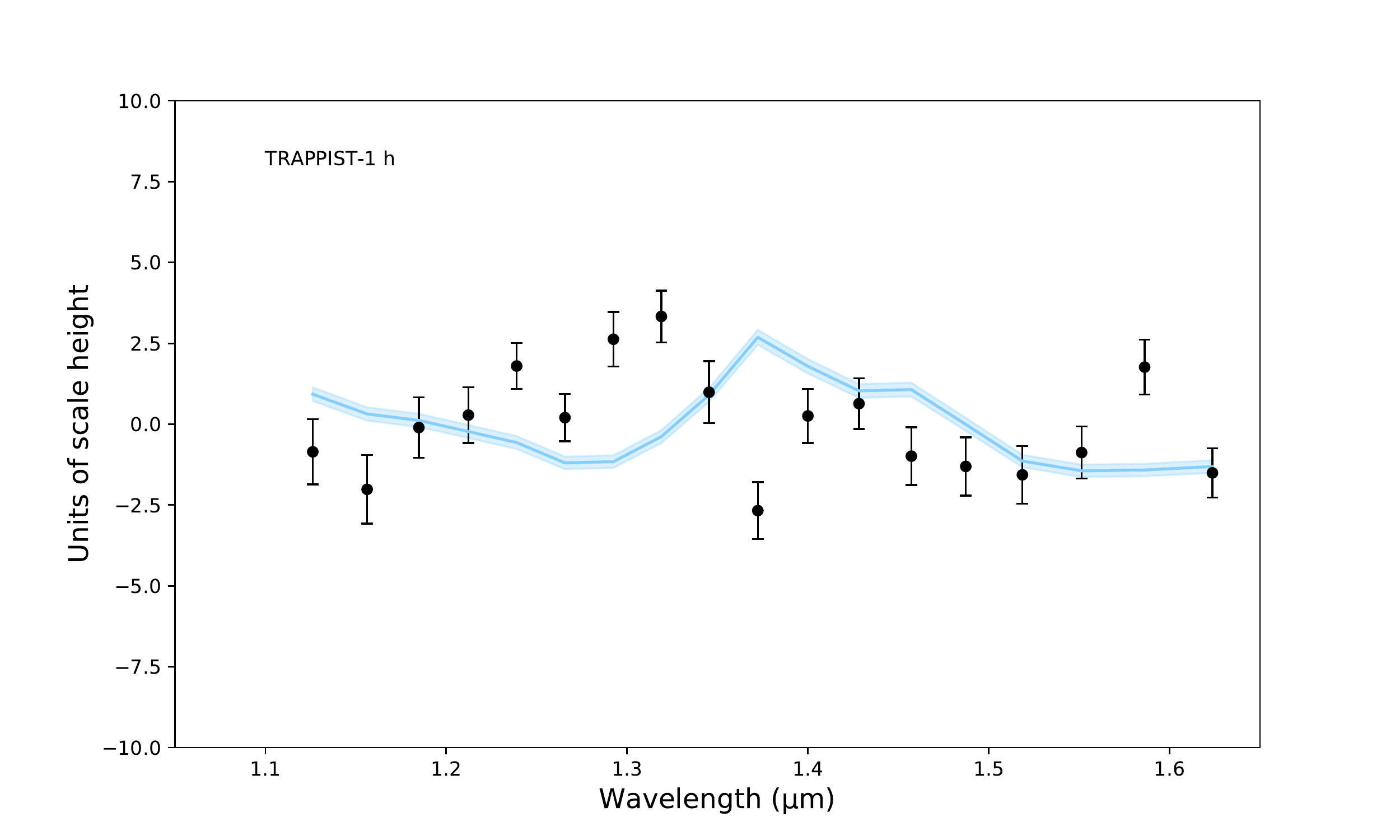}

    \caption{Comparison of the best-fit atmospheric results for TRAPPIST-1 planetary spectra in the case of a forced primary clear atmosphere with a volume mixing ratio of water fixed to 10$^{-3}$ in a H-dominated atmosphere. We used spectra from \citet{Zhang_2018} for the TRAPPIST-1 b to g retrievals and present the results in units of scale height.  }
    \label{fig:primary_T1system}
\end{figure*}

\subsection{Primary clear atmosphere}\label{sec:3.1}
We show in Sec. \ref{sec:2.1} that the HST/WFC3 extracted spectrum was compatible with either a secondary or a primary cloudy and hazy atmosphere if we retain the hypothesis of a presence of an atmosphere. In this section, we explore  the case of a primary clear atmosphere by fixing the molecular absorption of the different species to 10$^{-3}$, which forces spectral features. The temperature was allowed to vary between $\pm$20$\%$ of the equilibrium one (173K) and the radius was fitted between $\pm$50$\%$ of the published value. We tested six different opacity sources, H$_2$O, CO$_2$, CO, CH$_4$, NH$_3,$ and N$_2,$ separately by running a retrieval for each sources. We included collision-induced absorption and Rayleigh scattering  and fixed the He/H$_2$ ratio to 0.17. The atmosphere was simulated as in Sec. \ref{sec:2.1}, with 100 layers ranging between 10$^{-2}$ and 10$^{5}$ Pa.

We measured the size of a clear atmosphere in the case of TRAPPIST-1 h in those six configurations and show that a primary clear atmosphere is rejected in each case. We present in Table \ref{table:primary_clear} best-fit results from the six tested scenarios. We indicate the radius, the temperature, and the mean molecular weight, and we estimate the corresponding scale height. For comparison, we also indicate the results from the flat-line model of Sec. \ref{sec:2.1}. Statistical results, that is to say the logarithm evidence, from primary clear models are below the one of the flat line with an absolute difference of 3 or more, while including H$_2$O, CH$_4$, NH$_3$, or N$_2$. This result indicates that a primary clear atmosphere is rejected with high confidence (i.e. 3$\sigma$). Primary clear atmospheric scenarios with traces of CO or CO$_2$ have higher $\Delta$log(E), remaining below the one of the flat-line model, but they cannot be rejected as firmly as the others (see also Fig. \ref{fig:delta_logE_mmw} in Sec. \ref{sec:3.2}).

\begin{table*}[htpb]
\centering
\caption{Best-fit atmospheric results and derived parameters for a primary clear retrieval analysis. }
\label{table:primary_clear}
\begin{tabular}{ l| c c c c c c c c} \hline\hline
Model & R$_{\rm P}$(R$_{\rm \oplus}$) & T(K) & $\mu$(g/mol) & H(km) & $\chi^2$ &$\tilde{\chi}^2$ & log(E) & $\Delta$log(E)  \\
\hline
Flat-line & 0.61$\pm${0.110} & 296$\pm 225$ & 2.30 & 71.28 &64.95& 3.61  &110.55 & N/A  \\
H$_2$O & 0.70$\pm${0.003} & 140$\pm 2$ & 2.32 & 75.27 & 128.68 &7.15 & 74.78 & -35.77  \\
CO$_2$ & 0.71$\pm${0.003} & 157$\pm 26$ & 2.35 & 86.84 & 68.37&3.80 & 108.99 & -1.56\\
CO &  0.72$\pm${0.003} & 158$\pm 25$ & 2.33 & 88.84 &70.53 &3.92 & 107.93 & -2.62 \\
CH$_4$ & 0.69$\pm${0.003} & 139$\pm 2$ & 2.32 & 72.20 &146.35 &8.13 & 65.48 & -45.07\\
NH$_3$ & 0.68$\pm${0.003} & 140$\pm 3$ & 2.32 & 70.56 &107.58 & 5.98 & 85.38 & -25.17\\
N$_2$ & 0.72$\pm${0.003} & 156$\pm 26$ & 2.33 & 87.85 &69.71 & 3.87& 106.82 & -3.73\\ \hline
\end{tabular}
\tablefoot{The primary clear atmospheric scenario was simulated including the molecular absorption with a volume mixing ratio fixed to 10$^{-3}$ in a H-dominated atmosphere.}
\end{table*}
A primary clear atmosphere scenario was previously rejected for TRAPPIST-1 planets using HST/WFC3 G141 spectra \citep{de_Wit_2018, Zhang_2018, Wakeford_2019}. Performing the same exercise with a fixed 10$^{-3}$ water abundance for the others six TRAPPIST spectra from \citet{Zhang_2018}, we also confirm that a primary clear atmospheric model does not fit their spectra. We note that we simulated planet atmospheres with the same 100 layers between 10$^{-2}$ and 10$^{-5}$ even though they have a different size, radius, and mass. We present in Fig. \ref{fig:primary_T1system} the best-fit atmospheric retrieval results in the case of a Hydrogen-dominated atmosphere with water as a trace gas (the volume mixing ratio was fixed to 10$^{-3}$) for the seven planets of the TRAPPIST-1 system. The results are presented in number of scale height and we can see that TRAPPIST-1 planets are unlikely to possess a clear atmosphere dominated by hydrogen with water in a low quantity. The comparison of logarithm evidence between a flat-line model and a primary clear atmosphere for all seven planets is detailed in appendix \ref{appendix:primary_clear_T1}. This is in agreement with theoretical modelling as detailed in Sec. \ref{sec:1.3.1} and in \citet{Turbet_2020a} and \citet{Hori_2020}. 
\subsection{Steam atmosphere}\label{sec:3.2}
\begin{figure*}[htpb]
    \centering
    \includegraphics[width=\columnwidth]{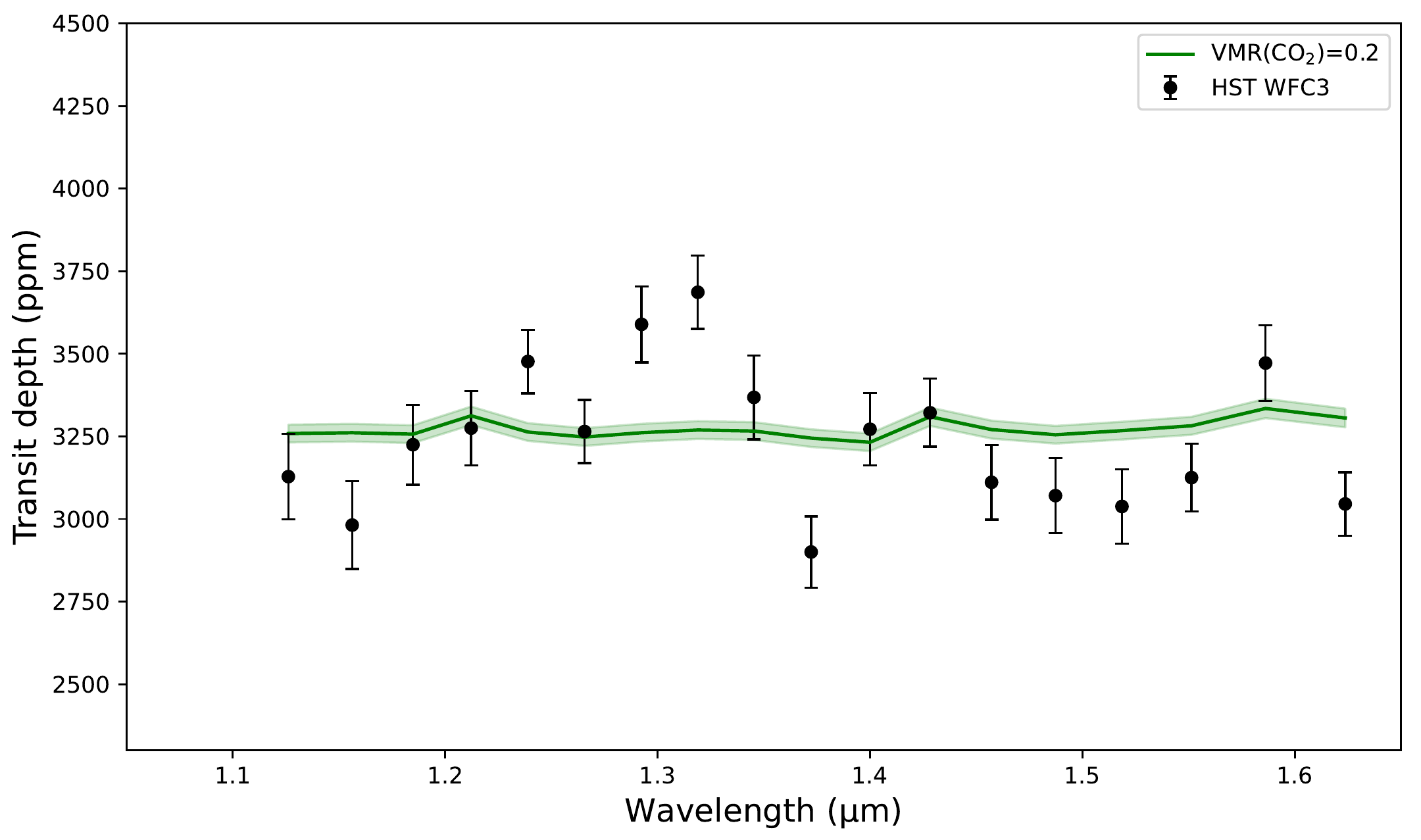}
    ~
    \includegraphics[width=\columnwidth]{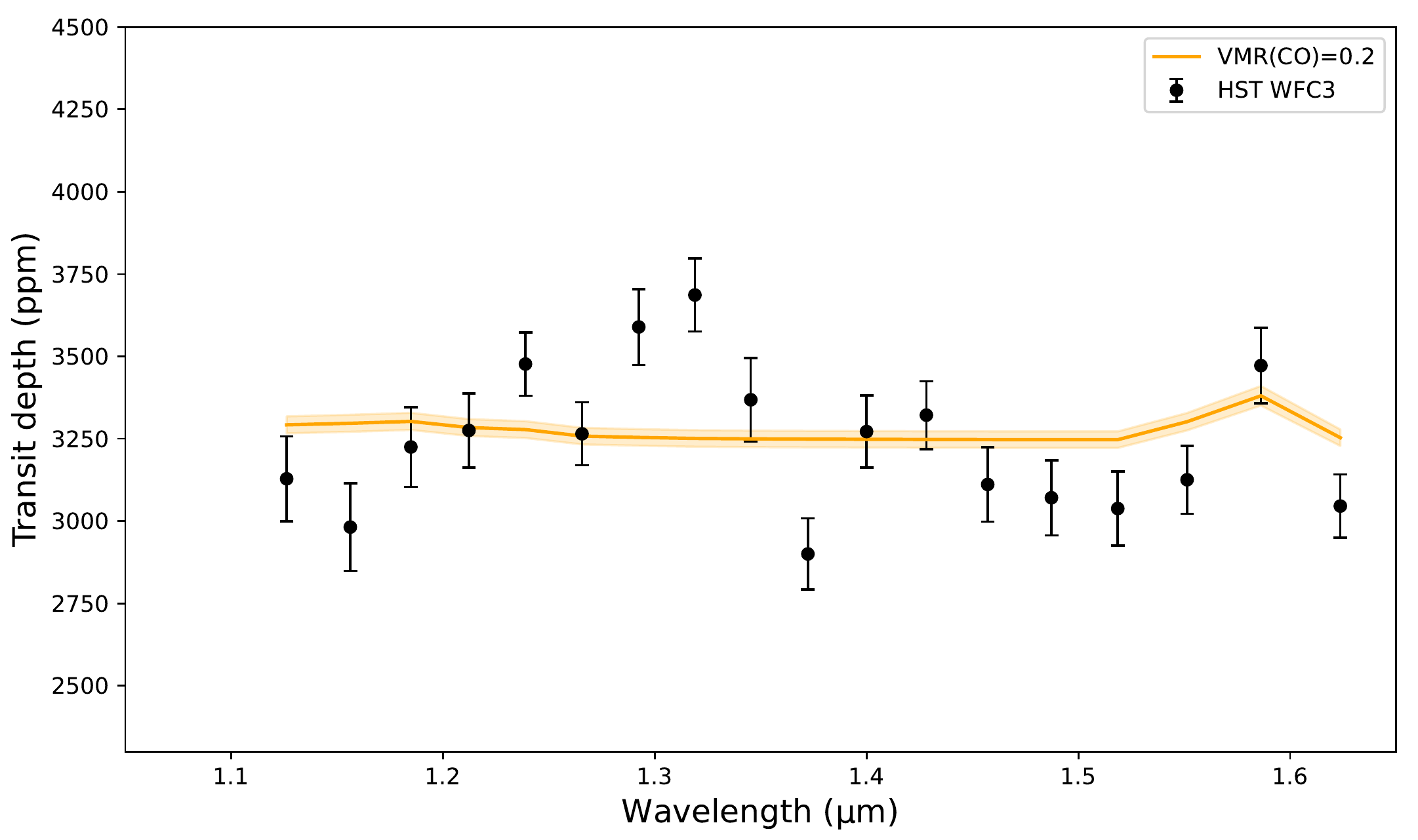}
    
    \includegraphics[width=\columnwidth]{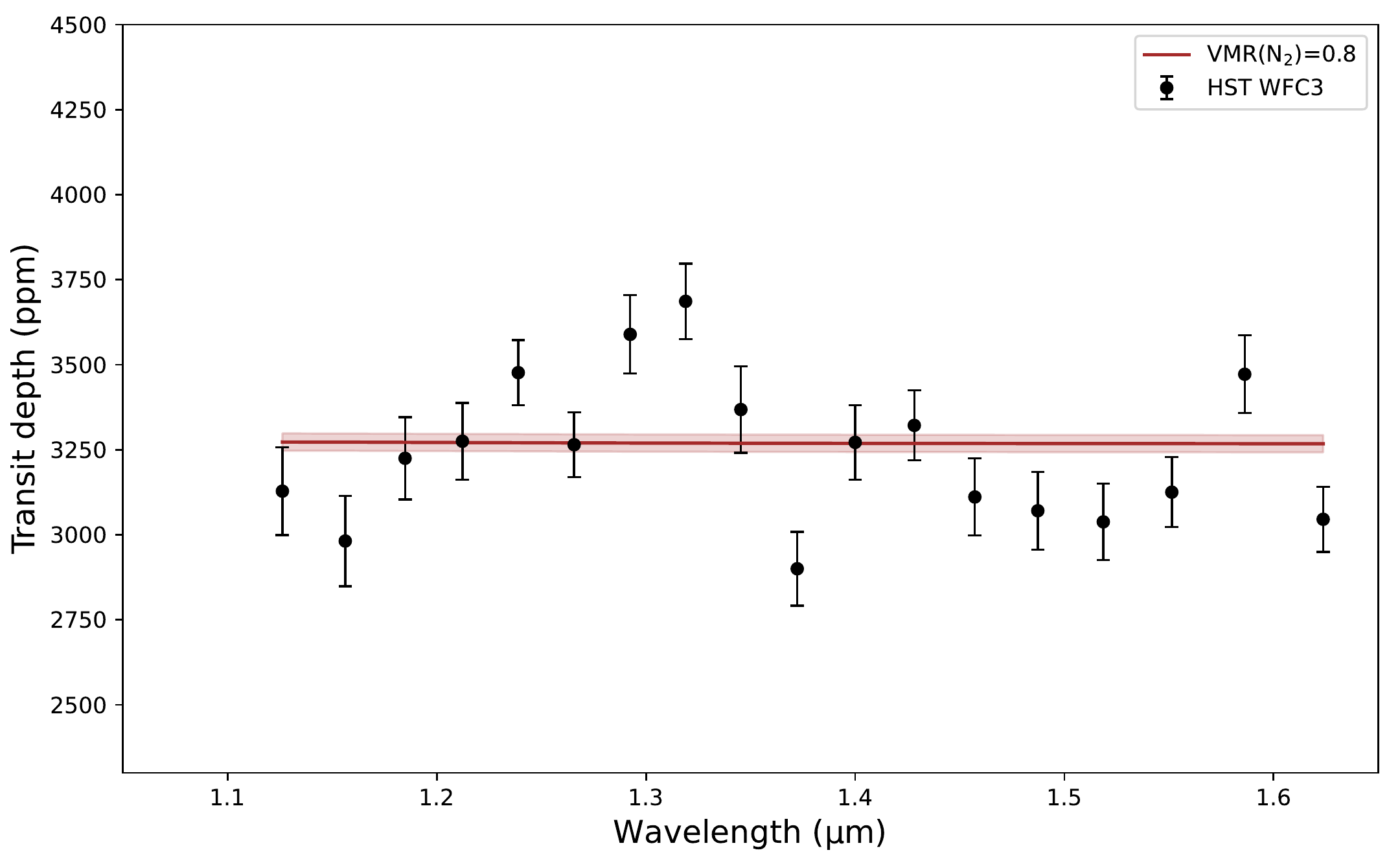}
    ~
    \includegraphics[width=\columnwidth]{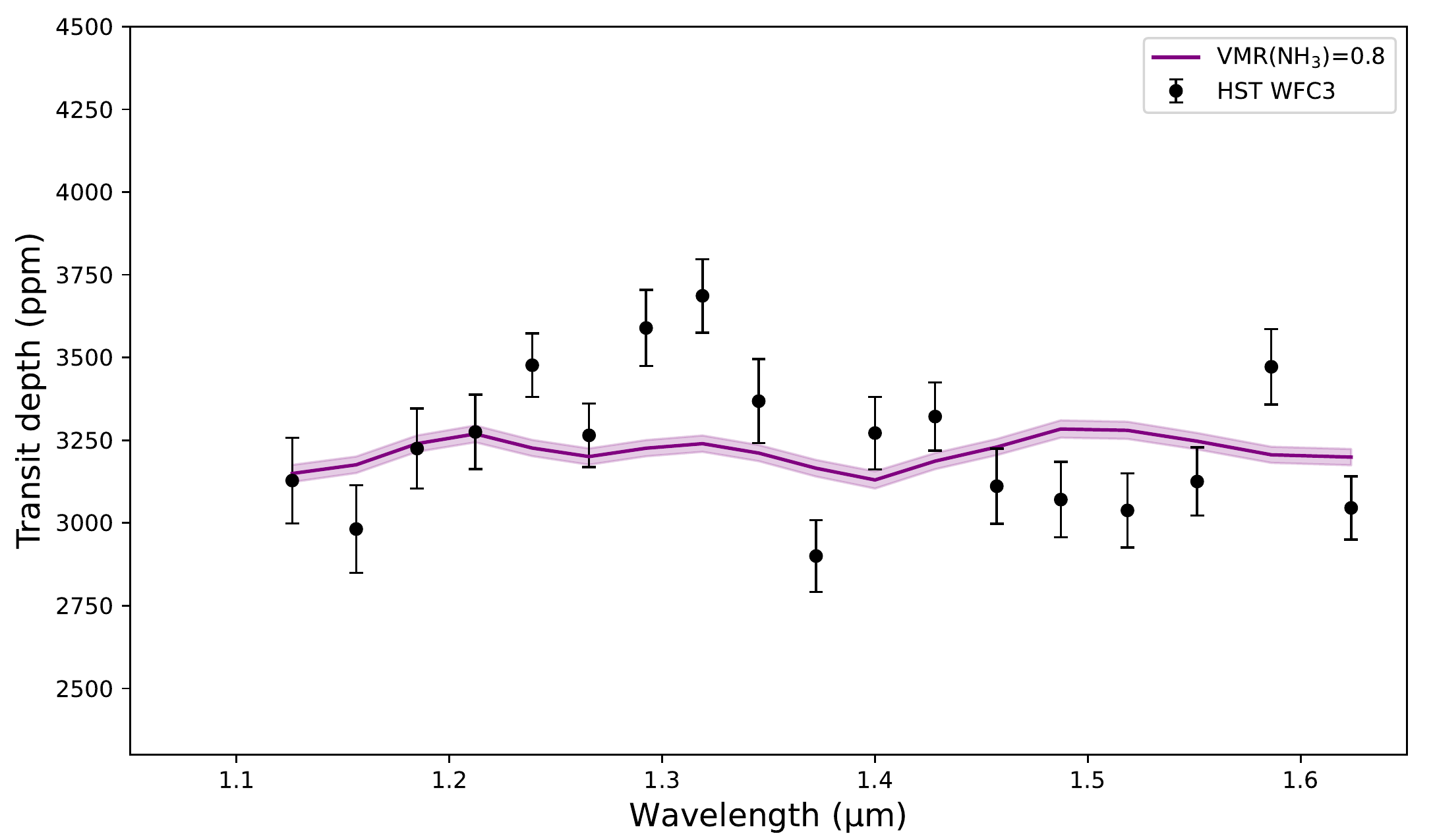}    
    \caption{Comparison of the best-fit atmospheric results for the TRAPPIST-1h spectrum in the case of four forced secondary clear atmospheres with a volume mixing ratio of CO$_2$ fixed to 0.2 (upper left),CO to 0.2 (upper right), N$_2$ to 0.8 (bottom left), and ammonia to 0.8 (bottom right). }
    \label{fig:steam_taurex}
\end{figure*} 
\begin{figure}
    \centering
    \resizebox{\hsize}{!}{\includegraphics{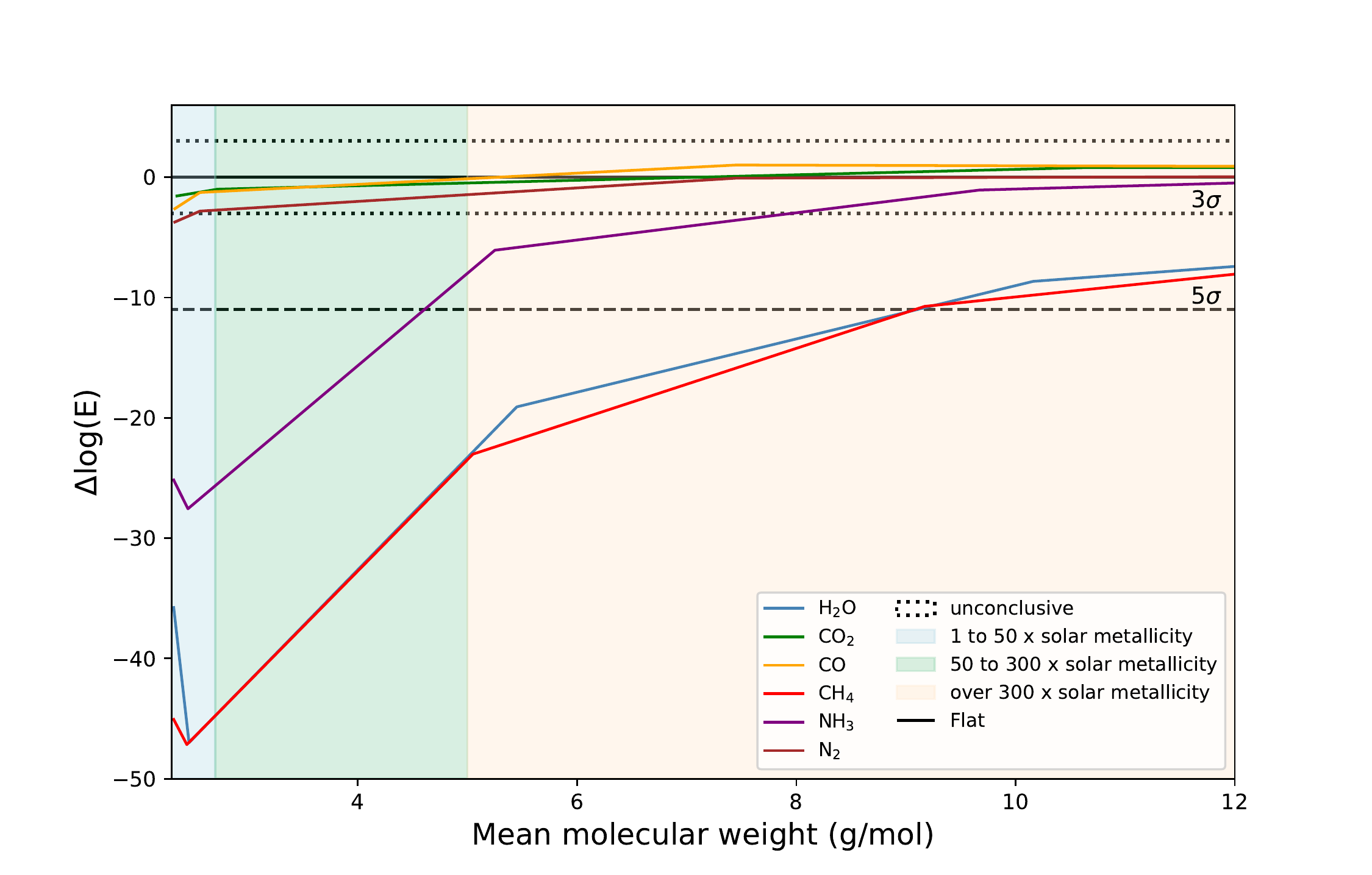}}
    \caption{Comparison of the log evidence for a flat line to that of single molecule retrievals where the abundance of the molecule is fixed and no clouds were included. We represent the Delta log(E) with respect to the mean molecular weight of the tested atmospheres. The region between dashes represents the set of Bayes factor values for which it is not possible to conclude compared to a flat line, that is with absolute $\Delta$log(E) below 3. Models below the large dashed lines are strongly disfavoured compared to the flat line. }
    \label{fig:delta_logE_mmw}
\end{figure}
From the review of the possible atmospheric scenario \citep{Turbet_2020a}, TRAPPIST-1h could have a water-, methane-, ammonia-, nitrogen-, or even a carbon-dioxide-rich atmosphere depending on the evolution of the planet, on the species collapses, and on the photo-chemistry. A steam atmosphere is unlikely for TRAPPIST-1h as it would require the planet to have retained its atmosphere and to be stable against stellar wind (see Sec. \ref{sec:1.3.1}). However, we tested those different hypothesis by using a similar approach as in Sec. \ref{sec:3.1}, but allowing for heavier atmospheres by increasing the volume mixing ratio of the tested molecular absorber from 0.01 to 0.8 progressively. Best-fit atmospheric results along with derived parameters and statistical criteria are detailed in Appendix \ref{table:secondary_clear}. We note that some forced secondary steam atmospheric models have log(E) equal to or slightly above the one of the flat-line model. The difference in log(E) is above one for one case, with the CO-rich atmosphere having a volume mixing ratio fixed to 0.2. This model has a $\Delta$log(E) of 1.01 corresponding to 2.1$\sigma$ confidence, hence a 'weak detection' in \citet{Benneke_2013} classification. The best-fit spectrum of the models presenting an elevated log(E) are plotted in Fig. \ref{fig:steam_taurex}. They correspond to the model of 20$\%$ CO$_2$ and CO and 80$\%$ of N$_2$ and NH$_3$. Carbon dioxide and carbon monoxide have similar absorption features in the HST/WFC3 wavelength range, which leads to similar best-fit results. Moreover, as the features are very small, we obtained the same volume mixing ratios for those species. We note that N$_2$ acts as a fill gas in the atmosphere as it does not have features; the best-fit spectrum is then similar to that of the flat line. We can add a CO-rich atmosphere to the possible atmospheric scenario for TRAPPIST-1h.

We note that $\Delta$log(E) remains below one for most of the other tested models, meaning that they are not statistically significant. We present in Fig. \ref{fig:delta_logE_mmw} the comparison of the log evidence for a flat line to that of single molecule retrievals from the primary clear analysis of Sec. \ref{sec:3.1} and the secondary models of Sec. \ref{sec:3.2} following the formalism of Fig.6 in \citet{Mugnai_2021}. We decided to represent the $\Delta$log(E) with respect to the mean molecular weight of the modelled atmospheres to compare the different scenarios as similar molecular abundances could lead to different weights and metallicities. Primary atmospheric models with a metallicity below 50 times solar (i.e. mmw=2.70 g/mol) are rejected with more than 5$\sigma$ confidence (i.e. the Bayes factor is inferior to -11), except for CO and CO$_2$. In addition, if the atmosphere was primary, it would be unlikely that it does not contain any water. The equivalence between abundances, the mean molecular weight, and solar metallicity is presented in Appendix Table \ref{table:secondary_clear} and a figure of all $\Delta$log(E) with respect to the abundances is presented in Appendix \ref{appendix:delta_logE}. The area between dashes represents the set of Bayes factor values for which it is not possible to conclude compared to a flat line, that is with absolute $\Delta$log(E) below 3. Models with a $\Delta$log(E) between -3 and -11 can be significantly rejected compared to a flat line, while the ones below -11 are strongly disfavoured.
\subsection{Impact of changing the spectral resolution}
\begin{figure}[htpb!]
    \centering
    \resizebox{\hsize}{!}{\includegraphics{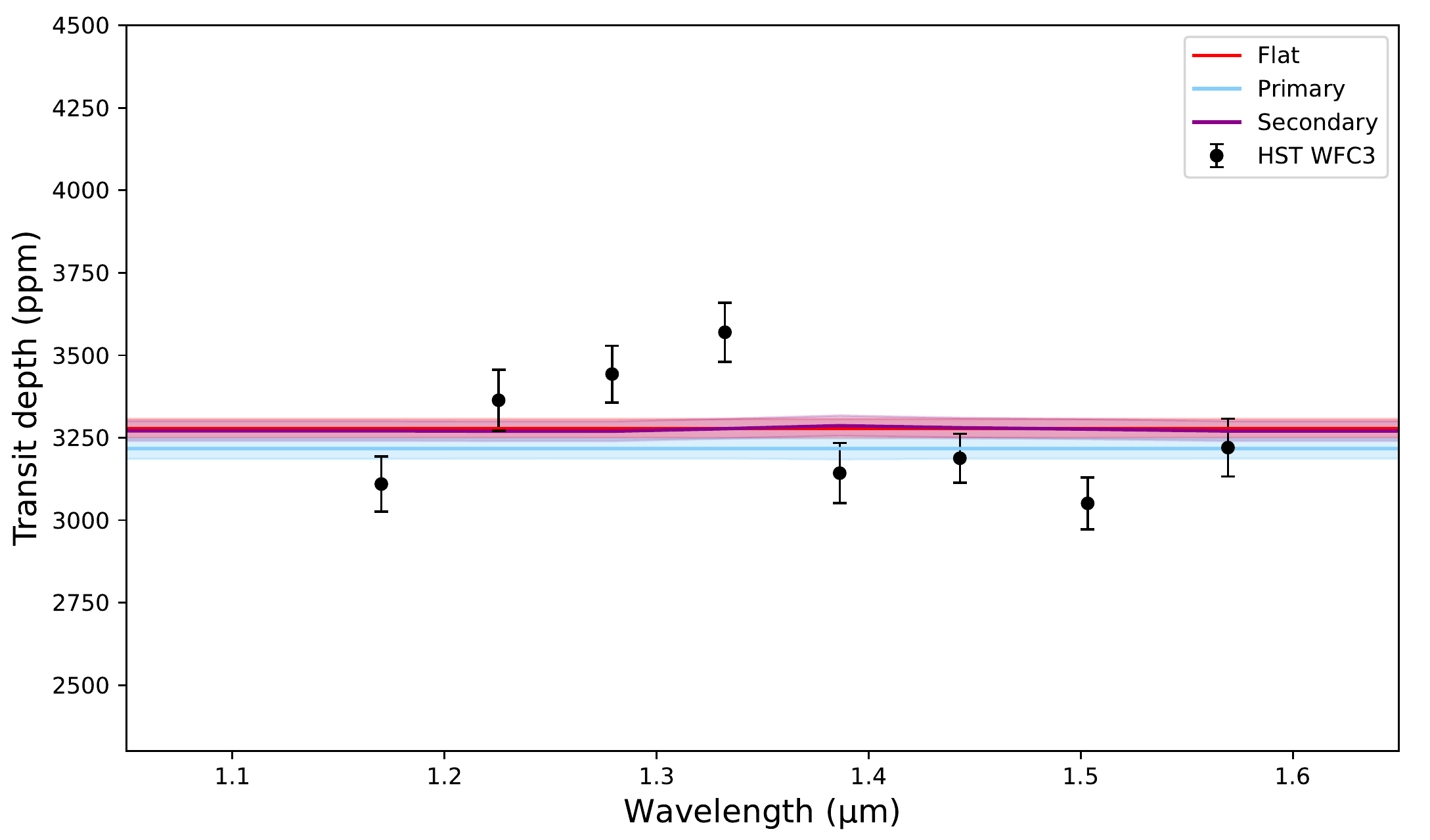}}
    \resizebox{\hsize}{!}{\includegraphics{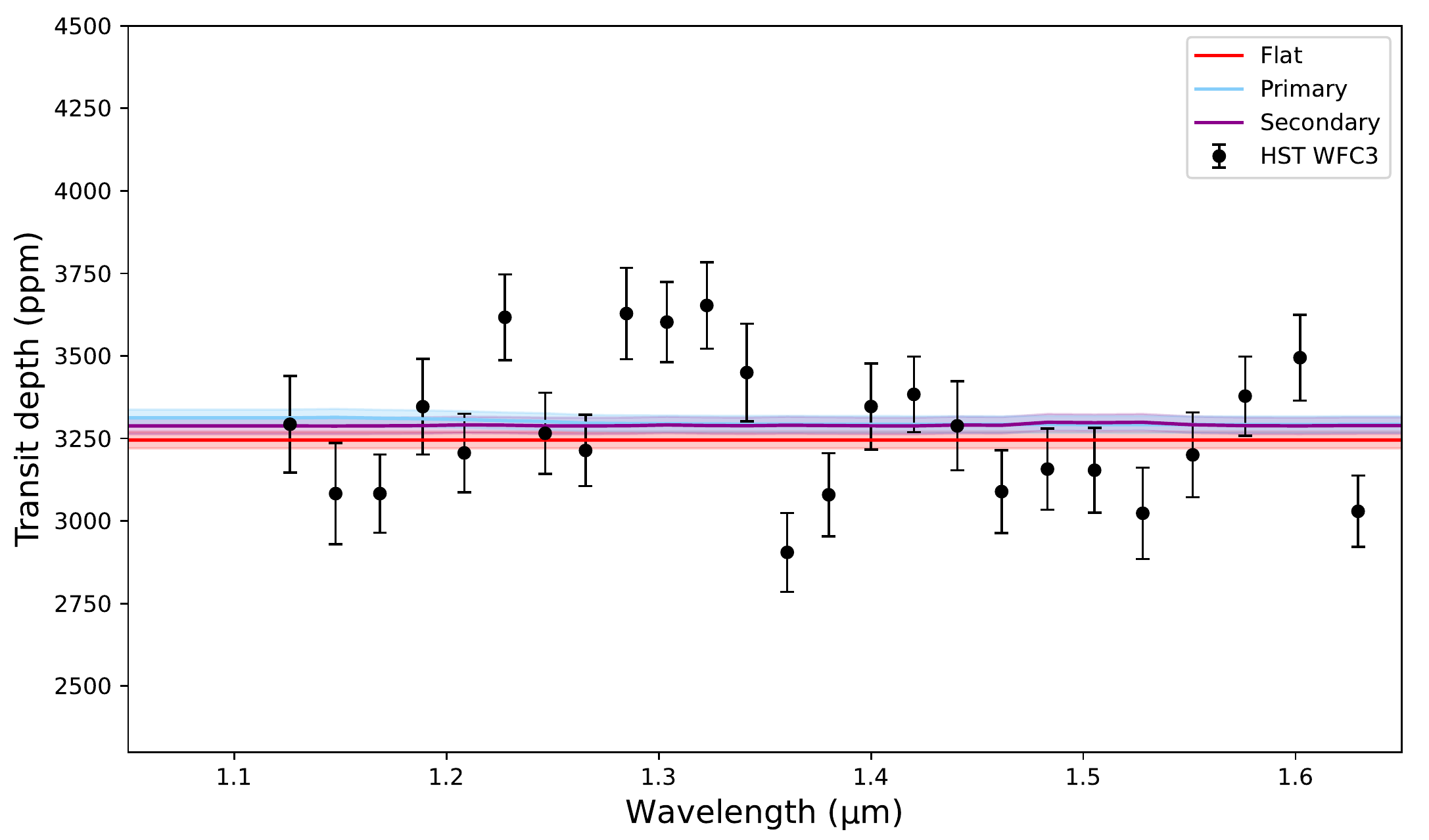}}
    \caption{Best fit models to TRAPPIST-1 h HST WFC3 G141 data using a very low (top) and a high (bottom) resolution with a resolving power of 25 and 70 around 1.4$\mu$m. }
    \label{fig:best_fit_resolution}
\end{figure}
Neither stellar contamination nor atmospheric absorption can explain the rise in the transit depth around 1.3 $\mu$m. This is probably due to scattering noise remaining after the extraction and the spectral light curve fitting. By changing the resolution of the data extraction, we investigated if the scattering at 1.3$\mu$m remains significant and if a single narrow band of the spectrum caused this 'feature'. We performed the same data analysis as in Sec. \ref{sec:2.1} using two other binning resolutions with a resolving power of 25 and 70 around 1.4$\mu$m, respectively. We also performed the same retrieval analysis using the primary, the secondary, and the flat-line set-ups on the two spectra. We obtained similar results; the flat-line model is the best fit according to the Bayes factor. 

We present in Fig.\ref{fig:best_fit_resolution} the best-fit atmospheric results on the two spectra of TRAPPIST-1 h. The log(E) of the flat line is 47.52 and 163.62 for the very low and the high resolution spectra whereas the log(E) of the primary model reaches 47.04 and 163.15, respectively. Log(E) of the secondary model are also below the one of the flat-line model, that is 47.42 and 163.42. Changing the resolution of the spectrum does not flatten or increase the rise of the transit depth at 1.3$\mu$m and it was recovered in each case. Results from Sec. \ref{sec:2.2} are confirmed while using different resolutions. A flat-line model remains the best-fit to the TRAPPIST-1 h spectrum.

\section{Conclusion}
Terrestrial planets with a secondary envelope are challenging to characterise especially given the low resolution and narrow wavelength coverage of HST. Here, we have presented a transmission spectrum of a 0.7 R${\rm \oplus}$ planet, TRAPPIST-1h, the seventh planet of the highly studied TRAPPIST-1 system. This planet is the furthest and the smallest planet of the system, yet we were able to obtain a spectrum by combining three different HST observations. We cannot make a strong claim from the analysis of the spectrum as it is better fitted using a flat line. However, we were able to rule out with more than 3$\sigma$ confidence a primary clear atmosphere, as for the other TRAPPIST planets. Given these observations, we are not yet able to distinguish between a featureless cloudy H-dominated atmosphere and a clear or cloudy secondary envelope with smaller spectral features. The two models have similar statistical significance and cannot be distinguished from retrieval analysis. We cannot completely rule out the possibility that TRAPPIST-1h has lost its atmosphere over its lifetime either as the evidence for a flat-line model is favoured. We tested secondary clear atmospheric scenarios and found that a CO-rich atmosphere with a volume mixing ratio of 0.2 in an hydrogen atmosphere obtained the best statistical result with a Bayes factor of 1.01 (i.e. a 2.1$\sigma$ detection). Yet this result is not significant enough and is mostly driven by the last points of the spectrum. This could be due to stellar activity even though all the stellar contamination models tested here were not able to reproduce those points and the rise of the transit depth around 1.3$\mu$m. Other absorbing species, such as H$_2$S or H$_2$CO, could also create features around 1.3$\mu$m, but they are unlikely to be produced in the TRAPPIST-1 h atmosphere with such a high level of absorption. The feature is likely caused by either stellar contamination, or by the planet. However, as previously stated in this paper, we cannot find an explanation for it. We note that the while these scatter data points will cause the atmospheric model to be poorly fit, the same is true of the flat and cloudy models. Therefore, as each will feature these points equally poorly, the evidence between the two will be independent of this and so not overly affected. Future observations with the James Webb Space Telescope (JWST) will hopefully remove the ambiguity; however, as shown in Fig. \ref{fig:delta_logE_mmw}, we can rule out clear H/He atmospheres with high confidence. It is then necessary to obtain more data on this planet and on the other six planets of the system to prove the presence of an atmosphere and better constrain the nature of this intriguing planetary system. 

\begin{acknowledgements} 
This study makes use of observations with the NASA/ESA Hubble Space Telescope, obtained at the
Space Telescope Science Institute (STScI) operated by the Association of Universities for Research in Astronomy. The publicly available HST observations presented here were taken as part of proposal 15304, led by Julien de Wit. These were obtained from the Hubble Archive which is part of the Mikulski Archive for Space Telescopes.
\end{acknowledgements}
 

\bibliographystyle{aa} 
\bibliography{main}

\begin{thebibliography}{91}
\expandafter\ifx\csname natexlab\endcsname\relax\def\natexlab#1{#1}\fi

\bibitem[{Abel {et~al.}(2011)Abel, Frommhold, Li, \& Hunt}]{abel_2011}
Abel, M., Frommhold, L., Li, X., \& Hunt, K.~L. 2011, The Journal of Physical
  Chemistry A, 115, 6805

\bibitem[{Abel {et~al.}(2012)Abel, Frommhold, Li, \& Hunt}]{abel_2012}
Abel, M., Frommhold, L., Li, X., \& Hunt, K.~L. 2012, The Journal of chemical
  physics, 136, 044319

\bibitem[{Al-Refaie {et~al.}(2021)Al-Refaie, Changeat, Venot, Waldmann, \&
  Tinetti}]{Alrefaie2021}
Al-Refaie, A.~F., Changeat, Q., Venot, O., Waldmann, I.~P., \& Tinetti, G.
  2021, A comparison of chemical models of exoplanet atmospheres enabled by
  TauREx 3.1

\bibitem[{{Al-Refaie} {et~al.}(2021){Al-Refaie}, {Changeat}, {Waldmann}, \&
  {Tinetti}}]{Alrefaie2019}
{Al-Refaie}, A.~F., {Changeat}, Q., {Waldmann}, I.~P., \& {Tinetti}, G. 2021,
  \apj, 917, 37

\bibitem[{Arney {et~al.}(2016)Arney, Domagal-Goldman, Meadows, Wolf,
  Schwieterman, Charnay, Claire, Hébrard, \& Trainer}]{Arney_2016}
Arney, G., Domagal-Goldman, S.~D., Meadows, V.~S., {et~al.} 2016, Astrobiology,
  16, 873–899

\bibitem[{Benneke \& Seager(2013)}]{Benneke_2013}
Benneke, B. \& Seager, S. 2013, The Astrophysical Journal, 778, 153

\bibitem[{Bolmont {et~al.}(2017)Bolmont, Selsis, Owen, Ribas, Raymond, Leconte,
  \& Gillon}]{Bolmont_2017a}
Bolmont, E., Selsis, F., Owen, J.~E., {et~al.} 2017, Monthly Notices of the
  Royal Astronomical Society, 464, 3728–3741

\bibitem[{Bourrier {et~al.}(2017{\natexlab{a}})Bourrier, Ehrenreich, Wheatley,
  Bolmont, Gillon, de~Wit, Burgasser, Jehin, Queloz, \&
  Triaud}]{Bourrier_2017b}
Bourrier, V., Ehrenreich, D., Wheatley, P.~J., {et~al.} 2017{\natexlab{a}},
  Astronomy \& Astrophysics, 599, L3

\bibitem[{Bourrier {et~al.}(2017{\natexlab{b}})Bourrier, Wit, Bolmont,
  Stamenković, Wheatley, Burgasser, Delrez, Demory, Ehrenreich, Gillon, \&
  et~al.}]{Bourrier_2017a}
Bourrier, V., Wit, J.~d., Bolmont, E., {et~al.} 2017{\natexlab{b}}, The
  Astronomical Journal, 154, 121

\bibitem[{Burdanov {et~al.}(2019)Burdanov, Lederer, Gillon, Delrez, Ducrot,
  de~Wit, Jehin, Triaud, Lidman, Spitler, \& et~al.}]{Burdanov_2019}
Burdanov, A.~Y., Lederer, S.~M., Gillon, M., {et~al.} 2019, Monthly Notices of
  the Royal Astronomical Society, 487, 1634–1652

\bibitem[{Burgasser \& Mamajek(2017)}]{Burgasser_2017}
Burgasser, A.~J. \& Mamajek, E.~E. 2017, The Astrophysical Journal, 845, 110

\bibitem[{Chubb {et~al.}(2021)Chubb, Rocchetto, Yurchenko, Min, Waldmann,
  Barstow, Mollière, Al-Refaie, Phillips, \& Tennyson}]{Chubb_2021}
Chubb, K.~L., Rocchetto, M., Yurchenko, S.~N., {et~al.} 2021, Astronomy \&
  Astrophysics, 646, A21

\bibitem[{Claret(2018)}]{Claret_2018}
Claret, A. 2018, Astronomy \& Astrophysics, 618, A20

\bibitem[{Claret {et~al.}(2012)Claret, Hauschildt, \& Witte}]{Claret_2012}
Claret, A., Hauschildt, P.~H., \& Witte, S. 2012, Astronomy \& Astrophysics,
  546, A14

\bibitem[{Claret {et~al.}(2013)Claret, Hauschildt, \& Witte}]{Claret_2013}
Claret, A., Hauschildt, P.~H., \& Witte, S. 2013, Astronomy \& Astrophysics,
  552, A16

\bibitem[{Coleman {et~al.}(2019)Coleman, Leleu, Alibert, \&
  Benz}]{Coleman_2019}
Coleman, G. A.~L., Leleu, A., Alibert, Y., \& Benz, W. 2019, Astronomy \&
  Astrophysics, 631, A7

\bibitem[{de~Wit {et~al.}(2016)de~Wit, Wakeford, Gillon, Lewis, Valenti,
  Demory, Burgasser, Burdanov, Delrez, Jehin, \& et~al.}]{de_Wit_2016}
de~Wit, J., Wakeford, H.~R., Gillon, M., {et~al.} 2016, Nature, 537, 69–72

\bibitem[{de~Wit {et~al.}(2018)de~Wit, Wakeford, Lewis, Delrez, Gillon, Selsis,
  Leconte, Demory, Bolmont, Bourrier, \& et~al.}]{de_Wit_2018}
de~Wit, J., Wakeford, H.~R., Lewis, N.~K., {et~al.} 2018, Nature Astronomy, 2,
  214–219

\bibitem[{Delrez {et~al.}(2018)Delrez, Gillon, Triaud, Demory, de~Wit, Ingalls,
  Agol, Bolmont, Burdanov, Burgasser, \& et~al.}]{Delrez_2018}
Delrez, L., Gillon, M., Triaud, A. H. M.~J., {et~al.} 2018, Monthly Notices of
  the Royal Astronomical Society, 475, 3577–3597

\bibitem[{Deming {et~al.}(2013)Deming, Wilkins, McCullough, Burrows, Fortney,
  Agol, Dobbs-Dixon, Madhusudhan, Crouzet, Desert, Gilliland, Haynes, Knutson,
  Line, Magic, Mandell, Ranjan, Charbonneau, Clampin, Seager, \&
  Showman}]{Deming_2013}
Deming, D., Wilkins, A., McCullough, P., {et~al.} 2013, The Astrophysical
  Journal, 774, 17

\bibitem[{Dencs \& Regály(2019)}]{Dencs_2019}
Dencs, Z. \& Regály, Z. 2019, Monthly Notices of the Royal Astronomical
  Society, 487, 2191–2199

\bibitem[{Dobos {et~al.}(2019)Dobos, Barr, \& Kiss}]{Dobos_2019}
Dobos, V., Barr, A.~C., \& Kiss, L.~L. 2019, Astronomy \& Astrophysics, 624, A2

\bibitem[{Dong {et~al.}(2019)Dong, Huang, \& Lingam}]{Dong_2019}
Dong, C., Huang, Z., \& Lingam, M. 2019, The Astrophysical Journal, 882, L16

\bibitem[{Dong {et~al.}(2018)Dong, Jin, Lingam, Airapetian, Ma, \& van~der
  Holst}]{Dong_2018}
Dong, C., Jin, M., Lingam, M., {et~al.} 2018, Proceedings of the National
  Academy of Sciences, 115, 260–265

\bibitem[{Dong {et~al.}(2017)Dong, Lingam, Ma, \& Cohen}]{Dong_2017}
Dong, C., Lingam, M., Ma, Y., \& Cohen, O. 2017, The Astrophysical Journal,
  837, L26

\bibitem[{Ducrot {et~al.}(2020)Ducrot, Gillon, Delrez, Agol, Rimmer, Turbet,
  Günther, Demory, Triaud, Bolmont, \& et~al.}]{Ducrot_2020}
Ducrot, E., Gillon, M., Delrez, L., {et~al.} 2020, Astronomy \& Astrophysics,
  640, A112

\bibitem[{Ducrot {et~al.}(2018)Ducrot, Sestovic, Morris, Gillon, Triaud,
  De~Wit, Thimmarayappa, Agol, Almleaky, Burdanov, \& et~al.}]{Ducrot_2018}
Ducrot, E., Sestovic, M., Morris, B.~M., {et~al.} 2018, The Astronomical
  Journal, 156, 218

\bibitem[{Edwards {et~al.}(2021)Edwards, Changeat, Mori, Anisman, Morvan, Yip,
  Tsiaras, Al-Refaie, Waldmann, \& Tinetti}]{Edwards_2021}
Edwards, B., Changeat, Q., Mori, M., {et~al.} 2021, The Astronomical Journal,
  161, 44

\bibitem[{Feroz {et~al.}(2009)Feroz, Hobson, \& Bridges}]{Feroz_2009}
Feroz, F., Hobson, M.~P., \& Bridges, M. 2009, Monthly Notices of the Royal
  Astronomical Society, 398, 1601–1614

\bibitem[{Fletcher {et~al.}(2018)Fletcher, Gustafsson, \&
  Orton}]{fletcher_2018}
Fletcher, L.~N., Gustafsson, M., \& Orton, G.~S. 2018, The Astrophysical
  Journal Supplement Series, 235, 24

\bibitem[{Foreman-Mackey {et~al.}(2013)Foreman-Mackey, Hogg, Lang, \&
  Goodman}]{Foreman_Mackey_2013}
Foreman-Mackey, D., Hogg, D.~W., Lang, D., \& Goodman, J. 2013, Publications of
  the Astronomical Society of the Pacific, 125, 306–312

\bibitem[{Gao {et~al.}(2015)Gao, Hu, Robinson, Li, \& Yung}]{Gao_2015}
Gao, P., Hu, R., Robinson, T.~D., Li, C., \& Yung, Y.~L. 2015, The
  Astrophysical Journal, 806, 12

\bibitem[{Gillon {et~al.}(2011)Gillon, Bonfils, Demory, Seager, Deming, \&
  Triaud}]{Gillon_2011}
Gillon, M., Bonfils, X., Demory, B.-O., {et~al.} 2011, Astronomy \&
  Astrophysics, 525, A32

\bibitem[{Gillon {et~al.}(2016)Gillon, Jehin, Lederer, Delrez, de~Wit,
  Burdanov, Van~Grootel, Burgasser, Triaud, Opitom, \& et~al.}]{Gillon_2016}
Gillon, M., Jehin, E., Lederer, S.~M., {et~al.} 2016, Nature, 533, 221–224

\bibitem[{Gillon {et~al.}(2017)Gillon, Triaud, Demory, Jehin, Agol, Deck,
  Lederer, de~Wit, Burdanov, Ingalls, \& et~al.}]{Gillon_2017}
Gillon, M., Triaud, A. H. M.~J., Demory, B.-O., {et~al.} 2017, Nature, 542,
  456–460

\bibitem[{Gillon {et~al.}(2013)Gillon, Triaud, Jehin, Delrez, Opitom, Magain,
  Lendl, \& Queloz}]{Gillon_2013}
Gillon, M., Triaud, A. H. M.~J., Jehin, E., {et~al.} 2013, Astronomy \&
  Astrophysics, 555, L5

\bibitem[{Grimm {et~al.}(2018)Grimm, Demory, Gillon, Dorn, Agol, Burdanov,
  Delrez, Sestovic, Triaud, Turbet, \& et~al.}]{Grimm_2018}
Grimm, S.~L., Demory, B.-O., Gillon, M., {et~al.} 2018, Astronomy \&
  Astrophysics, 613, A68

\bibitem[{Hori \& Ogihara(2020)}]{Hori_2020}
Hori, Y. \& Ogihara, M. 2020, The Astrophysical Journal, 889, 77

\bibitem[{Hu {et~al.}(2020)Hu, Peterson, \& Wolf}]{Hu_2020}
Hu, R., Peterson, L., \& Wolf, E.~T. 2020, The Astrophysical Journal, 888, 122

\bibitem[{Kass \& Raferty(1995)}]{Kass_1995}
Kass, R.~E. \& Raferty, A.~E. 1995, Journal of the American Statistical
  Association, 90, 773

\bibitem[{Kasting {et~al.}(1993)Kasting, Whitmire, \& Reynolds}]{Kasting_1993}
Kasting, J.~F., Whitmire, D.~P., \& Reynolds, R.~T. 1993, Icarus, 101

\bibitem[{{Kimura} \& {Ikoma}(2020)}]{Kimura_2020}
{Kimura}, T. \& {Ikoma}, M. 2020, \mnras, 496, 3755

\bibitem[{Kral {et~al.}(2018)Kral, Wyatt, Triaud, Marino, Thébault, \&
  Shorttle}]{Kral_2018}
Kral, Q., Wyatt, M.~C., Triaud, A. H. M.~J., {et~al.} 2018, Monthly Notices of
  the Royal Astronomical Society, 479, 2649–2672

\bibitem[{Kreidberg {et~al.}(2014)Kreidberg, Bean, Désert, Benneke, Deming,
  Stevenson, Seager, Berta-Thompson, Seifahrt, \& Homeier}]{Kreidberg_2014a}
Kreidberg, L., Bean, J.~L., Désert, J.-M., {et~al.} 2014, Nature, 505, 69–72

\bibitem[{Lammer {et~al.}(2003)Lammer, Selsis, Ribas, Guinan, Bauer, \&
  Weiss}]{lammer_2003}
Lammer, H., Selsis, F., Ribas, I., {et~al.} 2003, The Astrophysical Journal,
  598, L121

\bibitem[{Lincowski {et~al.}(2019)Lincowski, Lustig-Yaeger, \&
  Meadows}]{Lincowski_2019}
Lincowski, A.~P., Lustig-Yaeger, J., \& Meadows, V.~S. 2019, The Astronomical
  Journal, 158, 26

\bibitem[{Lincowski {et~al.}(2018)Lincowski, Meadows, Crisp, Robinson, Luger,
  Lustig-Yaeger, \& Arney}]{Lincowski_2018}
Lincowski, A.~P., Meadows, V.~S., Crisp, D., {et~al.} 2018, The Astrophysical
  Journal, 867, 76

\bibitem[{Luger {et~al.}(2017{\natexlab{a}})Luger, Lustig-Yaeger, \&
  Agol}]{Luger_2017a}
Luger, R., Lustig-Yaeger, J., \& Agol, E. 2017{\natexlab{a}}, The Astrophysical
  Journal, 851, 94

\bibitem[{Luger {et~al.}(2017{\natexlab{b}})Luger, Sestovic, Kruse, Grimm,
  Demory, Agol, Bolmont, Fabrycky, Fernandes, Van~Grootel, \&
  et~al.}]{Luger_2017b}
Luger, R., Sestovic, M., Kruse, E., {et~al.} 2017{\natexlab{b}}, Nature
  Astronomy, 1

\bibitem[{MacDonald \& Dawson(2018)}]{MacDonald_2018}
MacDonald, M.~G. \& Dawson, R.~I. 2018, The Astronomical Journal, 156, 228

\bibitem[{Makarov {et~al.}(2018)Makarov, Berghea, \& Efroimsky}]{Makarov_2018}
Makarov, V.~V., Berghea, C.~T., \& Efroimsky, M. 2018, The Astrophysical
  Journal, 857, 142

\bibitem[{Moran {et~al.}(2018)Moran, Hörst, Batalha, Lewis, \&
  Wakeford}]{Moran_2018}
Moran, S.~E., Hörst, S.~M., Batalha, N.~E., Lewis, N.~K., \& Wakeford, H.~R.
  2018, The Astronomical Journal, 156, 252

\bibitem[{Morello {et~al.}(2020)Morello, Claret, Martin-Lagarde, Cossou,
  Tsiaras, \& Lagage}]{Morello_2020}
Morello, G., Claret, A., Martin-Lagarde, M., {et~al.} 2020, The Astronomical
  Journal, 159, 75

\bibitem[{Morris {et~al.}(2018)Morris, Agol, Davenport, \&
  Hawley}]{Morris_2018a}
Morris, B.~M., Agol, E., Davenport, J. R.~A., \& Hawley, S.~L. 2018, The
  Astrophysical Journal, 857, 39

\bibitem[{Mugnai {et~al.}(2021)Mugnai, Modirrousta-Galian, Edwards, Changeat,
  Bouwman, Morello, Al-Refaie, Baeyens, Bieger, Blain, \& et~al.}]{Mugnai_2021}
Mugnai, L.~V., Modirrousta-Galian, D., Edwards, B., {et~al.} 2021, The
  Astronomical Journal, 161, 284

\bibitem[{Ormel {et~al.}(2017)Ormel, Liu, \& Schoonenberg}]{Ormel_2017}
Ormel, C.~W., Liu, B., \& Schoonenberg, D. 2017, Astronomy \& Astrophysics,
  604, A1

\bibitem[{Papaloizou {et~al.}(2018)Papaloizou, Szuszkiewicz, \&
  Terquem}]{Papaloizou_2018}
Papaloizou, J. C.~B., Szuszkiewicz, E., \& Terquem, C. 2018, Monthly Notices of
  the Royal Astronomical Society, 476, 5032–5056

\bibitem[{Pierrehumbert \& Gaidos(2011)}]{Pierrehumbert_2011}
Pierrehumbert, R. \& Gaidos, E. 2011, The Astrophysical Journal, 734, L13

\bibitem[{Polyansky {et~al.}(2018)Polyansky, Kyuberis, Zobov, Tennyson,
  Yurchenko, \& Lodi}]{Polyansky_2018}
Polyansky, O.~L., Kyuberis, A.~A., Zobov, N.~F., {et~al.} 2018, Monthly Notices
  of the Royal Astronomical Society, 480, 2597–2608

\bibitem[{Rackham {et~al.}(2018)Rackham, Apai, \& Giampapa}]{Rackham_2018}
Rackham, B.~V., Apai, D., \& Giampapa, M.~S. 2018, The Astrophysical Journal,
  853, 122

\bibitem[{Ramirez \& Kaltenegger(2014)}]{Ramirez_2014}
Ramirez, R.~M. \& Kaltenegger, L. 2014, The Astrophysical Journal, 797, L25

\bibitem[{Ramirez \& Kaltenegger(2017)}]{Ramirez_2017}
Ramirez, R.~M. \& Kaltenegger, L. 2017, The Astrophysical Journal, 837, L4

\bibitem[{Rothman {et~al.}(1987)Rothman, Gamache, Goldman, Brown, Toth,
  Pickett, Poynter, Flaud, Camy-Peyret, Barbe, Husson, Rinsland, \&
  Smith}]{Rothman_1987}
Rothman, L.~S., Gamache, R.~R., Goldman, A., {et~al.} 1987, Appl. Opt., 26,
  4058

\bibitem[{{Rothman} {et~al.}(2013){Rothman}, {Gordon}, {Babikov}, {Barbe},
  {Chris Benner}, {Bernath}, {Birk}, {Bizzocchi}, {Boudon}, {Brown},
  {Campargue}, {Chance}, {Cohen}, {Coudert}, {Devi}, {Drouin}, {Fayt}, {Flaud},
  {Gamache}, {Harrison}, {Hartmann}, {Hill}, {Hodges}, {Jacquemart}, {Jolly},
  {Lamouroux}, {Le Roy}, {Li}, {Long}, {Lyulin}, {Mackie}, {Massie},
  {Mikhailenko}, {M{\"u}ller}, {Naumenko}, {Nikitin}, {Orphal}, {Perevalov},
  {Perrin}, {Polovtseva}, {Richard}, {Smith}, {Starikova}, {Sung}, {Tashkun},
  {Tennyson}, {Toon}, {Tyuterev}, \& {Wagner}}]{Rothman_2013}
{Rothman}, L.~S., {Gordon}, I.~E., {Babikov}, Y., {et~al.} 2013, \jqsrt, 130, 4

\bibitem[{{Rothman} {et~al.}(2010){Rothman}, {Gordon}, {Barber}, {Dothe},
  {Gamache}, {Goldman}, {Perevalov}, {Tashkun}, \& {Tennyson}}]{Rothman_2010}
{Rothman}, L.~S., {Gordon}, I.~E., {Barber}, R.~J., {et~al.} 2010, \jqsrt, 111,
  2139

\bibitem[{Rugheimer {et~al.}(2015)Rugheimer, Segura, Kaltenegger, \&
  Sasselov}]{Rugheimer_2015}
Rugheimer, S., Segura, A., Kaltenegger, L., \& Sasselov, D. 2015, The
  Astrophysical Journal, 806, 137

\bibitem[{Sagan \& Chyba(1997)}]{Sagan_1997}
Sagan, C. \& Chyba, C. 1997, Science, 276, 1217

\bibitem[{Tamayo {et~al.}(2017)Tamayo, Rein, Petrovich, \&
  Murray}]{Tamayo_2017}
Tamayo, D., Rein, H., Petrovich, C., \& Murray, N. 2017, The Astrophysical
  Journal, 840, L19

\bibitem[{Tennyson \& Yurchenko(2018)}]{Tennyson_2018}
Tennyson, J. \& Yurchenko, S. 2018, Atoms, 6, 26

\bibitem[{Tennyson {et~al.}(2016)Tennyson, Yurchenko, Al-Refaie, Barton, Chubb,
  Coles, Diamantopoulou, Gorman, Hill, Lam, \& et~al.}]{Tennyson_2016}
Tennyson, J., Yurchenko, S.~N., Al-Refaie, A.~F., {et~al.} 2016, Journal of
  Molecular Spectroscopy, 327, 73–94

\bibitem[{Tian \& Ida(2015)}]{Tian_Ida_2015}
Tian, F.~D. \& Ida, S. 2015, Nature Geoscience, 8, 177

\bibitem[{Trotta(2008)}]{Trotta_2008}
Trotta, R. 2008, Contemporary Physics, 49, 71–104

\bibitem[{Tsiaras {et~al.}(2016{\natexlab{a}})Tsiaras, Rocchetto, Waldmann,
  Venot, Varley, Morello, Damiano, Tinetti, Barton, Yurchenko, \&
  et~al.}]{Tsiaras_2016a}
Tsiaras, A., Rocchetto, M., Waldmann, I.~P., {et~al.} 2016{\natexlab{a}}, The
  Astrophysical Journal, 820, 99

\bibitem[{Tsiaras {et~al.}(2016{\natexlab{b}})Tsiaras, Waldmann, Rocchetto,
  Varley, Morello, Damiano, \& Tinetti}]{Tsiaras_2016b}
Tsiaras, A., Waldmann, I.~P., Rocchetto, M., {et~al.} 2016{\natexlab{b}}, The
  Astrophysical Journal, 832, 17

\bibitem[{Tsiaras {et~al.}(2018)Tsiaras, Waldmann, Zingales, Rocchetto,
  Morello, Damiano, Karpouzas, Tinetti, McKemmish, Tennyson, \&
  et~al.}]{Tsiaras_2018}
Tsiaras, A., Waldmann, I.~P., Zingales, T., {et~al.} 2018, The Astronomical
  Journal, 155, 156

\bibitem[{Turbet {et~al.}(2020{\natexlab{a}})Turbet, Bolmont, Bourrier, Demory,
  Leconte, Owen, \& Wolf}]{Turbet_2020b}
Turbet, M., Bolmont, E., Bourrier, V., {et~al.} 2020{\natexlab{a}}, Space
  Science Reviews, 216

\bibitem[{Turbet {et~al.}(2020{\natexlab{b}})Turbet, Bolmont, Ehrenreich,
  Gratier, Leconte, Selsis, Hara, \& Lovis}]{Turbet_2020a}
Turbet, M., Bolmont, E., Ehrenreich, D., {et~al.} 2020{\natexlab{b}}, Astronomy
  \& Astrophysics, 638, A41

\bibitem[{Turbet {et~al.}(2018)Turbet, Bolmont, Leconte, Forget, Selsis, Tobie,
  Caldas, Naar, \& Gillon}]{Turbet_2018}
Turbet, M., Bolmont, E., Leconte, J., {et~al.} 2018, Astronomy \& Astrophysics,
  612, A86

\bibitem[{Turbet {et~al.}(2019{\natexlab{a}})Turbet, Ehrenreich, Lovis,
  Bolmont, \& Fauchez}]{Turbet_2019a}
Turbet, M., Ehrenreich, D., Lovis, C., Bolmont, E., \& Fauchez, T.
  2019{\natexlab{a}}, Astronomy \& Astrophysics, 628, A12

\bibitem[{Turbet {et~al.}(2019{\natexlab{b}})Turbet, Tran, Pirali, Forget,
  Boulet, \& Hartmann}]{Turbet_2019b}
Turbet, M., Tran, H., Pirali, O., {et~al.} 2019{\natexlab{b}}, Icarus, 321,
  189–199

\bibitem[{Vida {et~al.}(2017)Vida, Kővári, Pál, Oláh, \&
  Kriskovics}]{Vida_2017}
Vida, K., Kővári, Z., Pál, A., Oláh, K., \& Kriskovics, L. 2017, The
  Astrophysical Journal, 841, 124

\bibitem[{Vidal-Madjar {et~al.}(2003)Vidal-Madjar, Lecavelier~des Etangs,
  Désert, Ballester, Ferlet, Hébrard, \& Mayor}]{Vidal-Madjar_2003}
Vidal-Madjar, A., Lecavelier~des Etangs, A., Désert, J.~M., {et~al.} 2003,
  Nature, 422

\bibitem[{Wakeford {et~al.}(2019)Wakeford, Lewis, Fowler, Bruno, Wilson, Moran,
  Valenti, Batalha, Filippazzo, Bourrier, \& et~al.}]{Wakeford_2019}
Wakeford, H.~R., Lewis, N.~K., Fowler, J., {et~al.} 2019, The Astronomical
  Journal, 157, 11

\bibitem[{Wheatley {et~al.}(2017)Wheatley, Louden, Bourrier, Ehrenreich, \&
  Gillon}]{Wheatley_2017}
Wheatley, P.~J., Louden, T., Bourrier, V., Ehrenreich, D., \& Gillon, M. 2017,
  Monthly Notices of the Royal Astronomical Society: Letters, 465, L74–L78

\bibitem[{Wolf \& Toon(2010)}]{Wolf_2010}
Wolf, E.~T. \& Toon, O.~B. 2010, Science, 328, 1266

\bibitem[{Wordsworth {et~al.}(2017)Wordsworth, Kalugina, Lokshtanov, Vigasin,
  Ehlmann, Head, Sanders, \& Wang}]{Wordsworth_2017}
Wordsworth, R., Kalugina, Y., Lokshtanov, S., {et~al.} 2017, Geophysical
  Research Letters, 44, 665–671

\bibitem[{Wordsworth {et~al.}(2018)Wordsworth, Schaefer, \&
  Fischer}]{Wordsworth_2018}
Wordsworth, R.~D., Schaefer, L.~K., \& Fischer, R.~A. 2018, The Astronomical
  Journal, 155, 195

\bibitem[{Yip {et~al.}(2020)Yip, Changeat, Edwards, Morvan, Chubb, Tsiaras,
  Waldmann, \& Tinetti}]{Yip_2020}
Yip, K.~H., Changeat, Q., Edwards, B., {et~al.} 2020, The Astronomical Journal,
  161, 4

\bibitem[{Yurchenko {et~al.}(2011)Yurchenko, Barber, \&
  Tennyson}]{Yurchenko_2011}
Yurchenko, S.~N., Barber, R.~J., \& Tennyson, J. 2011, Monthly Notices of the
  Royal Astronomical Society, 413, 1828–1834

\bibitem[{Yurchenko \& Tennyson(2014)}]{Yurchenko_2014}
Yurchenko, S.~N. \& Tennyson, J. 2014, Monthly Notices of the Royal
  Astronomical Society, 440, 1649–1661

\bibitem[{Zhang {et~al.}(2018)Zhang, Zhou, Rackham, \& Apai}]{Zhang_2018}
Zhang, Z., Zhou, Y., Rackham, B.~V., \& Apai, D. 2018, The Astronomical
  Journal, 156, 178

\end{thebibliography}

\begin{appendix}
\section{Additional tables}
\begin{table}[htpb]
\centering
\caption{Best-fit logarithm evidence of a flat-line and a primary clear model for the seven planets of the TRAPPIST-1 system.}
\label{appendix:primary_clear_T1}
\begin{tabular}{ l |c c } \hline\hline
Model & Flat-line & Primary clear \\
\hline
TRAPPIST-1 b & 76.54 & 45.37 \\
TRAPPIST-1 c & 77.42 & 68.28 \\
TRAPPIST-1 d & 71.92 & 37.84 \\
TRAPPIST-1 e & 82.16 & 58.06 \\
TRAPPIST-1 f & 79.58 & 65.72 \\
TRAPPIST-1 g & 79.73 & 75.73 \\
TRAPPIST-1 h & 110.55 & 74.78 \\ \hline
\end{tabular}
\tablefoot{The primary clear atmospheric scenario was simulated including H$_2$O with a volume mixing ratio fixed to 10$^{-3}$ in a H-dominated atmosphere for all seven planets of the TRAPPIST-1 system. }
\end{table}

\begin{table}[htpb!]
\centering
\caption{Statistical results of the stellar contamination modelling on combined TRAPPIST-1h HST WFC3 G141 data and the K2 photometry value from \citet{Luger_2017a}.}
\label{appendix:stats_K2}
\begin{tabular}{ l|c c  } \hline\hline
Model & $\chi^2$ &$\tilde{\chi}^2$\\ \hline
Flat-line & 72.14  & 3.80\\
Stellar \citet{Zhang_2018} &73.74  & 3.81\\
Stellar \citet{Wakeford_2019} & 197.75 & 10.15\\
Stellar \citet{Morris_2018a} &99.37  &5.23 \\

\end{tabular}
\tablefoot{Chi-squared ($\chi^2$) and reduced chi-squared ($\tilde{\chi}^2$2) were computed using the result of the stellar contamination models.}
\end{table}

\begin{table*}[htpb!]
\centering
\caption{Best-fit atmospheric results and derived parameters for secondary retrieval analysis.}
\label{table:secondary_clear}
\begin{tabular}{ l |c c c c c c c c c c } \hline\hline
Model & R$_{\rm P}$(R$_{\rm \oplus}$) & T(K) & $\mu$(g/mol) & H(km) & met (x solar)& $\chi^2$ &$\tilde{\chi}^2$ & log(E) & $\Delta$log(E)\\
\hline
Flat-line & 0.61$\pm${0.11} & 296$\pm 225$ & 2.30 & 71.28 & 1 &64.95 & 3.61 & 110.55 & N/A \\
\hline
VMR(H$_2$O) & \\
0.01 & 0.69$\pm${0.003} & 139$\pm 2$ & 2.46 & 69.09 & 20 & 150.67 & 8.37 & 63.55 & -47.00 \\
0.2 & 0.71$\pm${0.003} & 142$\pm 6$ & 5.45 & 32.98 & 350 &97.25 & 5.40 & 91.46 & -19.09 \\
0.5 & 0.72$\pm${0.003} & 147$\pm 15$ & 10.16 & 18.78 & 900 &78.43 & 4.35 &101.89& -8.66 \\
0.8 & 0.72$\pm${0.003} & 153$\pm 24$ & 14.87 & 13.50 & 1400 &72.92 & 4.05& 105.07 & -5.48\\
\hline

VMR(CO$_2$) & \\
0.01 & 0.71$\pm${0.003} & 155$\pm 22$ & 2.72 & 73.68 &50 & 64.22 & 3.57 &109.55 & -1.00\\
0.2 & 0.72$\pm${0.003} & 173$\pm 24$ & 10.65 & 21.48& 950 &63.69 & 3.54& 111.35 & 0.80\\
0.5 & 0.72$\pm${0.003} & 174$\pm 24$ & 23.16 & 10.03 &2400 &63.93 &3.55& 111.26 & 0.71 \\
0.8 & 0.73$\pm${0.003} & 174$\pm 25$ & 35.67 & 6.54 &3800 &64.30 & 3.57& 110.98 & 0.43\\
\hline

VMR(CO) & \\
0.01 & 0.72$\pm${0.003} & 159$\pm 27$ & 2.56 & 81.36 & 30 &68.36 &3.80 &109.28 & -1.27\\
0.2 & 0.72$\pm${0.003} & 174$\pm 24$ & 7.45 & 31.19 & 600 & 63.23 & 3.51 &111.56 & 1.01\\
0.5 & 0.73$\pm${0.003} & 175$\pm 26$ & 15.16 & 15.45 &1500 & 63.64 & 3.54 &111.38 & 0.83 \\
0.8 & 0.73$\pm${0.003} & 174$\pm 24$ & 22.87 & 10.21 &2300 & 63.97 & 3.55 &111.15 & 0.60\\
\hline

VMR(CH$_4$) & \\
0.01 & 0.68$\pm${0.003} & 139$\pm 3$ & 2.44 & 66.32 & 20 & 170.07 & 9.45 &63.39 & -47.16\\
0.2 & 0.69$\pm${0.003} & 141$\pm 5$ & 5.05 & 33.60 & 300 &104.28 & 5.79 & 87.53 & -23.02\\
0.5 & 0.70$\pm${0.003} & 145$\pm 12$ & 9.17 & 19.87 & 800 &81.64&4.54& 99.81 & -10.74\\
0.8 & 0.71$\pm${0.004} & 150$\pm 18$ & 13.30 & 14.43 & 1300 & 74.92 & 4.16 & 103.72 & -6.83\\
\hline

VMR(NH$_3$) & \\
0.01 & 0.66$\pm${0.003} & 140$\pm 3$ & 2.45 & 63.82 & 20 & 112.31 & 6.24 & 82.99 & -27.56\\
0.2 & 0.69$\pm${0.004} & 145$\pm 11$ & 5.25 & 33.95 & 350 &72.48 &4.03& 104.48 & -6.07\\
0.5 & 0.70$\pm${0.005} & 158$\pm 26$ & 9.67 & 20.39 & 850 &67.57 & 3.75 & 109.48 & -1.07\\
0.8 & 0.71$\pm${0.004} & 168$\pm 26$ & 14.09 & 15.11 & 1500 & 65.14 & 3.62 &110.59 & 0.04\\
\hline

VMR(N$_2$) &\\
0.01 & 0.72$\pm${0.003} & 159$\pm 27$ & 2.56 & 81.54 & 30 & 71.67 & 3.98 & 107.73 & -2.82\\
0.2 & 0.73$\pm${0.003} & 171$\pm 25$ & 7.45 & 30.78 & 600 & 65.33 & 3.63 & 110.47 & -0.08\\
0.5 & 0.73$\pm${0.003} & 174$\pm 24$ & 15.16 & 15.36 & 1500 &64.96 & 3.61 & 110.67 & 0.12\\
0.8 & 0.73$\pm${0.003} & 171$\pm 25$ & 22.87 & 10.05 & 2300 & 64.72 & 3.59 & 110.72 & 0.17\\

\hline
\end{tabular}
\tablefoot{Secondary atmospheric scenarios were simulated including molecular absorption with a fixed volume mixing ratio increasing progressively from 0.01 to 0.8. }
\end{table*}

\section{Additional figures}
\begin{figure*}
    \centering
    \includegraphics[width=\columnwidth]{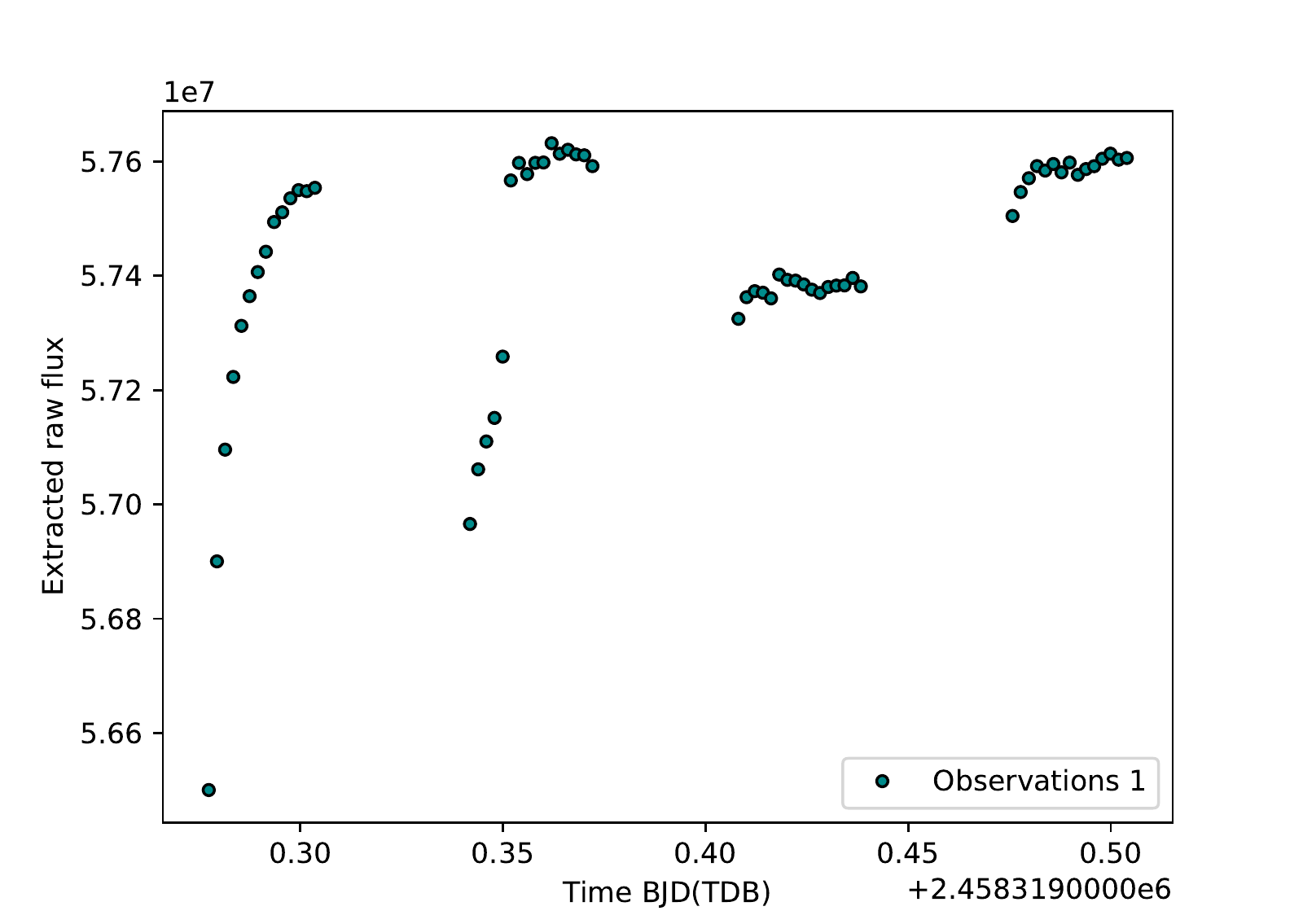}
    ~
    \includegraphics[width=\columnwidth]{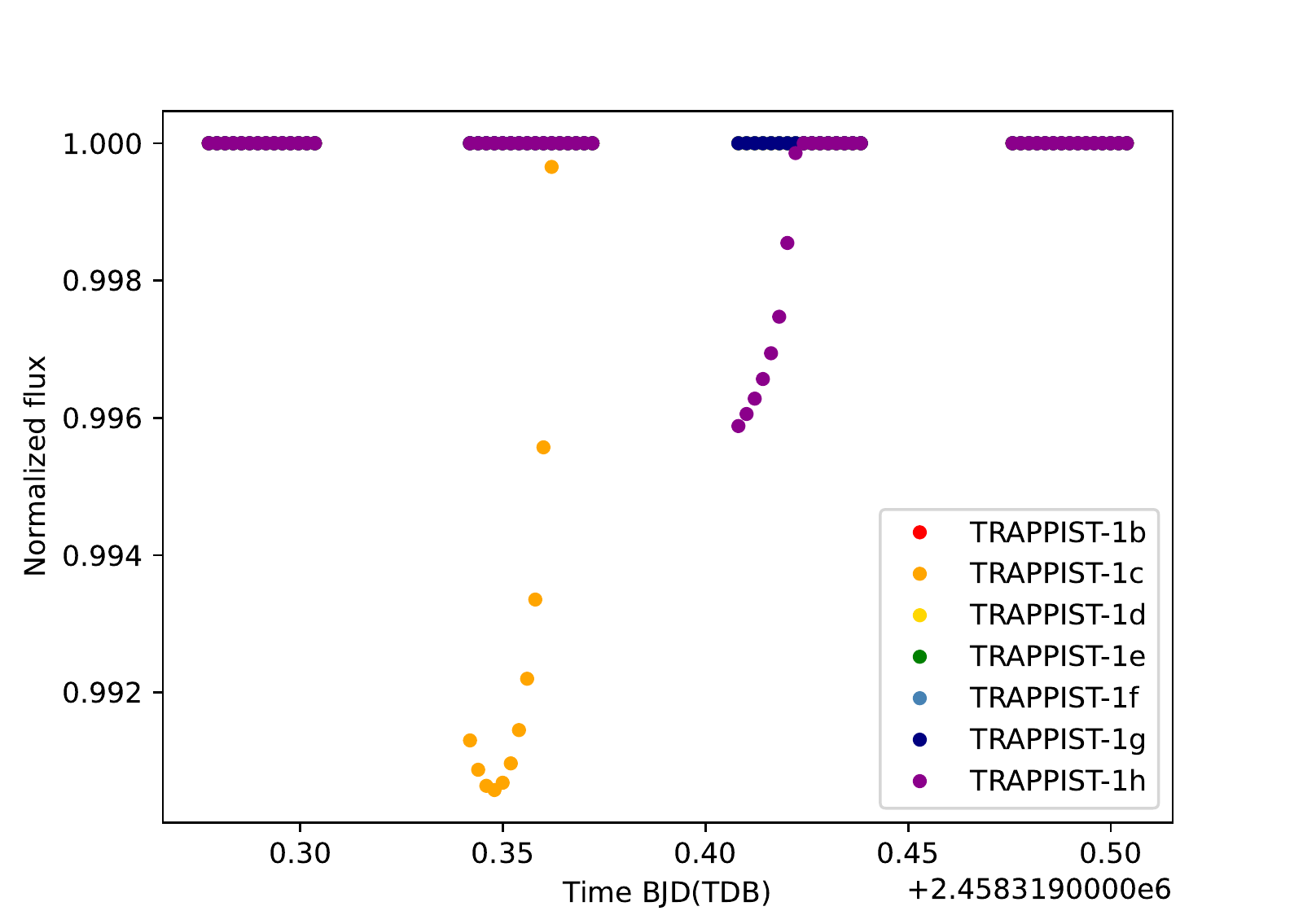}
    
    \includegraphics[width=\columnwidth]{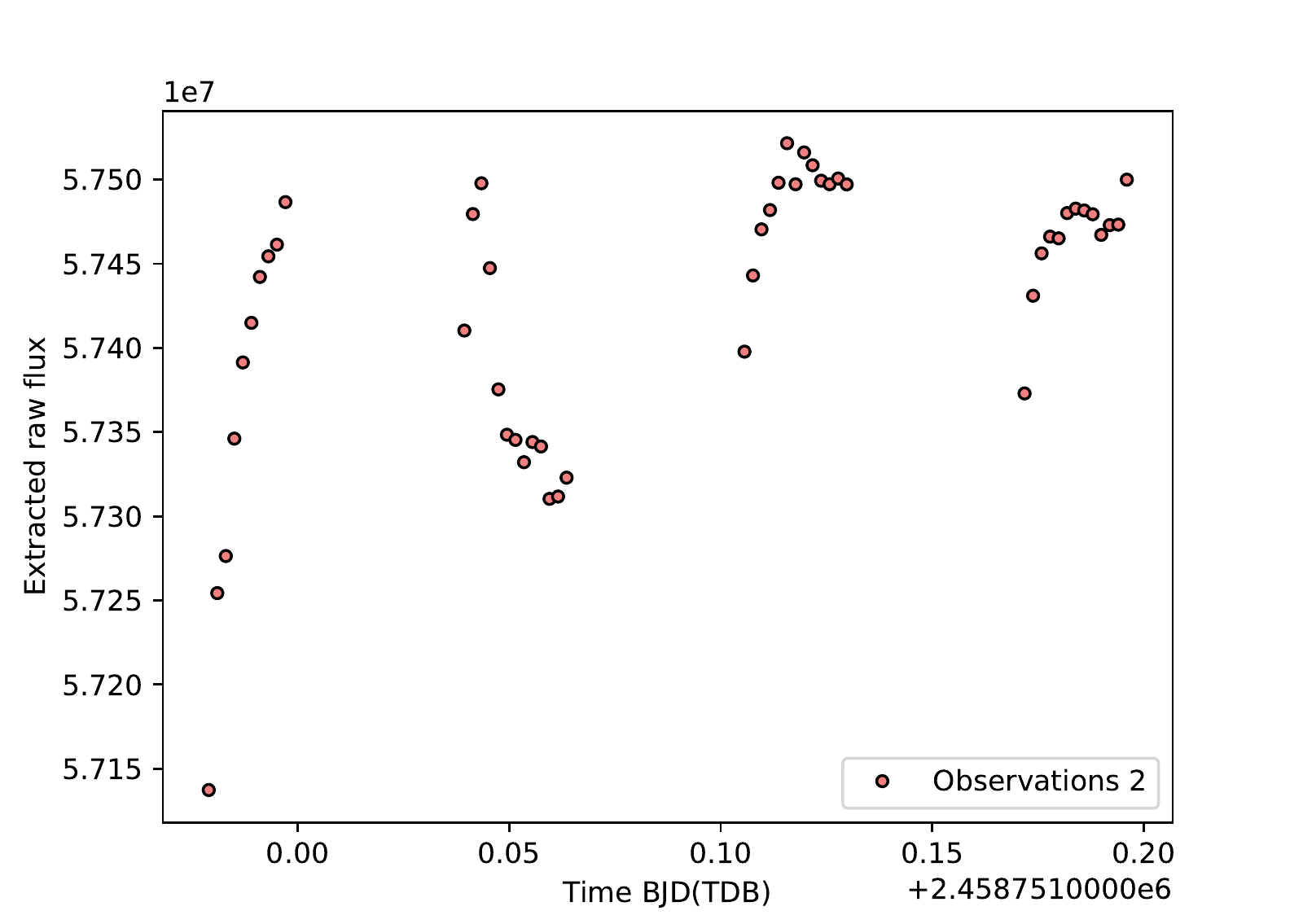}
    ~
    \includegraphics[width=\columnwidth]{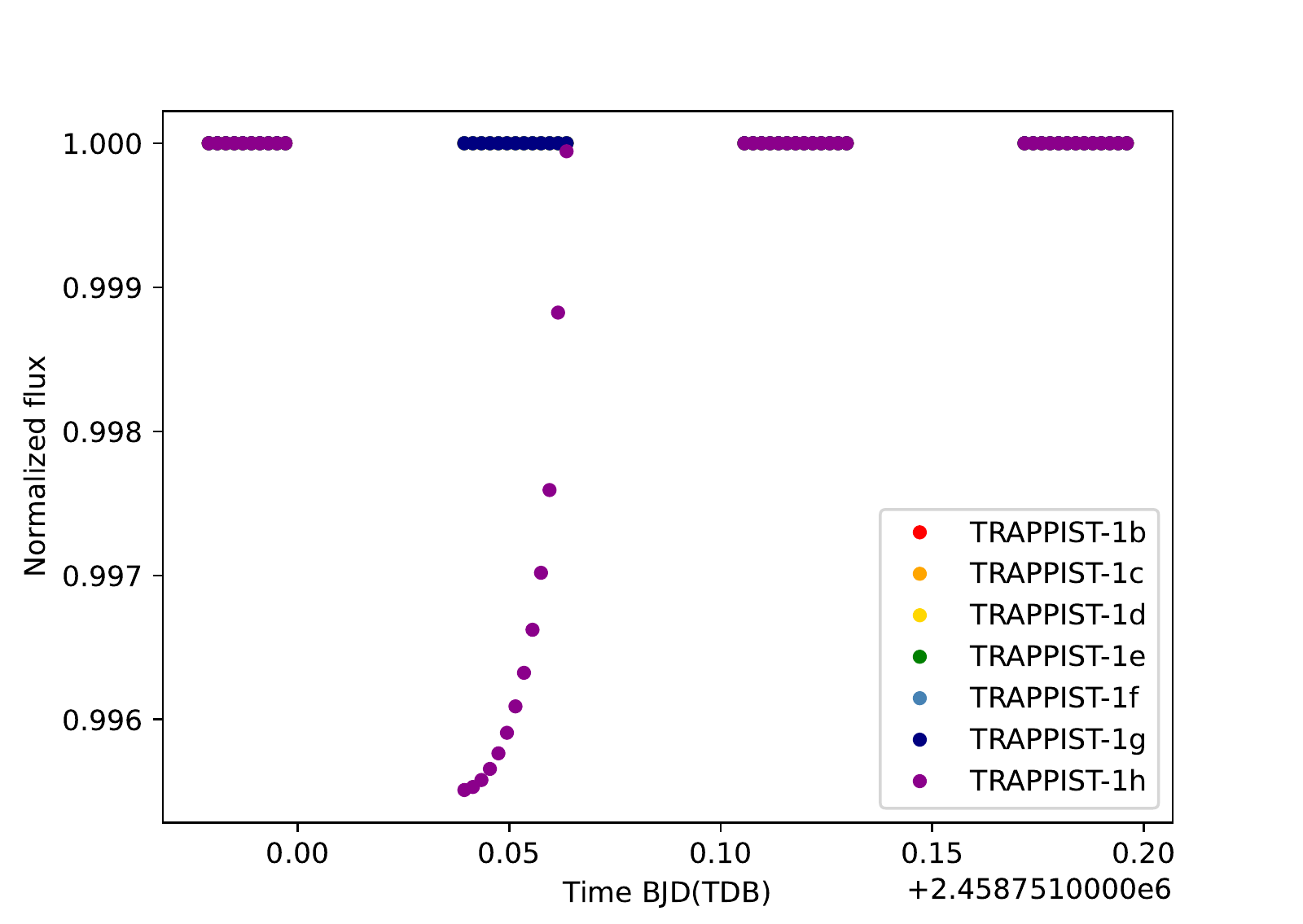}
    
    \includegraphics[width=\columnwidth]{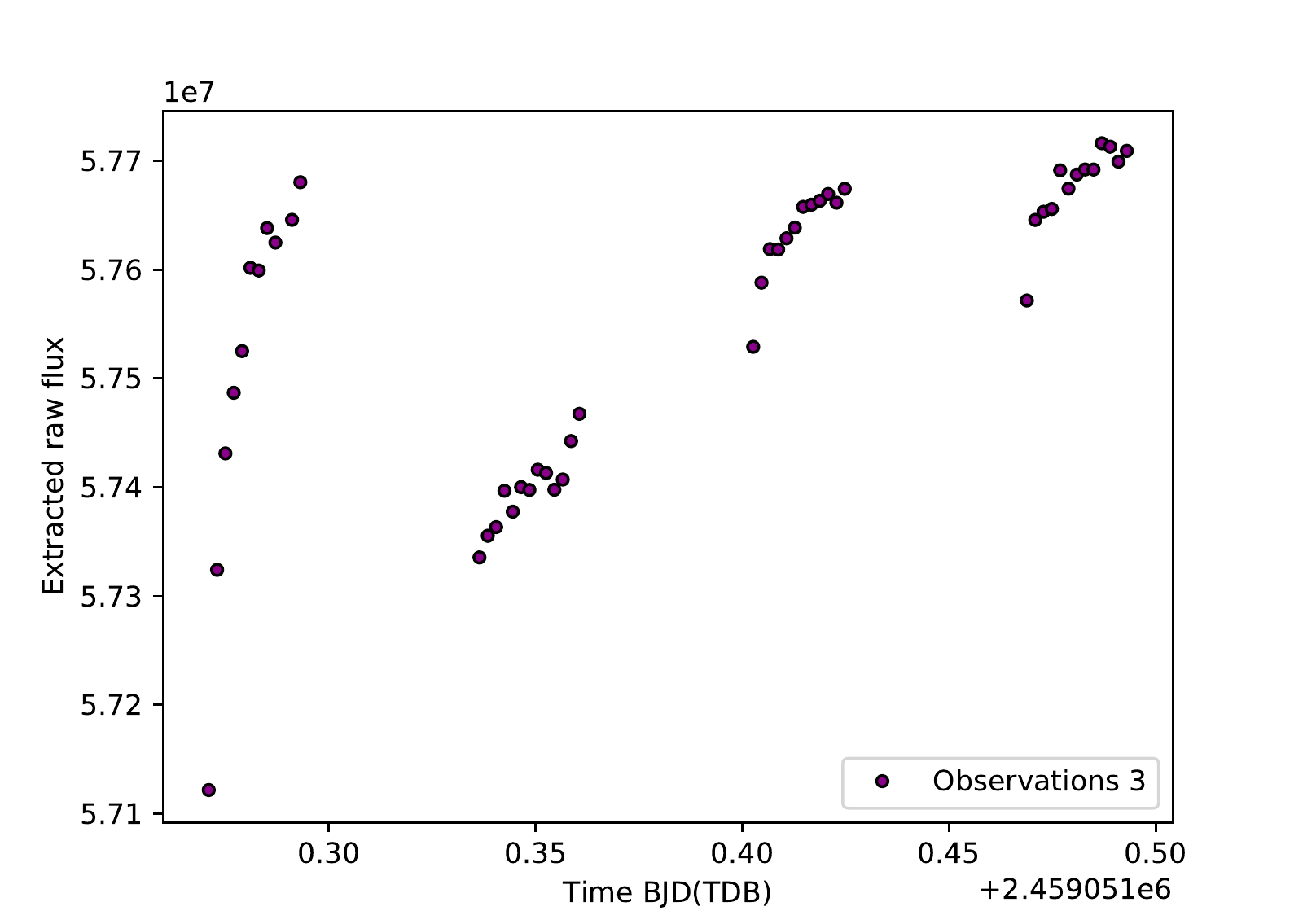}
    ~
    \includegraphics[width=\columnwidth]{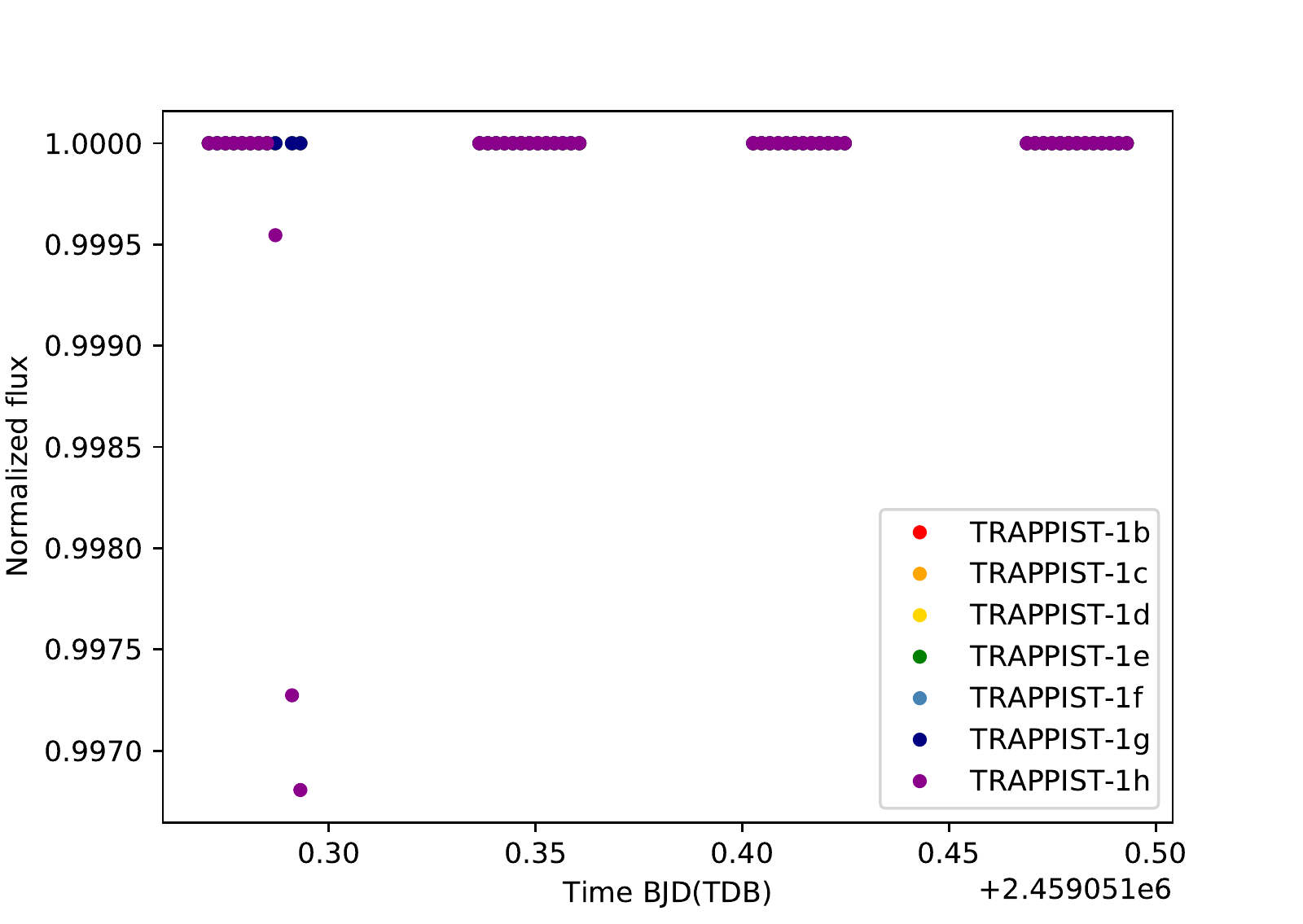}
    \caption{Left: Raw extracted light curves for the TRAPPIST-1h observations (top: Observations 1 July 2017, middle: Observations 2 September 2019, and bottom Observations 3: July 2020). Right: Predicted planetary transits using PyLightcurve transits model and \citet{Gillon_2017} system parameters at the time of the observations.}
    \label{appendix:predicted_transits}
\end{figure*}

\begin{figure*}
    \centering
    \includegraphics[width=12cm]{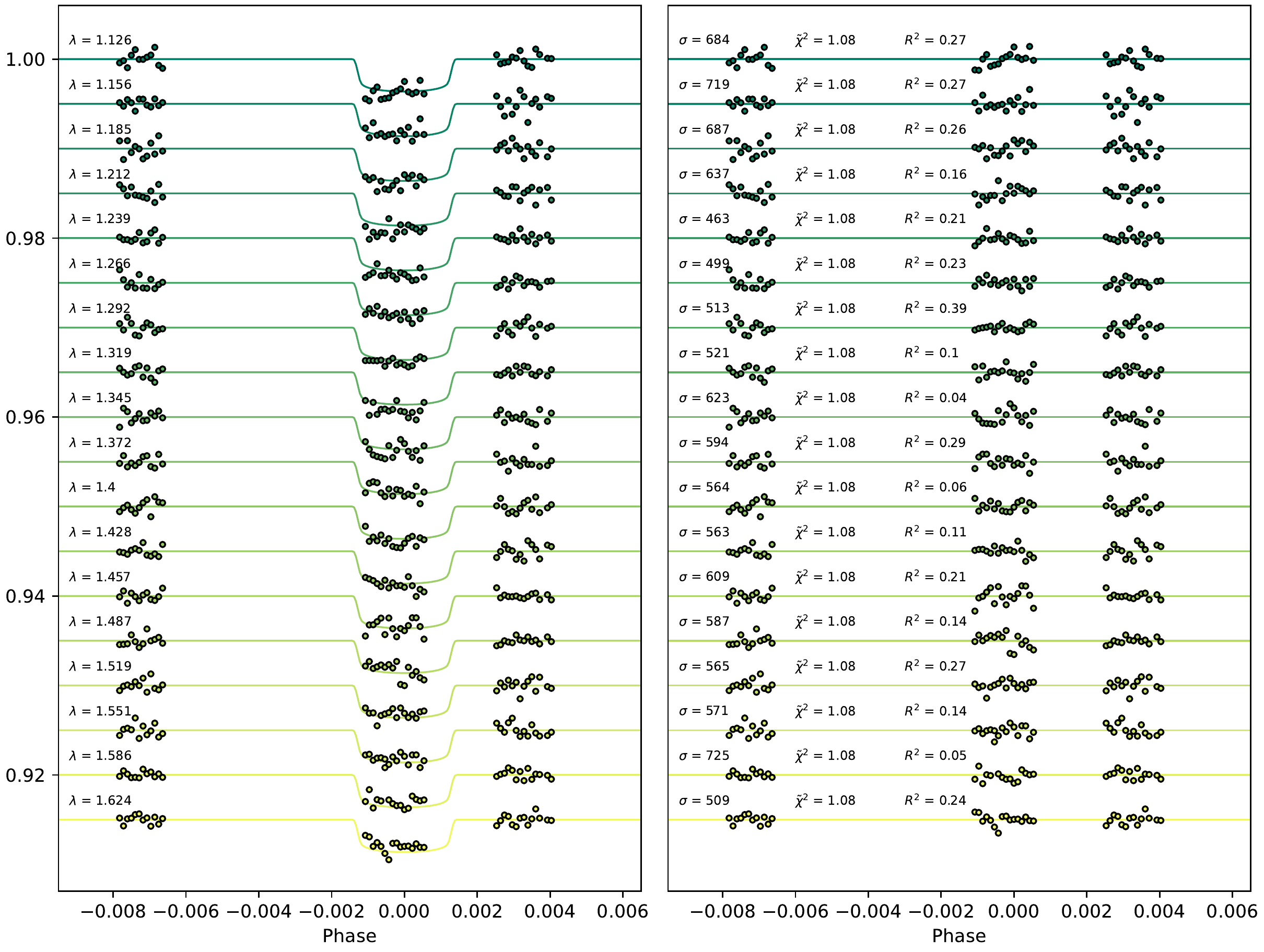}
    
    \includegraphics[width=12cm]{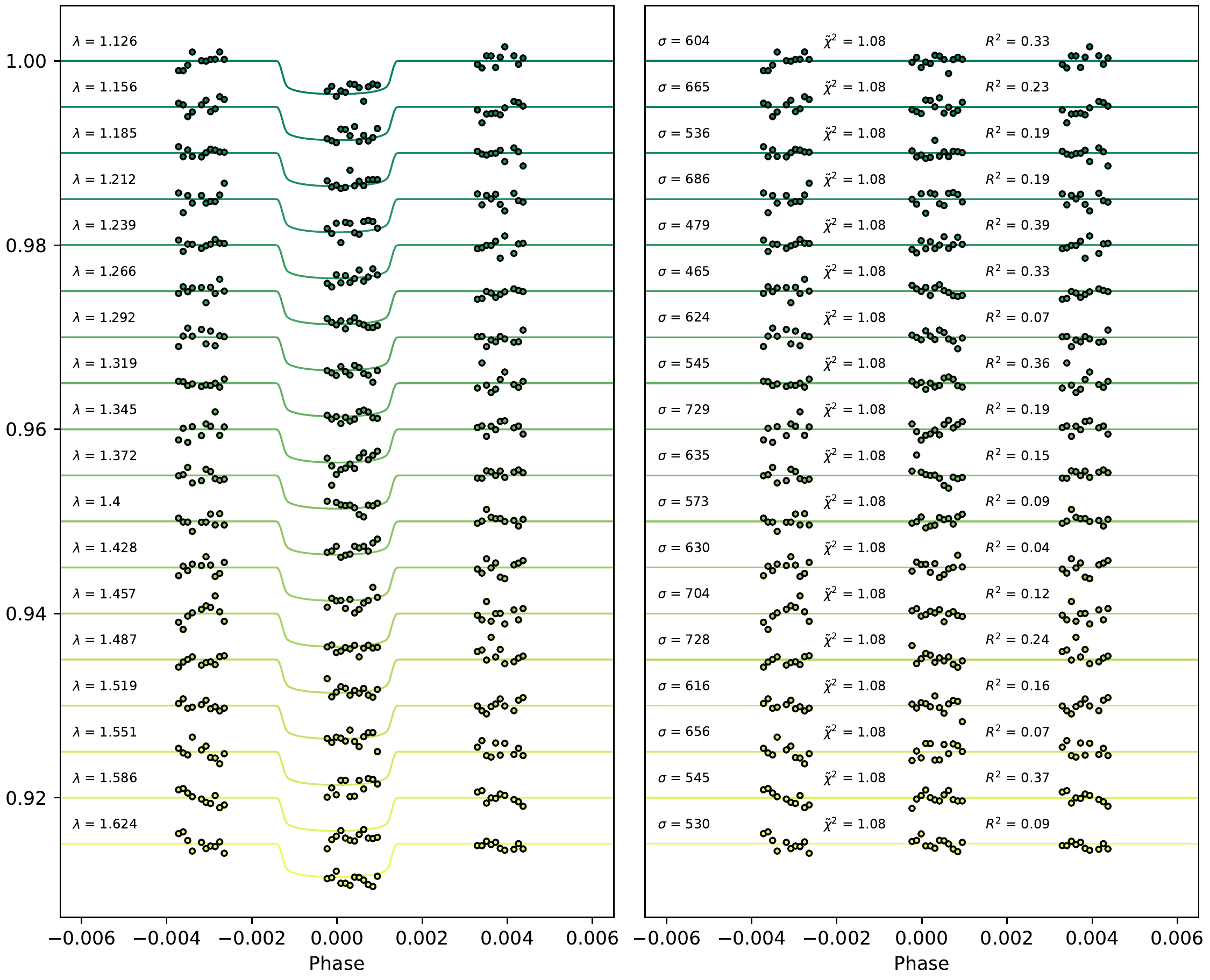}

    \caption{Spectral light curve fits of the July 2017 (top) and 2020 (bottom) visits (Observation 1 and 3) for the transmission spectra of TRAPPIST- 1H. An artificial offset in the y-axis has been applied for clarity. For each light curve, the left panel shows the de-trended spectral light curves with the best fit model in dotted lines with the centred wavelength, and the right panel shows the residuals and values for the standard deviation ($\sigma$) in ppm, the reduced Chi-squared ($\tilde{\chi}^2$), and the auto-correlation (R$^2$).}
    \label{appendix:spcl_obs2_obs3}
\end{figure*}

\begin{figure*}
    \centering
    \includegraphics[width=12cm]{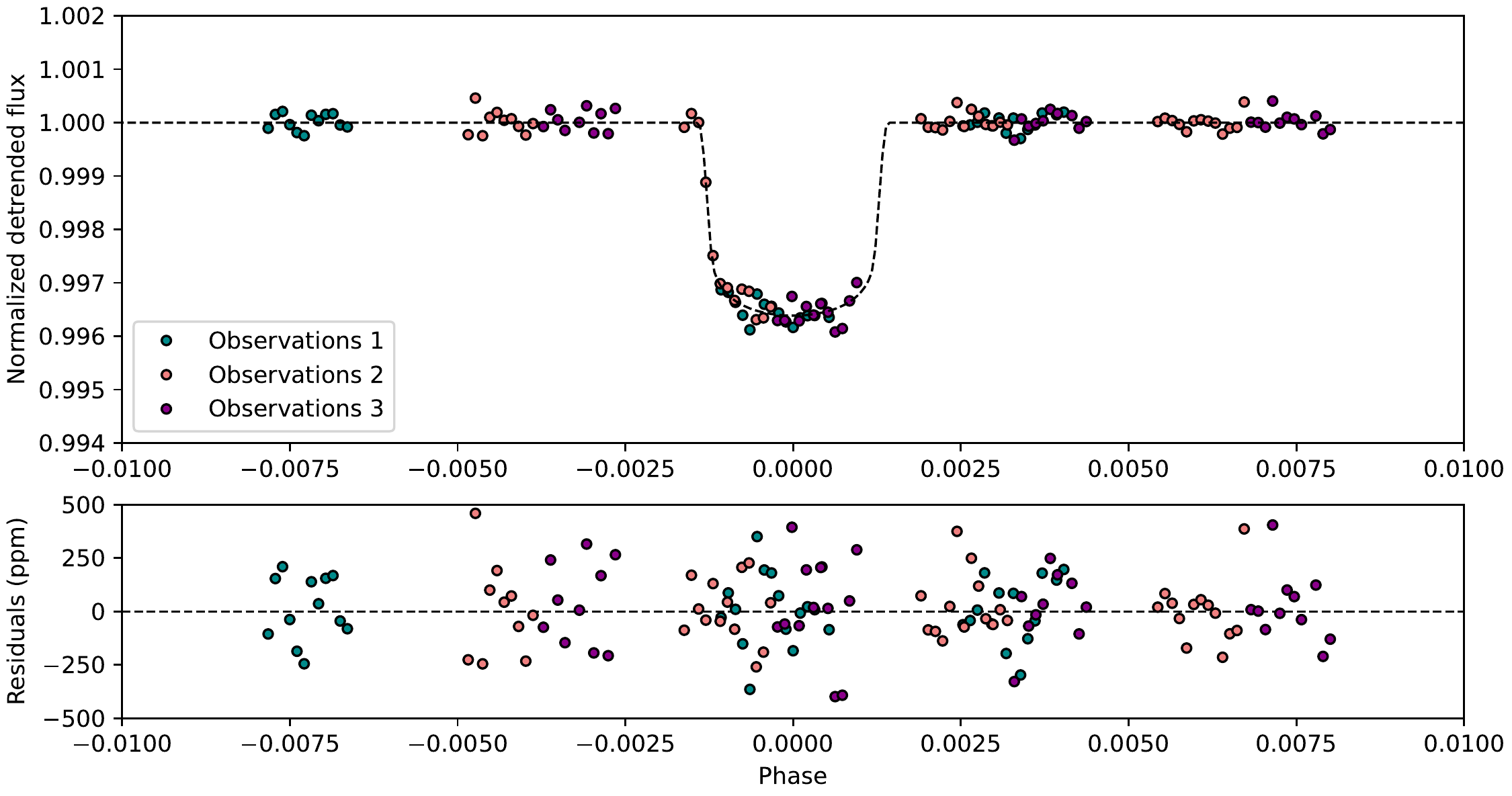}

    \caption{White light curve fits for the three visits on TRAPPIST-1h (top). The transit model (dotted line) was simulated using a weighted mean of the three observations' planet-to-star radius ratios, i.e. 0.0572. }
    \label{appendix:wlc_spcl_all}
\end{figure*}

\begin{figure*}[htpb!]
    \centering
    \includegraphics[width=17cm]{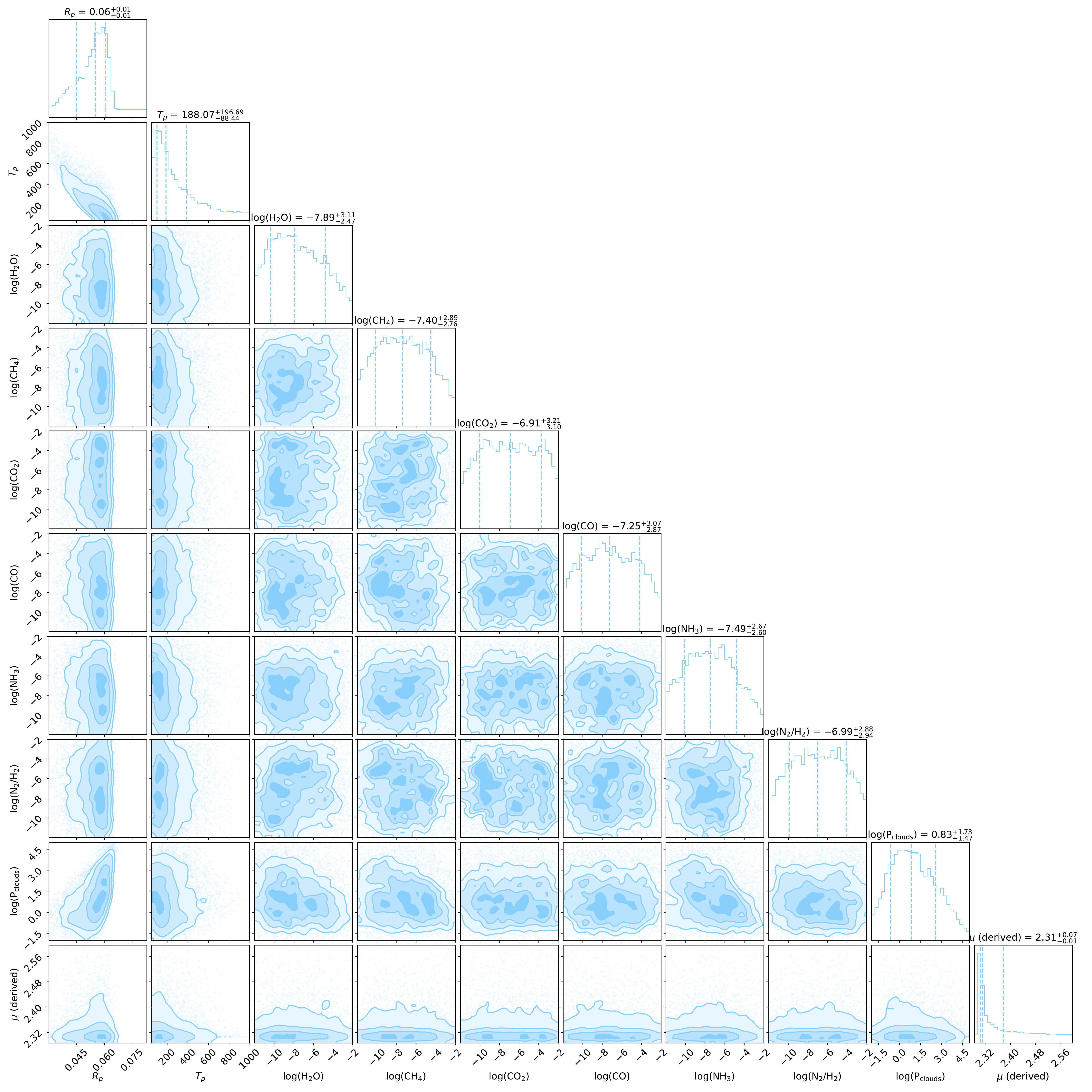}
    \caption{Posterior distributions for the primary atmospheric retrieval on the extracted TRAPPIST-1h spectrum.}
    \label{appendix:posteriors_primary}
\end{figure*}

\begin{figure*}[htpb!]
    \centering
    \includegraphics[width=12cm]{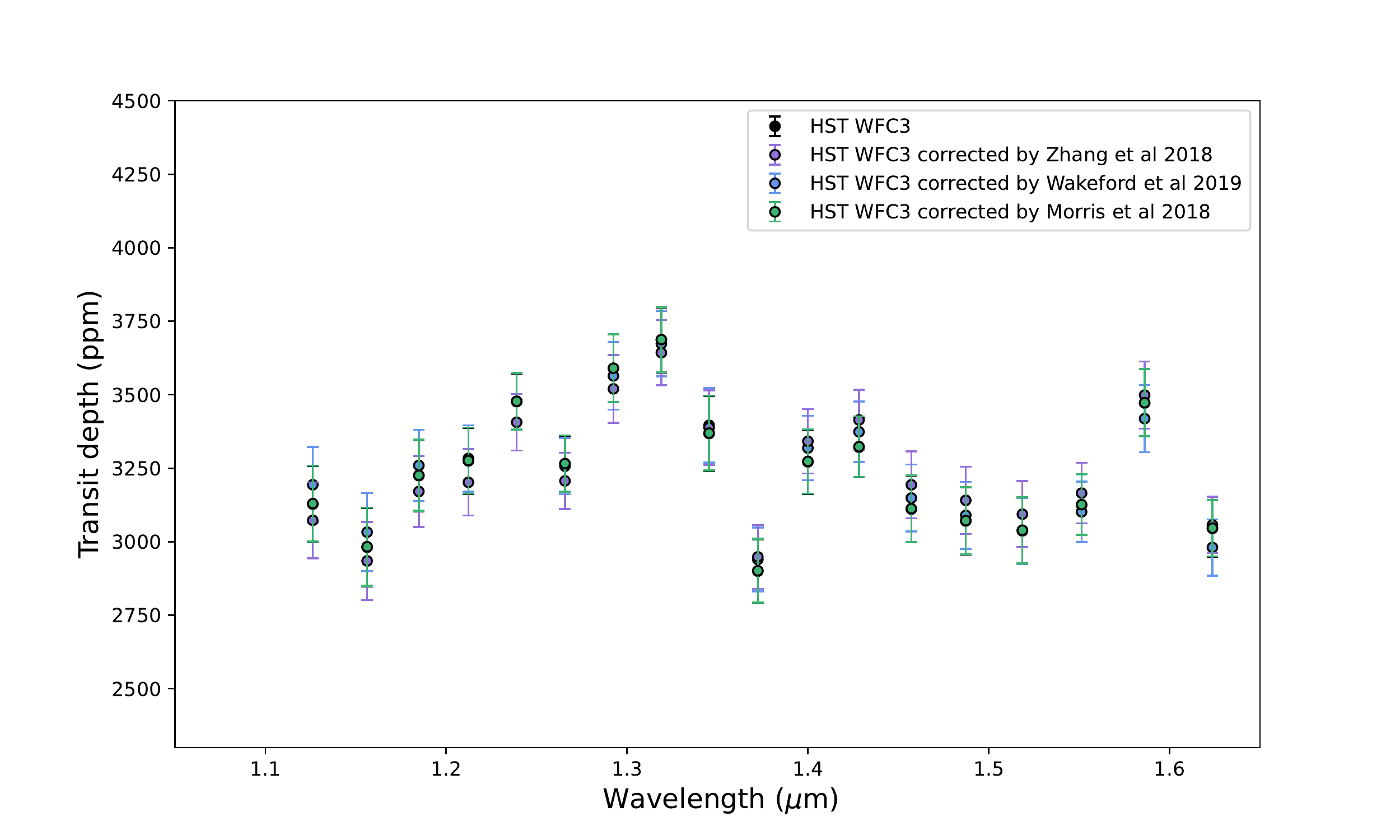}
    \caption{Combined transit depth in ppm using the three HST/WFC3 G141 transit observations and corrected transit depth using stellar contamination models from \citet{Zhang_2018} (purple), \citet{Wakeford_2019} (blue) , and \citet{Morris_2018a} (green). The latter (in green) and the raw extracted spectrum (in black) are almost similar because the stellar
contribution of \citet{Morris_2018a} is flat in the HST/WFC3 NIR wavelength range. }
    \label{appendix:corrected_spectra}
\end{figure*}

\begin{figure*}[htpb!]
    \centering
    \includegraphics[width=12cm]{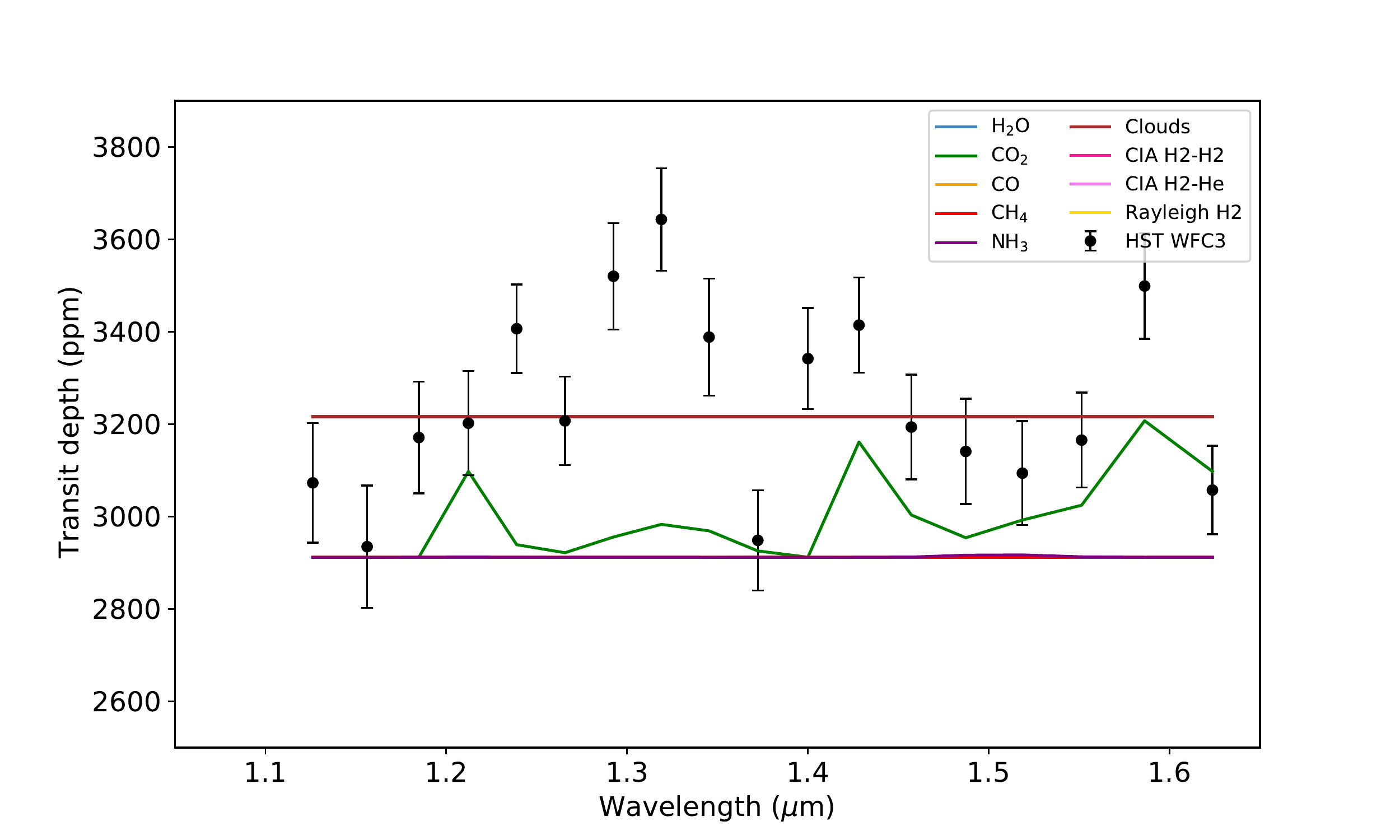}
    
    \includegraphics[width=12cm]{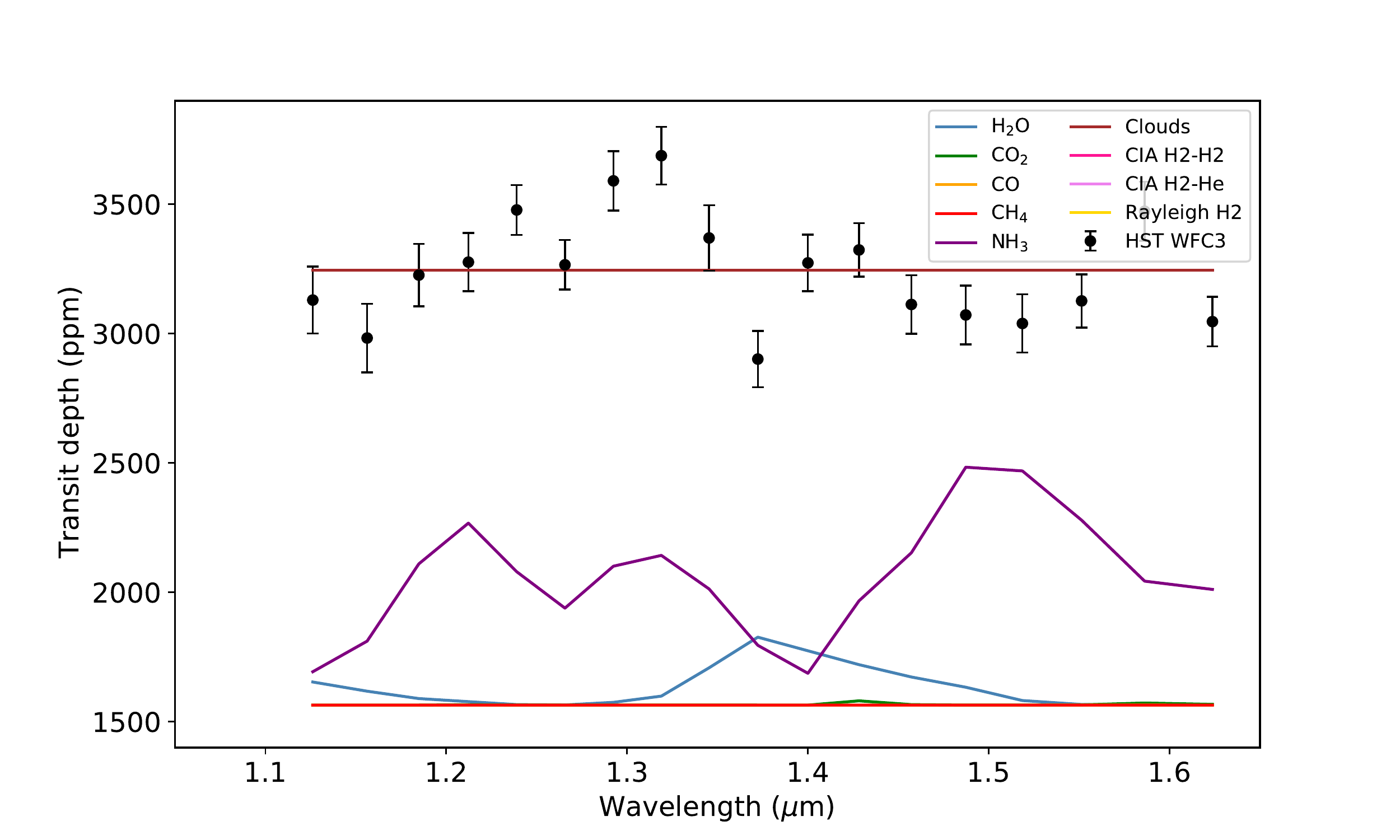}
    \caption{ Opacity contributions from the primary best-fit retrieval model on the TRAPPIST-1h spectrum corrected by \citet{Zhang_2018} (top) and by \citet{Morris_2018a} (bottom). We omitted Rayleigh scattering opacity contributions from others species than H$_2$ for clarity.}
    \label{appendix:contributions}
\end{figure*}

\begin{figure*}[htpb!]
    \centering
    \includegraphics[width=11cm]{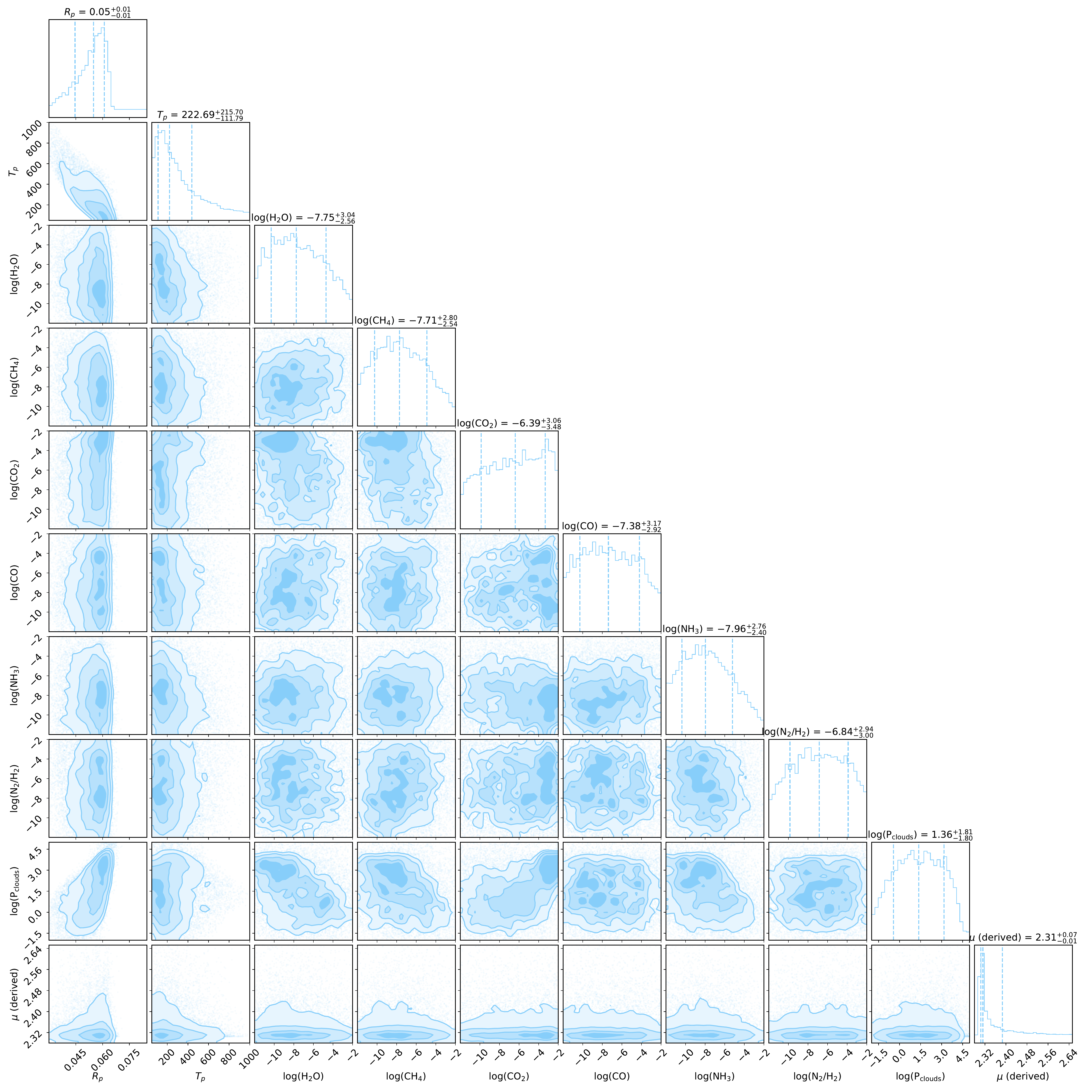}
    \includegraphics[width=11cm]{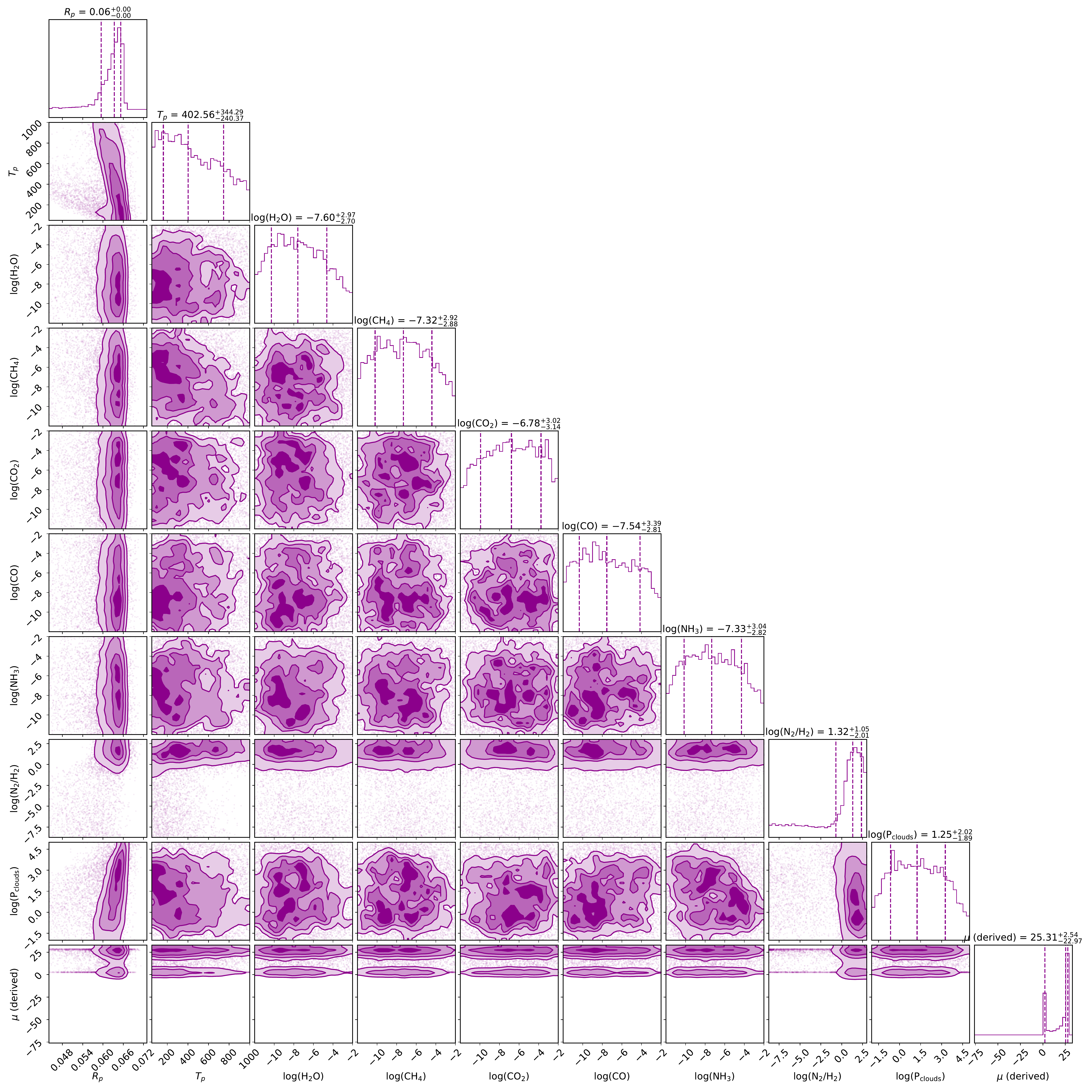}
    \caption{Posterior distributions for the primary atmospheric retrieval (top) and the secondary retrieval (bottom) on TRAPPIST-1h spectra corrected by the  stellar contamination model of \citet{Zhang_2018}.}
    \label{appendix:posteriors_taurex_Zhang}
\end{figure*}

\begin{figure*}[htpb!]
    \centering
    \includegraphics[width=11cm]{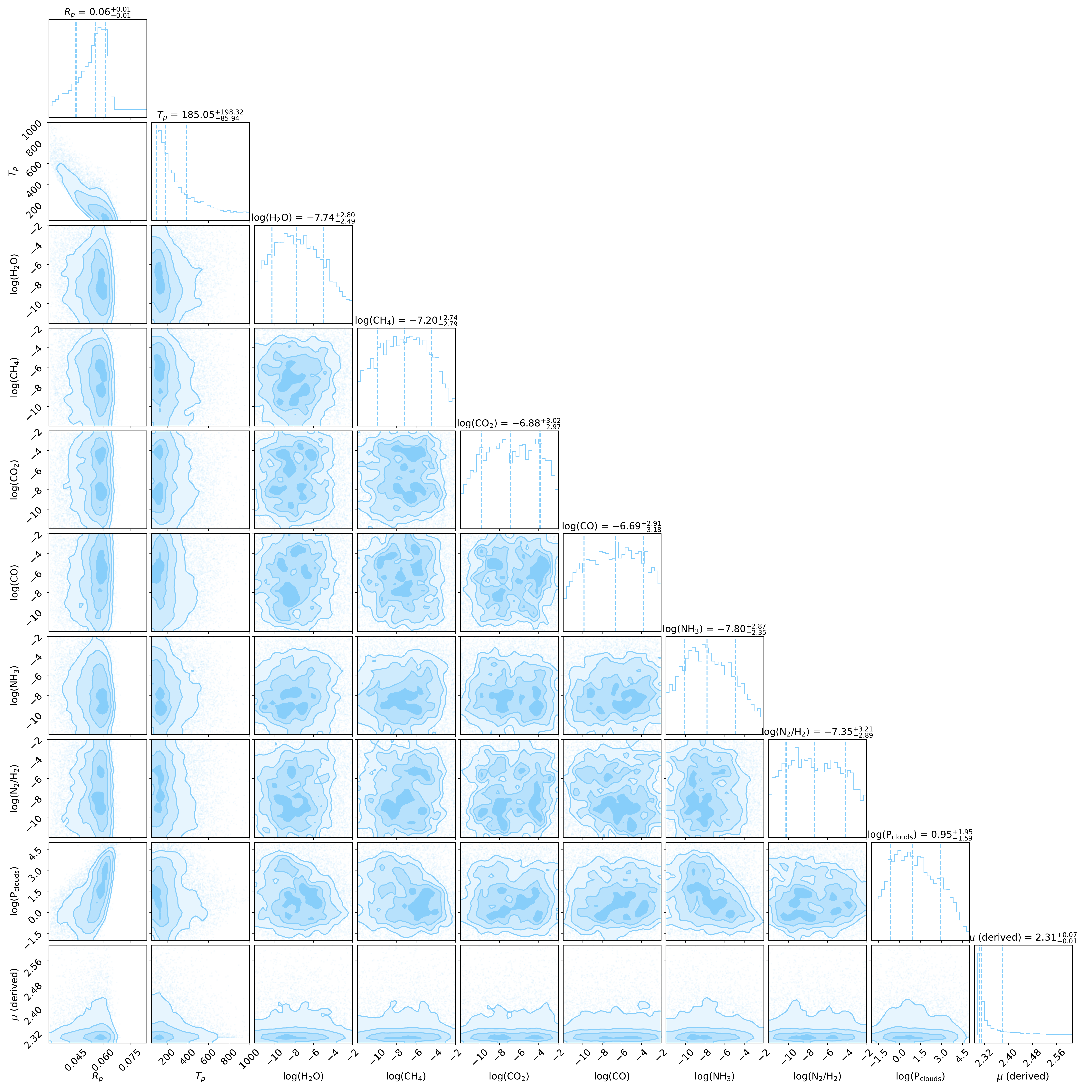}
    
    \includegraphics[width=11cm]{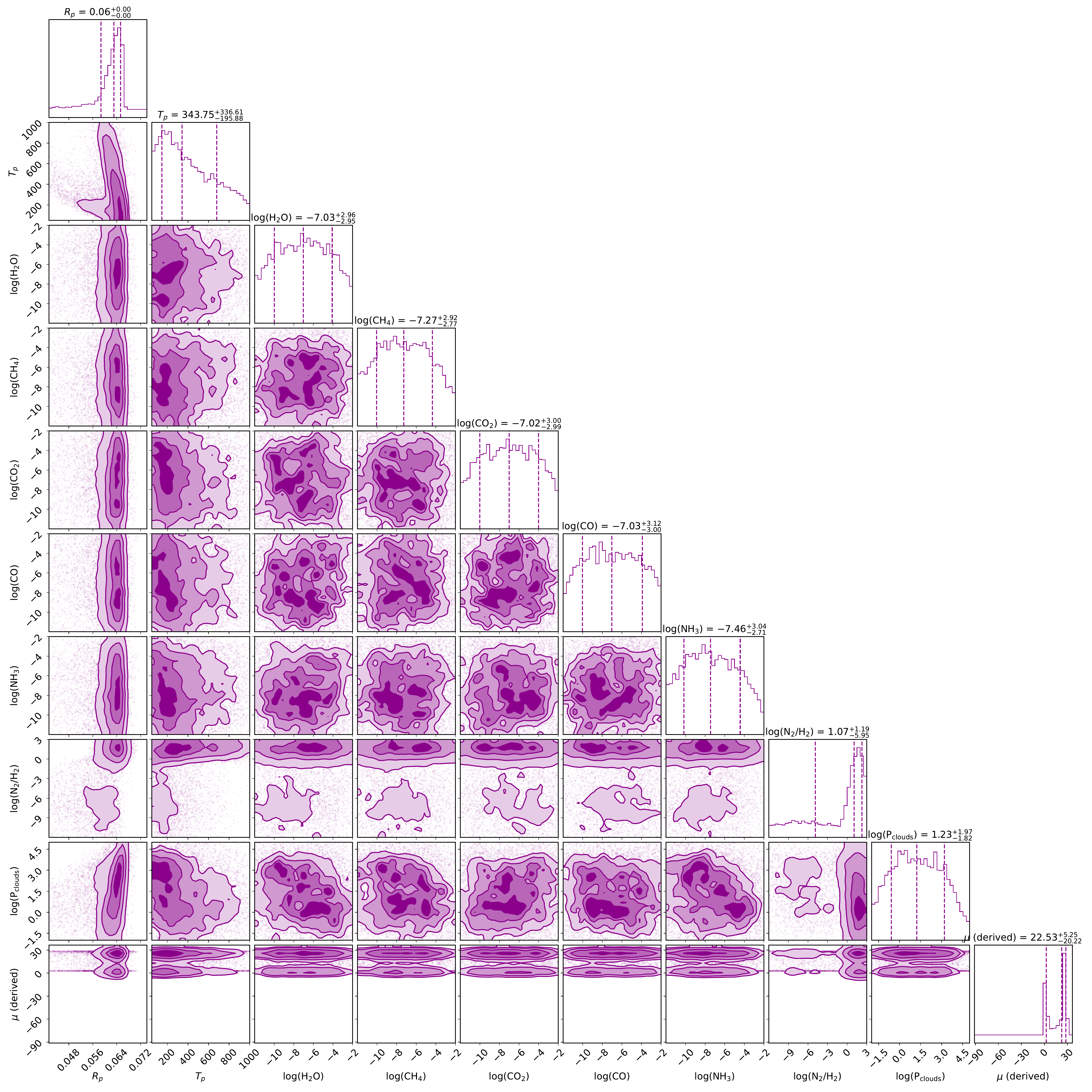}
    \caption{Posterior distributions for the primary atmospheric retrieval (top) and the secondary retrieval (bottom) on TRAPPIST-1h spectra corrected by the  stellar contamination model of \citet{Wakeford_2019}.}
    \label{appendix:posteriors_taurex_Wakeford}
\end{figure*}

\begin{figure*}[htpb!]
    \centering
    \includegraphics[width=11cm]{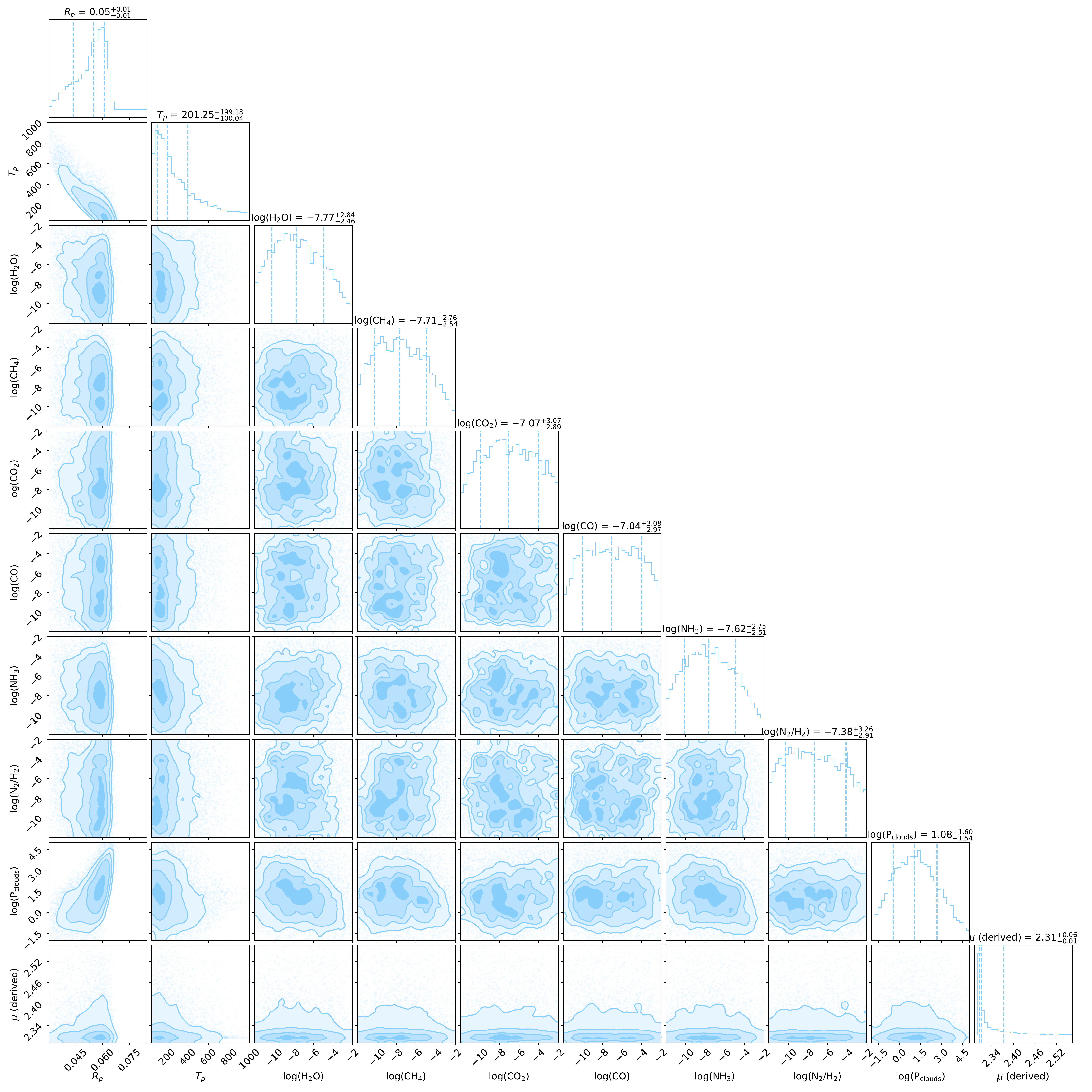}
    
    \includegraphics[width=11cm]{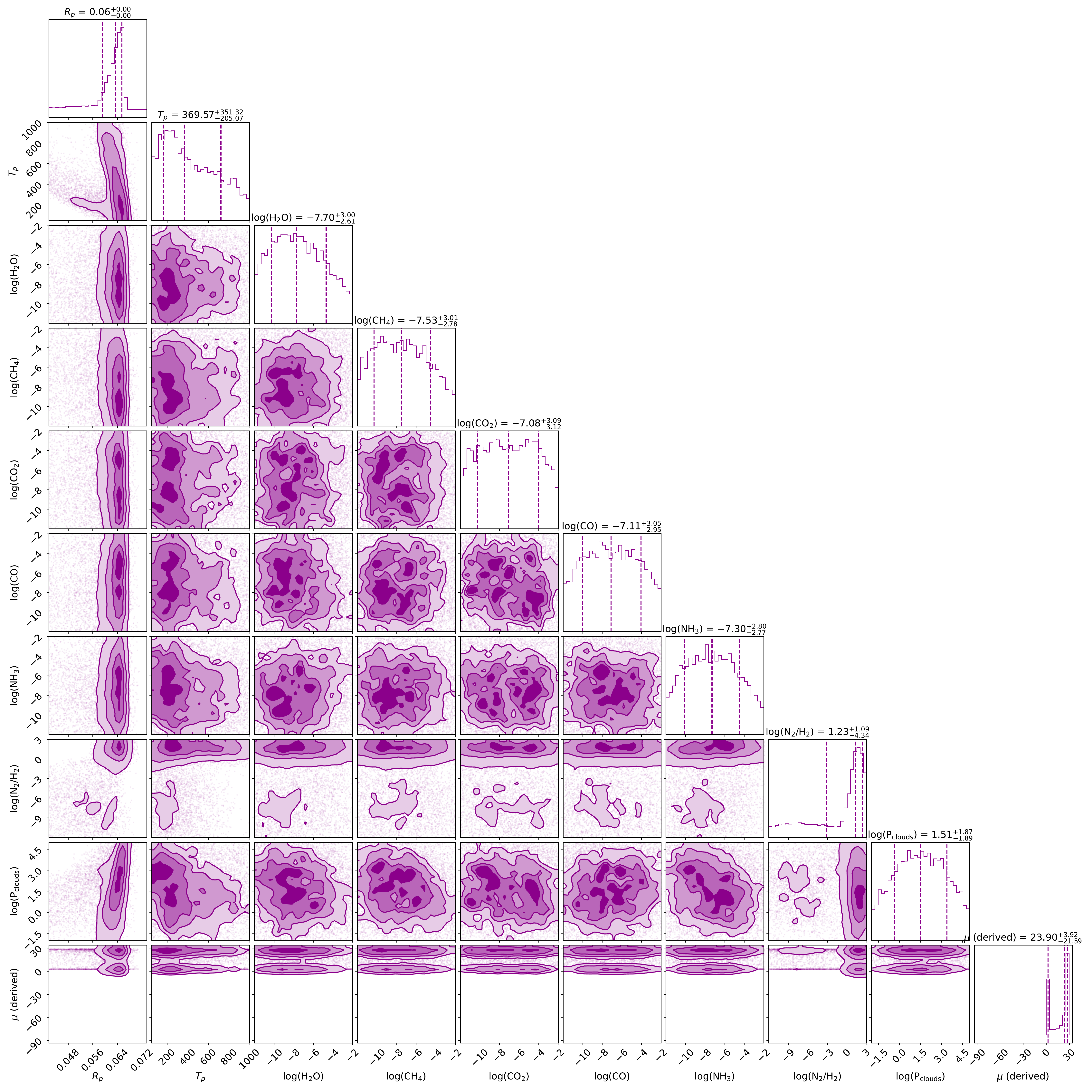}
    \caption{Posterior distributions for the primary atmospheric retrieval (top) and the secondary retrieval (bottom) on TRAPPIST-1h spectra corrected by the  stellar contamination model of \citet{Morris_2018a}.}
    \label{appendix:posteriors_taurex_Morris}
\end{figure*}

\begin{figure*}[htpb!]
    \centering
    \includegraphics[width=12cm]{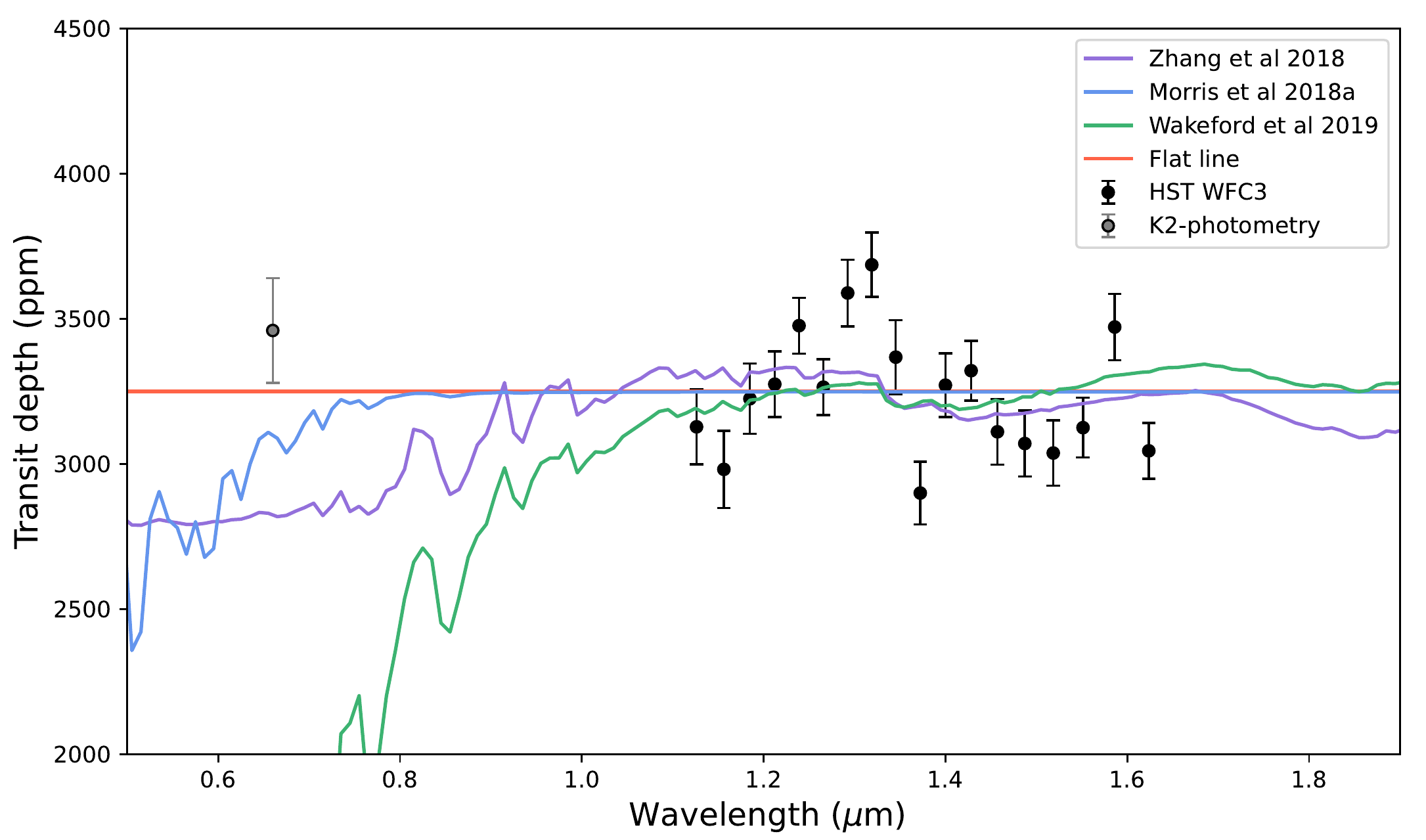}
    
    \caption{Best fit models to TRAPPIST-1 h HST/WFC3 G141 data (black) and K2 optical value \citep{Luger_2017a} (grey) from stellar contamination models based on \citet{Zhang_2018},\citet{Wakeford_2019}, and \citet{Morris_2018a} .}
    \label{appendix:stellar_contamination_K2}
\end{figure*}

\begin{figure*}
    \centering
    \includegraphics[width =12cm]{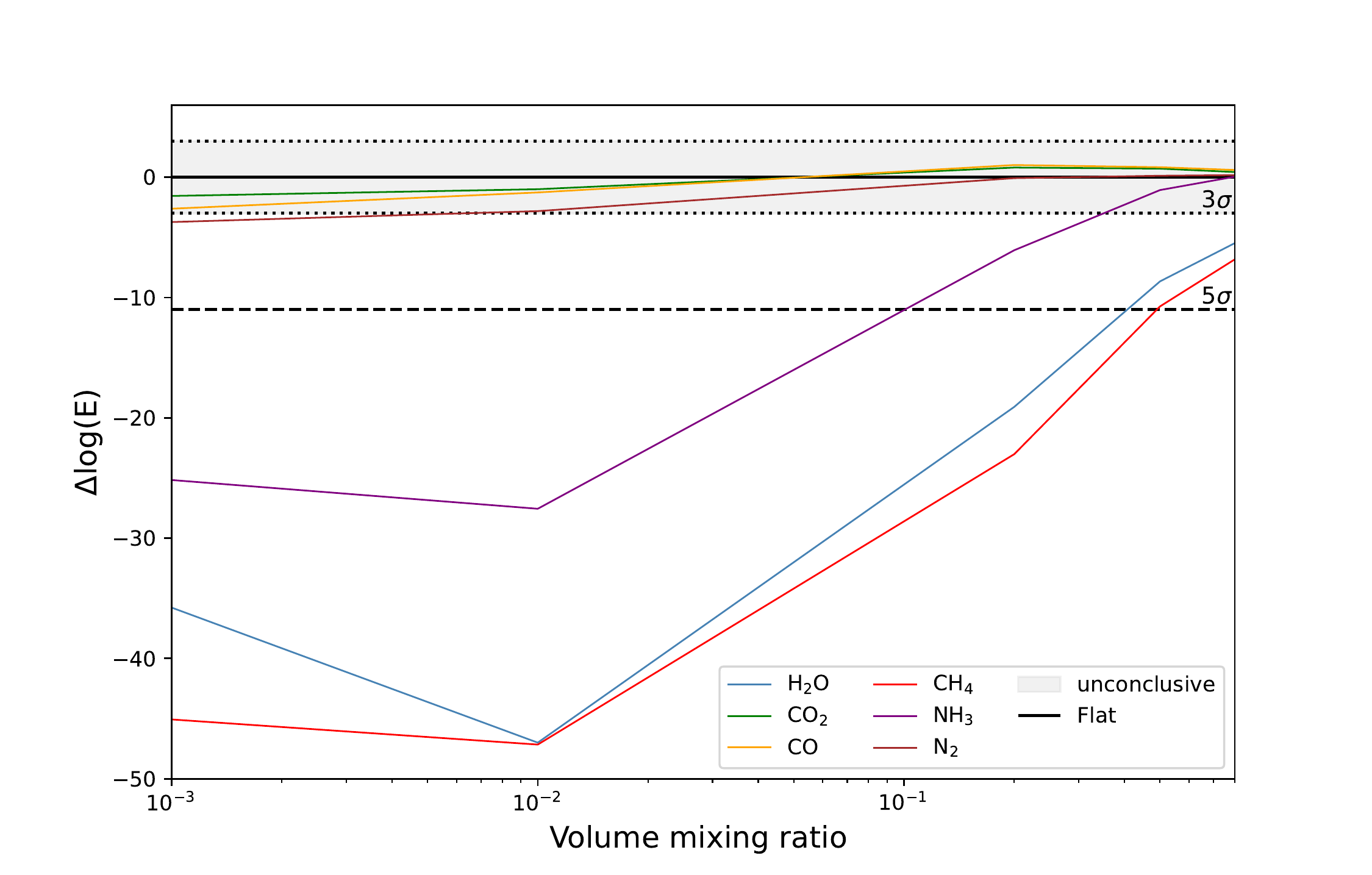}
    \caption{Comparison of the log evidence for a flat line to that of single molecule retrievals where the abundance of the molecule is fixed and no clouds were included. The shaded region represents the set of Bayes factor values for which it is not possible to conclude compared to a flat line, that is with absolute $\Delta$log(E) below 3. Models below the dashed lines are strongly disfavoured compared to the flat line. }
    \label{appendix:delta_logE}
\end{figure*}
\end{appendix}

\end{document}